%% file: main_paper.tex
\documentclass[journal,final,letterpaper,twoside,twocolumn]{IEEEtran}

\input{preamble_article}

\usepackage{calc}
\usepackage[dvipsnames]{xcolor}

\usepackage{tikz}
\tikzset{data/.style={draw,rectangle,rounded corners = 3pt,thick,fill=black!25}}
\tikzset{hp/.style={draw,rectangle}} 
\tikzset{mlink/.style={very thick,->,>=stealth’}}
\tikzset{hlink/.style={<->,>=stealth’,thick,dashed}}

\newcommand{\pa}[1]{\textcolor{black}{#1}}

\input{notations}

\title{A Hierarchical Bayesian Model Accounting for Endmember Variability and Abrupt Spectral Changes to Unmix Multitemporal Hyperspectral Images}

\author{Pierre-Antoine Thouvenin,~\IEEEmembership{Member,~IEEE}, Nicolas Dobigeon,~\IEEEmembership{Senior Member,~IEEE} and Jean-Yves~Tourneret,~\IEEEmembership{Senior Member,~IEEE}

\thanks{This work was supported in part by the Hypanema ANR Project no. ANR-12-BS03-003, by the MapInvPlnt ERA-NET MED Project no. ANR-15-NMED-0002-02, by the Thematic Trimester on Image Processing of the CIMI Labex under Grant ANR-11-LABX-0040-CIMI within the Program ANR-11-IDEX-0002-02 and by the Direction G{\'e}n{\'e}rale de l'Armement, French Ministry of Defence.}
\thanks{The authors are with the University of Toulouse, IRIT/INP-ENSEEIHT, 31071 Toulouse, France. (e-mail: \{pierreantoine.thouvenin, Nicolas.Dobigeon, Jean-Yves.Tourneret\}@enseeiht.fr}
\thanks{This paper has supplementary downloadable material available at http://ieeexplore.ieee.org., provided by the authors. The material includes complementary illustrations to the main paper, as well as further details on the values chosen for the hyperparameters involved in the experiments. Contact pierreantoine.thouvenin@enseeiht.fr for further questions about this work.}}

\begin{document}
\setlength{\textfloatsep}{5pt} 
\setlength{\intextsep}{5pt}
\setlength{\abovedisplayskip}{4pt}
\setlength{\belowdisplayskip}{4pt}
\maketitle
\begin{abstract}
Hyperspectral unmixing is a blind source separation problem which consists in estimating the reference spectral signatures contained in a hyperspectral image, as well as their relative contribution to each pixel according to a given mixture model. In practice, the process is further complexified by the inherent spectral variability of the observed scene and the possible presence of outliers. More specifically, multi-temporal hyperspectral images, i.e., sequences of hyperspectral images acquired over the same area at different time instants, are likely to simultaneously exhibit moderate endmember variability and abrupt spectral changes either due to outliers or to significant time intervals between consecutive acquisitions. Unless properly accounted for, these two perturbations can significantly affect the unmixing process. In this context, we propose a new unmixing model for multitemporal hyperspectral images accounting for smooth temporal variations, construed as spectral variability, and abrupt spectral changes interpreted as outliers. The proposed hierarchical Bayesian model is inferred using a Markov chain Monte-Carlo (MCMC) method allowing the posterior of interest to be sampled and Bayesian estimators to be approximated. A comparison with unmixing techniques from the literature on synthetic and real data allows the interest of the proposed approach to be appreciated.
\end{abstract}
\begin{IEEEkeywords}
Hyperspectral imagery, multitemporal images, endmember variability, Markov chain Monte-Carlo (MCMC) methods.
\end{IEEEkeywords}

\section{Introduction} \label{sec:intro}
\IEEEPARstart{A}{cquired} in hundreds of contiguous spectral bands, hyperspectral (HS) images have received an increasing interest due to the significant spectral information they convey, which is somewhat mitigated by their lower spatial resolution in remote sensing applications. This limitation, combined with possibly complex interactions between the incident light and the observed materials, implies that the observed spectra are mixtures of several signatures corresponding to distinct materials. Spectral unmixing then consists in identifying a limited number of reference spectral signatures composing the data -- referred to as \emph{endmembers} -- and their abundance fractions in each pixel according to a predefined mixture model. The choice of a specific model generally reflects the practitioners' prior knowledge on the environmental factors possibly affecting the acquisitions, such as declivity or multiple reflections. Traditionally, a linear mixing model (LMM) is adopted since it is appropriate to describe hyperspectral data when the declivity of the scene and microscopic interactions between the observed materials are negligible \cite{Bioucas2012jstars}. Depending on the applications, various models have also been investigated to capture higher order interactions (i.e., nonlinearities) between the incident light and the observed materials (see \cite{Dobigeon2014,Heylen2014b} for recent reviews on this topic). However, varying acquisition conditions, such as local illumination variations or the natural evolution of the scene, may significantly alter the shape and the amplitude of the acquired signatures \cite{Somers2011,Zare2014IEEESPMAG}, thus affecting the extracted endmembers. Endmember variability has hitherto been extensively considered within a single HS image, either in a deterministic \cite{Somers2012jstars2,Vengazones2014WIHSPERS,Uezato2016,Thouvenin2015} or a statistical setting \cite{Eches2010ip,Du2014,Halimi2015}.

Recent works also considered temporal variability by exploiting the possibilities offered by multitemporal HS (MTHS) images \cite{Halimi2015eusipco,Thouvenin2015b}. From a hyperspectral unmixing perspective, MTHS images, \ie, sequences of HS images acquired over the same area at different time instants, can be of interest to exploit information redundancy between consecutive images (\eg, through features exhibiting moderate or smooth temporal variations as in \cite{Halimi2015icassp,Halimi2016tgrs}) while allowing the endmember temporal evolution to be characterized. For instance, MTHS have been recently exploited to improve endmember unmixing results \cite{Henrot2016,Halimi2015eusipco,Thouvenin2015b} and used in a change detection problem involving two HS images \cite{Erturk2015,Liu2016}.

Even though the approaches proposed in \cite{Henrot2016,Halimi2015eusipco,Thouvenin2015b} specifically allow smooth temporal variations of some of the mixture parameters to be considered, they do not account for abrupt spectral changes either due to outliers or to possibly significant time intervals between two consecutive images. In practice, such situations can be reasonably expected, depending on the acquisition dates and possible climatic hazards, \eg, when vegetation or water is present in the observed scene. Unless specifically accounted for, this situation frequently observed in real datasets has a significant impact on the recovered endmembers, which motivates the present work.
Inspired by \cite{Halimi2015eusipco,Altmann2015tci,Chenot2015}, and based on an original interpretation of the unmixing problem under study, our contribution consists in jointly accounting for smooth endmember variations -- construed as temporal endmember variability -- and abrupt changes interpreted as outliers (\eg, significant variability within a single image or presence of non-linearities) using a carefully designed hierarchical Bayesian model. More precisely, we focus our analysis on scenes in which mostly the same materials are expected to be observed from an image to another. In this context, using the endmembers extracted from the reference scene as a starting point to unmix the whole MTHS image constitutes a reasonable attempt to generalize the analyses previously conducted for a single image. On the one hand, the endmembers identified in each single image can \emph{in fine} be considered as time-varying instances of reference signatures shared by the different images, thus justifying the use of a modified version of the perturbed linear mixing model (PLMM) proposed in \cite{Thouvenin2015b}. This formulation will notably allow smooth spectral variations occurring over time to be captured, leading to competitive results when compared to methods analyzing the images individually. On the other hand, the signatures corresponding to materials appearing in only a few images, which induce abrupt spectral changes, can be regarded as outliers with respect to the commonly shared endmembers. This paper studies a new Bayesian model allowing both spectral variability and presence of outliers to be considered in the unmixing of MTHS images. The resulting unmixing task is solved using a Markov chain Monte-Carlo (MCMC) allowing the posterior of interest to be sampled and Bayesian estimators to be approximated.

The paper is organized as follows. The mixing model considered in this paper is introduced in Section \ref{sec:PLMM}, and the associated hierarchical Bayesian model is developed in Section \ref{sec:bayes}.
Section \ref{sec:gibbs} investigates a Gibbs sampler to solve the resulting mixed integer non-linear problem. The performance of the proposed approach on synthetic and real data is studied in Sections \ref{sec:exp} and \ref{sec:expr}. In particular, the results obtained with the proposed method are compared to those of the VCA/FCLS algorithm \cite{Nascimento2005,Bioucas2010}, the SISAL/FCLS algorithm \cite{Bioucas2009}, the algorithm associated with the robust LMM (RLMM) proposed in \cite{Fevotte2015} and the MTHS optimization method \cite{Thouvenin2015b}. Finally, Section \ref{sec:conclusion} concludes this work and outlines further research perspectives. \vspace{-0.2cm}

\section{Problem statement} \label{sec:PLMM}

We consider a sequence of HS images acquired at $\ntime$ different time instants over the same area, where mostly the same materials are expected to be observed over time. In the following, at most $\nendm$ endmembers are assumed to be shared between the $\ntime$ images composing the sequence, where $\nendm$ is \emph{a priori} known. Since the observed instances of a given endmember can be reasonably expected to vary from an image to another, we propose to account for smooth endmember spectral variations via a modified version of the perturbed linear mixing model (PLMM) proposed in \cite{Thouvenin2015,Thouvenin2015b}. Inspired by the total least squares problem \cite{Golub1980}, the PLMM consists in representing each pixel $\mathbf{y}_{nt}$ by a linear combination of the $\nendm$ endmembers -- denoted by $\mathbf{m}_r$ -- affected by an additive error term $\mathbf{dm}_{r,t}$ accounting for temporal endmember variability. However, this model shows notable limitations when the vector $\mathbf{y}_{n,t}$ is affected by abrupt changes. Consequently, this paper investigates a new unmixing model jointly accounting for endmember variability and abrupt changes possibly affecting MTHS images. To this end, the proposed model is a generalized PLMM, which includes an additional term $\x_{n,t}$ to capture significant deviations from the LMM, \ie, significant spatial variability or non-linearities within each image \cite{Fevotte2015,Altmann2015tci}. The resulting observation model can thus be written
\begin{equation}
\y_{n,t}  =  \sum_{r=1}^\nendm a_{r,n,t}\Bigl(\m_r + \dm_{r,t} \Bigr) + \x_{n,t} +  \mathbf{b}_{n,t}
\end{equation}%
for $n = 1,\dotsc , \nbpix$ and $t = 1, \dotsc,\ntime$, where $\y_{n,t}$ denotes the $n$th image pixel at time $t$, $\mathbf{m}_r$ is the $r$th endmember, $a_{r,n,t}$ is the proportion of the $r$th endmember in the $n$th pixel at time $t$, $\dm_{r,t}$ denotes the perturbation of the $r$th endmember at time $t$, and $\x_{n,t}$ denotes the contribution of outliers in the $n$th pixel at time $t$. Finally, $\mathbf{b}_{n,t}$ represents an additive noise resulting from the data acquisition and the modeling errors. The so-called robust PLMM can be written
\begin{equation} \label{eq:model}
\Y_t  = (\M+\dM_t)\A_t + \X_t + \mathbf{B}_t
\end{equation}%
where $\Y_t = \left[ \mathbf{y}_{1,t},\dotsc,\mathbf{y}_{N,t} \right]$ is an $\nband \times \nbpix$ matrix containing the pixels of the $t$th image, $\M$ denotes an $\nband \times \nendm$ matrix containing the endmembers that are common to all the images of the sequence, $\A_t$ is an $\nendm \times \nbpix$ matrix composed of the abundance vectors $\mathbf{a}_{n,t}$, $\dM_t$ is an $\nband \times \nendm$ matrix whose columns contain the variability inherent to the $t$th image, $\X_t$ is an $\nband \times \nbpix$ matrix whose columns are the outliers present in the image $t$, and $\mathbf{B}_t$ is an $\nband \times \nbpix$ matrix accounting for the noise at time $t$. The constraints considered to reflect physical considerations are
\begin{align} \label{eq:constraints}
\begin{split}
\A_t & \succeq \mathbf{0}_{\nendm ,\nbpix}, \quad  \A_t^T \mathbf{1}_\nendm  = \mathbf{1}_\nbpix, \, \forall t \in  \{1,\dotsc , \ntime  \}  \\%
\M & \succeq \mathbf{0}_{\nband ,\nendm}, \quad \M + \dM_t \succeq \mathbf{0}_{\nband ,\nendm}, \, \forall t \in \{ 1,\dotsc , \ntime \} \\
\X_t & \succeq \mathbf{0}_{\nband ,\nbpix},  \quad \forall t \in \{ 1,\dotsc , \ntime \}
\end{split}
\end{align}{}%
where $\succeq$ denotes a term-wise inequality. Note that the outlier term $\X_t$ is intended to describe abrupt changes due for instance to the appearance of one or several new endmembers that were not present in the reference image. This justifies the corresponding non-negativity constraint, similar to the one imposed on the other endmembers. Note however that different phenomena not considered in this work, possibly represented by the terms $\X_t$, can induce a decrease in the total reflectance, e.g., shadowing effects or some nonlinearities as detailed in \cite{Heylen2016b}. To address this case, the non-negativity constraint on the outlier terms $\X_t$ should be removed.

Given the mixture model \eqref{eq:model}, the unmixing problem considered in this work consists in inferring the abundances $\A_t$, the endmembers $\M$, the variability $\dM_t$ and the outliers $\X_t$ from the observations $\Y_t$, $t = 1, \dotsc, \ntime$. In the next section, this problem is tackled in a Bayesian framework to easily incorporate all the prior knowledge available on the mixture parameters.

\section{Bayesian model} \label{sec:bayes}

This section details the specific structure imposed on the parameters to be inferred \emph{via} appropriate prior distributions. Note that dependencies \wrt constant parameters are omitted in the following paragraphs to simplify the notations.

\subsection{Likelihood}
Assuming the additive noise $\mathbf{b}_{n,t}$ is distributed according to a Gaussian distribution $\mathbf{b}_{n,t} \sim \mathcal{N}(\zero{\nband},\sigma_t^2 \I_\nband)$, the observation model \eqref{eq:model} leads to
\begin{equation*} \label{eq:likelihood}
\y_{n,t} \mid \M,\dM_t,\A_t,\X_t,\sigma_t^2 \sim \mathcal{N} \Bigl( (\M + \dM_t)\a_{n,t} + \x_{n,t},\sigma_t^2 \I_\nband \Bigr).
\end{equation*}

In addition, assuming prior independence between the pixels within each image and between the images $\Y_t$ themselves, the likelihood function of all images $\Yb = [\Y_1,\dotsc,\Y_\ntime]$ is
\begin{equation}
\begin{split}
p(\Yb \mid &\Theta) \propto \prod_{t = 1}^\ntime (\sigma_t^2)^{-\nbpix\nband/2} \times \\
&\exp \Bigl( -\frac{1}{2\sigma_t^2} \Fnorm{\Y_t - (\M + \dM_t)\A_t - \X_t}  \Bigr)
\end{split}
\end{equation}
where the underline notation stands for the overall set of the corresponding parameters, $\fnorm{\cdot}$ is the Frobenius norm, $\Theta = \{ \Theta_p, \Theta_h \}$ and
\begin{equation}
\Theta_p = \{ \M, \dMb, \Ab, \Xb, \sig, \Z \}, \; \Theta_h = \{ \bPsi, \s, \bbeta \}
\end{equation}
denote the parameters and hyperparameters whose priors are defined in the following paragraphs. Note that the independence assumption between the observed images conditionally on the unknown parameters is justified by the fact that the sequence of images has been acquired by possibly different sensors at different time instants.
\begin{remark}
The proposed method can easily accommodate different structures for the noise covariance matrix (e.g., diagonal or full covariance matrix) in case the correlation between the spectral bands is significant (see, e.g., \cite{Dobigeon2008icassp}). However, this modification would increase the computational and memory cost of the estimation algorithm introduced in Section~\ref{sec:gibbs}.
\end{remark}

\subsection{Parameter priors}

	\subsubsection{Abundances} \label{sec:abundance_prior}

We propose to promote smooth temporal variations of the abundances between successive time instants for pixels that are not classified as outliers. To this end, we first introduce the binary latent variables $\z_t \in \{0,1\}^\nbpix$ to describe the support of the outliers (i.e., $z_{n,t}=0$ in the absence of outliers in the pixel $(n,t)$, 1 otherwise). With this notation, we introduce a new abundance prior defined for $n = 1,\dotsc \nbpix$ as
\begin{align}
&\a_{n,1} \mid z_{n,1} = 0 \sim \mathcal{U}_{\Sk} \\
&\a_{n,t} \mid z_{n,t} = 1 \sim \mathcal{U}_{\widetilde{\Sk}},  \; \text{for } t = 1, \dotsc, \ntime \\
\begin{split}
&p \left( \a_{n,t} \mid z_{n,t} = 0, \Ab_{\setminus \{\a_{n,t}\}} \right) \propto \exp \biggl\{ -\frac{1}{2\varepsilon_n^2} \times \\
& \Bigl( [\Tf \neq \emptyset] \Enorm{a_{n,t} - a_{n,\tf}} \Bigr) \biggr\} \id_{\Sk} (\a_{n,t}), \; \text{for } t \geq 2
\end{split}
\end{align}
where $\mathcal{U}_{\Sk}$ denote the uniform distribution on the set $\Sk$, $\id_{\Sk}$ is the indicator function of the set $\Sk$, $[\mathscr{P}]$ denotes the Iverson bracket applied to the logical proposition $\mathscr{P}$, i.e.,
\begin{equation*}
  [\mathscr{P}] = \left\{
     \begin{array}{ll}
       1, & \hbox{if $\mathscr{P}$ is true;} \\
       0, & \hbox{otherwise}
     \end{array}
   \right.
\end{equation*}
and
\begin{equation}
\Sk = \{\x \in \mathbb{R}^\nendm \mid \forall i, x_i \geq 0 \text{ and } \t{\x}\1{\nendm} = 1 \} 
\end{equation}
\begin{equation}
\widetilde{\Sk} = \{\x \in \mathbb{R}^\nendm \mid \forall i, x_i \geq 0 \text{ and } \t{\x}\1{\nendm} \leq 1 \} 
\end{equation}
\begin{equation}
\Tf = \left\{ \tau < t \mid z_{n,\tau} = 0 \right\}, \quad \tf = \max_{\tau \in \Tf} \tau.
\end{equation}
By convention, we set $\Tf = \emptyset$ when $t = 1$. To be more explicit, consider an image at time $t$ and a pixel $n$ within this image which is not corrupted by outliers (i.e., $z_{n,t} = 0$). For $t = 1$, a uniform distribution defined in the unit simplex is selected to reflect the absence of specific prior knowledge while accounting for the related constraints in \eqref{eq:constraints}. For $t > 1$, smooth variations of $\a_{n,t}$ are promoted \emph{via} a one-dimensional Gaussian Markov field \cite{Mazet2015,Halimi2015} penalizing the Euclidean distance between $\a_{n,t}$ and the abundance of the last corresponding outlier-free pixel in the preceding images of the sequence, i.e., at time instant $\tf$. On the contrary, when outliers are present in the pixel $(n,t)$ ($\x_{n,t} = 1$), the usual abundance sum-to-one constraint is relaxed ($\a_{n,t}^T \mathbf{1}_\nendm  \leq 1$) so that the prior allows cases in which the linear model does not exhaustively describe the data to be addressed. Note that the \emph{a priori} independence assumptions between the abundance vectors $\a_{n,t}$ (conditionally to the labels $z_{n,t}$) is reasonable from a physical point of view, since they can evolve independently from a pixel to another. In the following, the joint abundance prior is denoted by
\begin{equation}
\begin{split}
p(\Ab \mid \Z) &= \prod_{n = 1}^\nbpix \biggl[ \underset{t_j: z_{n,t_j} = 1}{\prod_{j = 1}^{J_n}} p(\a_{n,t_j} \mid z_{n,t_j} = 1) \biggr] \\
& \times \biggl[ \underset{t_i: z_{n,t_i} = 0}{\prod_{i=1}^{I_n}} p(\a_{n,t_i} \mid \a_{n,t_{i-1}}, z_{n,t_i} = 0) \biggr]
\end{split}
\end{equation}
with $I_n = \sharp \{t: z_{n,t} = 0 \}$, $J_n = \ntime - I_n$ and $\sharp$ denotes the cardinal operator. Note that the events $[z_{n,t} = 0]$ and $[\x_{n,t} = \zero{\nband}]$ (respectively $[z_{n,t} = 1]$ and $[\x_{n,t} \neq \zero{\nband}]$) are equivalent, which allows $p(\Ab \mid \Xb)$ to be defined. 

In the following paragraph, the latent variables $z_{n,t}$ are assigned a specific prior to reflect the fact that outliers are \emph{a priori} assumed to represent a limited number of pixels within the sequence of image.

	\subsubsection{Outliers $\Xb$ and label maps $\Z$} \label{sec:outlier_prior}

Similarly to \cite{Altmann2015tci}, outliers are \ap assumed to be spatially sparse. Different approaches have been proposed in the literature to include this prior knowledge, either relying on the $\ell_1$ penalty (such as the LASSO \cite{Tibshirani1996}) or on mixtures of probability distributions involving a Dirac mass at zero and a continuous probability distribution \cite{Vila2013} (such as the Bernoulli-Laplace \cite{Dobigeon2009ip} or Bernoulli-Gaussian distributions \cite{Kormylo1982,Lavielle1993}, extensively used in the literature \cite{Bourguignon2005ssp,Bazot2011icassp,Chaari2015eusipco}). In this work, we propose to assign the following prior to the outliers $\x_{n,t}$ to promote spatial sparsity
\begin{equation} \label{eq:prior_labels}
p(\x_{n,t} \mid z_{n,t}, s_t^2) = (1 - z_{n,t})\delta(\x_{n,t}) + z_{n,t} \, \mathcal{N}_{\mathbb{R}_+^\nband} (\zero{\nband}, s_t^2)
\end{equation}
where $\mathcal{N}_{\mathbb{R}_+^\nband}$ denotes a Gaussian distribution truncated to the set $\mathbb{R}_+^\nband$. Note that $z_{n,t} = 1$ if an outlier is present in the corresponding pixel, and $0$ otherwise. The proposed prior notably allows outliers to be \emph{a priori} described by a truncated Gaussian distribution when $z_{n,t} = 1$, since the outliers $\x_{n,t}$ are mainly due to the appearance of new endmembers (i.e., that were not present in the reference image). With this context in mind, we further propose to promote spatial correlations between the outliers' support, since new materials are likely to appear in multiple contiguous pixels. The binary label maps $\z_t \in \mathbb{R}^\nbpix$ ($t=1,\ldots,T$) are consequently modeled as Ising-Markov random fields \cite{Eches2011tgrs,Altmann2014a,Halimi2015}, for which the Hammersley-Clifford theorem yields
\begin{equation}
p(\z_t \mid \beta_t) = \frac{1}{C(\beta_t)} \exp \biggl( \beta_t \sum_{n=1}^\nbpix \sum_{k \in \mathcal{V}(n)} \delta(z_{n,t} - z_{k,t}) \biggr)
\end{equation}
where $\mathcal{V}(n)$ denotes the 4-neighbourhood of the pixel $n$, and $C(\beta_t)$ is the partition function \cite{Giovannelli2010}. In practice, the outlier terms $\x_{n,t}$ can be assumed to be \emph{a priori} independent conditionally on $\z_{n,t}$ (since the values of outliers are not \emph{a priori} correlated over space and time from a physical point of view). A similar assumption can be made on the labels $\z_t$, which leads to
\begin{equation}
p(\Xb \mid \Z, \s) = \prod_{n,t} p(\x_{n,t} \mid z_{n,t}, s_t^2)
\end{equation}
\begin{equation}
p(\Z \mid \bbeta) = \prod_t p(\z_t \mid \beta_t)
\end{equation}
with $\Z \in \mathbb{R}^{\nbpix \times \ntime}$ and $\bbeta \in \mathbb{R}^{\ntime}$. Note that the prior \eqref{eq:prior_labels} leads to the following result, which will be useful to sample the label maps in Section~\ref{sec:sample_X_Z}
\begin{equation} \label{eq:cond_x}
\begin{split}
p(\x_{n,t} \mid & \z_{\setminus n,t}, s^2_t, \beta_t) \nonumber \\
& = (1 - \omega_{n,t}) \delta(\x_{n,t}) + \omega_{n,t} \, \mathcal{N}_{\mathbb{R}_+^\nband} (\zero{\nband},s_t^2 \I_\nband)
\end{split}
\end{equation}
%
%
where $\z_{\setminus n,t}$ denotes the label map $\z_t$ whose $n$th entry has been removed, and
\begin{equation}
\omega_{n,t} = \frac{1}{C} \exp \biggl( \beta_t \sum_{k \in \mathcal{V}(n)} \delta(1 - z_{k,t}) \biggr).
\end{equation}
with $C = \sum_{i=0}^1 \exp \Bigl( \beta_t \sum_{k \in \mathcal{V}(n)} \delta(i - z_{k,t}) \Bigr)$.

	\subsubsection{Endmembers}
\label{sec:endmember_prior}
A non-informative prior is adopted for the endmember matrix $\M$ to reflect the absence of specific prior knowledge about the spectral signatures contained in the image. More precisely, as in previous studies related to hyperspectral unmixing \cite{Altmann2015tci,Halimi2015}, we consider the following truncated multivariate Gaussian distribution
\begin{equation}
\m_r \sim \mathcal{N}_{\mathbb{R}_+^\nband} (\zero{\nband},\xi \I_\nband), \; \text{for } r = 1,\dotsc, \nendm
\end{equation}
where $\xi$ is set to a sufficiently large value to ensure an uninformative prior (\eg, $\xi = 1$). Assuming the endmembers $\m_r$ are independent (which is physically reasonable since the endmembers characterize different materials), the joint prior for the endmembers can be written as
\begin{equation}
p(\M) = \prod_{r = 1}^\nendm p(\m_r).
\end{equation}
In addition, the endmembers can be \emph{a priori} assumed to live in a subspace of dimension $K = \nendm - 1$ \cite{Dobigeon2009} whose practical determination can be performed by a principal component analysis (PCA) or a robust PCA (rPCA) \cite{Candes2009}. This dimensionality reduction step is essentially aimed at reducing the computational complexity of the proposed approach. More explicitly, the PCA applied to the original data $\Yb$ leads to a decomposition which can be expressed as \cite{Dobigeon2009}
\begin{equation}
\label{eq:PCA}
\mathbf{m}_r = \U \e_r + \check{\y}, \quad \check{\y} = (\I_\nband - \U\t{\U})\bar{\y}, \quad \t{\U}\U = \I_K
\end{equation}
where $\U$ denotes a basis of the subspace of dimension $K$ and $\bar{\y}$ denotes the average spectral signature obtained from $\Yb$. Note that using an rPCA would result in similar expressions (modulo a simple change of notations). The projected endmembers $\e_r$ are then assigned the following truncated Gaussian prior, which ensures the non-negativity of the endmembers
\begin{equation}
\e_r \sim \mathcal{N}_{\mathcal{E}_r} (\mathbf{0}_K,\xi \I_K), \; \text{for } r = 1,\dotsc, \nendm
\end{equation}
with
\begin{equation}
\mathcal{E}_r = [e_{1,r}^-,e_{1,r}^+] \times \dotsc \times [e_{K,r}^-,e_{K,r}^+]
\end{equation}
\begin{equation}
e_{k,r}^- =  \max_{\ell \in \mathcal{U}_k^+} \left( - \frac{\check{y}_\ell + \sum_{j \neq k} u_{\ell,j} e_{j,r}}{u_{\ell,k}} \right)
\end{equation}
\begin{equation}
e_{k,r}^+ = \min_{\ell \in \mathcal{U}_k^-} \left( - \frac{\check{y}_\ell + \sum_{j \neq k} u_{\ell,j} e_{j,r}}{u_{\ell,k}} \right)
\end{equation}
\begin{equation}
\mathcal{U}_k^- = \{r: u_{k,r} < 0 \}, \quad \mathcal{U}_k^+ = \{r: u_{k,r} > 0 \}.
\end{equation}

	\subsubsection{Endmember variability}
\label{sec:variability_prior}

We consider a prior for the vectors $\dm_{r,t}$ (associated with the endmember variability) promoting smooth temporal variations while accounting for the term-wise non-negativity of the observed endmembers (\ie, $\m_r + \dm_{r,t} \succeq \zero{\nband, \nendm}$), expressed as
\begin{align}
&\dm_{r,1} \mid \m_r \sim \mathcal{N}_{\mathcal{I}_r} (\zero{\nband},\nu \I_\nband) \\
%
%
&dm_{\ell,r,t} \mid m_{\ell,r}, dm_{\ell,r,(t-1)}, \psi_{\ell,r}^2  \sim \mathcal{N}_{\mathcal{I}_{\ell,r}} \left( dm_{\ell,r,(t-1)},\psi_{\ell,r}^2 \right)
\end{align}
for $\ell = 1, \dotsc,\nband$, $r = 1, \dotsc,\nendm$, $t = 1,\dotsc, \ntime$, where $\mathcal{I}_r = \mathcal{I}_{1,r} \times \dotsc \times \mathcal{I}_{\nband,r}$ and $\mathcal{I}_{\ell,r} = [-m_{\ell,r}, +\infty)$. Assuming \emph{a priori} independence between the different endmember variabilities (since the variability can be independent from a material to another), the joint variability prior can finally be expressed as
\begin{equation}
\begin{split}
p(\dMb \mid \M,& \bPsi) = \prod_{r=1}^\nendm \biggl[ p(\dm_{r,1} \mid \m_r) \biggr. \\
& \times \biggl. \prod_{t = 2}^\ntime p(\dm_{r,t} \mid \m_r,\dm_{r,(t-1)},\bbpsi_r) \biggr].
\end{split}
\end{equation}

	\subsubsection{Noise variance} \label{sec:noise_prior}

A non-informative inverse-gamma conjugate prior is selected for the noise variance
\begin{equation}
\sigma_t^2 \sim \mathcal{IG} (\asig,\bsig)
\end{equation}
for $t = 1,\dotsc,\ntime$, with $\asig = \bsig = 10^{-3}$ in order to ensure a weakly informative prior. The noise variances $\sigma_t^2$ can be assumed to be \emph{a priori} independent (given the absence of \emph{a priori} correlation between the noise in different images), thus leading to
\begin{equation}
p(\sig) = \prod_t p(\sigma_t^2).
\end{equation}

\subsection{Hyperparameters} \label{sec:hyperparameter_prior}

In order to complete the description of the proposed hierarchical Bayesian model, we consider the following generic priors for the different hyperparameters. Note that the \emph{a priori} independence assumptions made in this section are more of a computational nature, i.e., aimed at simplifying the estimation procedure detailed in the next section.
\begin{enumerate}[label=(\roman*)]
\item Non-informative conjugate inverse-gamma priors for the variability variances $\bPsi$ and the outlier variances $\s$, \ie, for $\ell = 1, \dotsc, \nband$, $r = 1, \dotsc, \nendm$ and $t = 1 ,\dotsc, \ntime$
%
\begin{equation}
\psi_{\ell,r}^2 \sim \mathcal{IG}(\apsi,\bpsi), \quad s_t^2 \sim \mathcal{IG}(\as,\bs)
\end{equation}
where $\mathcal{IG}(\apsi,\bpsi)$ denotes the inverse gamma distribution and $\apsi = \bpsi = \as = \bs = 10^{-3}$. Classical independence assumptions for the different hyperparameters lead to
\begin{equation}
p(\bPsi) = \prod_{\ell,r} p(\psi_{\ell,r}^2), \quad
p(\s) = \prod_t p(s_t^2).
\end{equation}
%
\item A uniform prior for the granularity parameter of a Potts-Markov random field (\emph{a fortiori} of an Ising-Markov random field). Previous studies have shown that it is reasonable to constrain the granularity parameter to belong to the interval $[0,2]$ \cite{Giovannelli2011}, leading to
%
\begin{equation}
\beta_t \sim \mathcal{U}_{[0,2]}, \;
\text{for} \; t = 1,\dotsc, \ntime.
\end{equation}
Assuming the granularity parameters are \emph{a priori} independent for different time instants finally yields
\begin{equation}
p(\bbeta) = \prod_t p(\beta_t).
\end{equation}
\end{enumerate}

\input{dag2}

\subsection{Joint posterior distribution}

Applying Bayes' theorem, the joint posterior distribution of the parameters of interest is given by
\begin{equation} \label{eq:posterior}
\begin{split}
p(& \Theta \mid \Yb) \propto  p(\Yb \mid \Theta) p(\Ab \mid \Xb) p(\Xb \mid \Z,\s) p(\s) \\
&  \times p(\Z \mid \bbeta) p(\bbeta) p(\dMb \mid \M, \bPsi) p(\M) p(\bPsi) p(\sig).
\end{split}
\end{equation}

The complexity of the proposed Bayesian model summarized in the directed acyclic graph of Fig.~\ref{fig:dag} and its resulting posterior \eqref{eq:posterior} prevent a simple computation of the maximum \emph{a posteriori} (MAP) or minimum mean square (MMSE) estimators. For instance, the optimization problem associated with the determination of the MAP estimator of $\Theta$ is clearly complex, since the  negative log-posterior is non-convex and parameterized by mixed continuous and discrete variables. In this context, classical matrix factorization techniques such as \cite{Fevotte2015} cannot be used efficiently. An MCMC method is consequently adopted to sample the posterior \eqref{eq:posterior} and to build estimators of the parameters involved in the proposed Bayesian model using the generated samples.

\section{Gibbs sampler} \label{sec:gibbs}

This section studies a Gibbs sampler, which is guaranteed to produce samples asymptotically distributed according to the target distribution \eqref{eq:posterior}. This sampler described in Algo.~\ref{alg:gibbs} consists in generating samples distributed according to the conditional distribution of each parameter of interest. Section~\ref{sec:bayesian_inference} introduces the proposed sampling method, and the conditional distributions of all the parameters of interest (see Fig.~\ref{fig:dag}) are detailed in the following paragraphs.

	\subsection{Bayesian inference and parameter estimation} \label{sec:bayesian_inference}

The main steps of the proposed Gibbs sampler are summarized in Algo. \ref{alg:gibbs}. Similarly to \cite{Altmann2015tci}, the sequence $\{\Theta^{(q)}\}_{q = \nbi + 1}^{\nmc}$ generated by the proposed sampler (\ie, after $\nbi$ burn-in iterations) is used to approximate the MMSE estimators of the different unknown parameters $\M$, $\A_t$, $\dM_t$ and $\X_t$ by replacing the expectations by empirical averages.
\begin{equation}
\Mmse \simeq \frac{1}{\nmc - \nbi} \sum_{q = \nbi + 1}^{\nmc} \M^{(q)}
\end{equation}
\begin{equation}
\Amse_t \simeq \frac{1}{\nmc - \nbi} \sum_{q = \nbi + 1}^{\nmc} \A_t^{(q)}
\end{equation}
\begin{equation}
\dMmse_t \simeq \frac{1}{\nmc - \nbi} \sum_{q = \nbi + 1}^{\nmc} \dM_t^{(q)}
\end{equation}
\begin{equation}
\Xmse_t \simeq \frac{1}{\nmc - \nbi} \sum_{q = \nbi + 1}^{\nmc} \X_t^{(q)}.
\end{equation}
This choice is justified by the fact that MAP estimators computed using MCMC algorithms are often less accurate when the number of unknown parameters is relatively large \cite{Doucet2002}. Finally, the following \emph{marginal maximum a posterior} (mMAP) estimator is considered for the label maps
\begin{equation}
\zmap = \argmax{z_{n,t} \in \{0,1\}} p \left( z_{n,t} \mid \y_{n,t}, \Theta_{\setminus \{z_{n,t}\}} \right).
\end{equation}
It is approximated by
\begin{equation}
\zmap \simeq \left\{
\begin{array}{l}
0, \text{ if } \sharp \{q > \nbi: z_{n,t}^{(q)} = 0 \} \leq \frac{\nmc - \nbi}{2} \\
1, \text{ otherwise}.
\end{array} \right.
\end{equation}

\input{gibbs}

	\subsection{Sampling the abundances $\Ab$} \label{sec:sample_A}

The likelihood function \eqref{eq:likelihood} combined with the prior given in Section~\ref{sec:abundance_prior} leads to the following conditional distribution for the abundances

\begin{equation}
\a_{n,t} \mid \y_{n,t}, \Theta_{\setminus \{ \a_{n,t} \}} \sim \mathcal{N}_{\Sk} (\bmu_{n,t}^{(\A)}, \bLambda_{n,t})
\end{equation}
\begin{equation}
\bLambda_{n,t}^{-1} = \frac{1}{\sigma_t^2}\t{\M_t}\M_t + \frac{1}{\varepsilon_n^2} \Bigl( [\Tf \neq \emptyset] + [\Ts \neq \emptyset] \Bigr)\I_\nendm 
\end{equation}
\begin{equation}
\M_t \triangleq \M + \dM_t
\end{equation}
\begin{equation}
\begin{split}
\bmu_{n,t}^{(\A)} & = \bLambda_{n,t} \biggl[ \frac{1}{\sigma_t^2}\t{\M_t}(\y_{n,t} - \x_{n,t}) \\
& + \frac{1}{\varepsilon_n^2} \Bigl( [\Tf \neq \emptyset]\a_{n,\tf} + [\Ts \neq \emptyset]\a_{n,\ts} \Bigr) \biggr]
\end{split}
\end{equation}
where $\mathcal{N}_{\Sk}(\bmu,\bLambda)$ denotes a Gaussian distribution truncated to the set $\Sk$ and
\begin{equation}
\Ts = \left\{ \tau > t \mid z_{n,\tau} = 0 \right\}, \quad \ts = \min_{\tau \in \Ts} \tau
\end{equation}
with the convention $\Ts = \emptyset$ if $t = \ntime$.

Samples distributed according to the above truncated multivariate Gaussian distributions can be generated by a Gibbs sampler described in \cite[Section IV.B.]{DobigeonTR2007b} \cite{Altmann2014ssp}, by an Hamiltonian Monte-Carlo procedure \cite{Altmann2014b,Pakman2014} or by the general method recently proposed in \cite{Botev2016}. In this work, the Gibbs sampler \cite[Section IV.B.]{DobigeonTR2007b} has been adopted to sample the parameters of interest. Note that the abundance vectors $\a_{n,t}$ can be sampled in parallel to accelerate the algorithm.

	\subsection{Sampling the endmembers $\M$} \label{sec:sample_M}

Combining \eqref{eq:likelihood} and the endmember prior given in Section~\ref{sec:endmember_prior} leads to
\begin{align}
&m_{\ell,r} \mid \Yb, \Theta_{\setminus \{m_{\ell,r}\}} \sim \mathcal{N}_{[b_{\ell,r},+ \infty)} (\mu_{\ell,r}^{(\M)}, \kappa_{\ell,r}^2) \\
& b_{\ell,r} = \max \Bigl\{ 0, \max_t (-dm_{\ell,r,t} ) \Bigr\}
\end{align}
\begin{equation}
\begin{split}
\mu_{\ell,r}^{(\M)} = \kappa_{\ell,r}^2 \sum_{t} \frac{1}{\sigma_t^2} \bigl[ \yl_{\ell,t} - \xl_{\ell,t} &- \ml_{l,\setminus r} \A_{\setminus r,t} \\
& - \dml_{\ell,t} \A_t  \bigr] \t{\ar_{r,t}}
\end{split}
\end{equation}
\begin{equation}
\kappa_{\ell,r}^2 = \left[ \sum_{n,t} \frac{a_{r,n,t}^2}{\sigma_t^2} + \frac{1}{\xi} \right]^{-1} 
\end{equation}
where $\dml_{\ell,t}$ is the $\ell$th row of $\dM_t$, $\ml_{\ell,\setminus r}$ is the $\ell$th row of $\M$ whose $r$th entry has been removed, $\A_{\setminus r, t}$ denotes the matrix $\A_t$ without its $r$th row and $\ar_{r,t}$ is the $r$th row of $\A_t$. Samples distributed according to the above truncated Gaussian distributions can be efficiently generated using the algorithm described in \cite{Chopin2011}. When using a PCA as a preprocessing step \eqref{eq:PCA}, the projected endmembers $\e_r$, for $r = 1, \dotsc, \nendm$ have a truncated multivariate Gaussian distribution \cite{Dobigeon2009}
\begin{equation}
\e_r \mid \Yb, \Theta_{\setminus \{ \e_r \}} \sim \mathcal{N}_{\mathscr{E}_r} (\bmu_r^{(\E)}, \bLambda_r )
\end{equation}
where $\mathscr{E}_r$, $\bmu_r^{(\E)}$ and $\bLambda_r$ have been reported in Appendix \ref{app:subspace}. Note that the rows of $\M$ (resp. of the projected endmember matrix $\E$) can be sampled in parallel to decrease the computational time required by the algorithm.

	\subsection{Sampling the variability terms $\dMb$} \label{sec:sample_dM}

Similarly, the likelihood function \eqref{eq:likelihood} and the prior given in Section~\ref{sec:variability_prior} lead to
\begin{align}
dm_{\ell,r,t} \sim \mathcal{N}_{[-m_{\ell,r}, +\infty)} (\mu_{\ell,r,t}^{(\dM)}, \eta_{\ell,r,t}^2)
\end{align}
with
\begin{equation}
\begin{split}
\frac{1}{\eta_{\ell,r,t}^2} = \frac{1}{\sigma_t^2} \sum_n   &a_{r,n,t}^2 + \frac{1}{\nu} \bigl[ t=1 \bigr] \\
& + \frac{1}{\psi_{\ell,r}^2} \Bigl( 1 + \bigl[ 1 < t < \ntime \bigr] \Bigr)
\end{split}
\end{equation}

\begin{equation}
\begin{split}
&\mu_{\ell,r,t}^{(\dM)} = \left[ \frac{1}{\sigma_t^2} \bigl( \yl_{\ell,t} - \dml_{\ell,\setminus r, t} \A_{\setminus r,t} - \ml_\ell \a_{n,t} - \x_{\ell,n,t} \bigr) \tilde{\a}_{r,t}^{\text{T}} \right. \\
& \left.  + \frac{1}{\psi_{\ell,r}^2} \Bigl( [t<\ntime]dm_{\ell,r,(t+1)} + [t>1]dm_{\ell,r,(t-1)} \Bigr) \right] \eta_{\ell,r,t}^2
\end{split}
\end{equation}
where $\dml_{\ell,\setminus r, t}$ denotes the $\ell$th row of $\dM_t$ whose $r$th element has been removed, $\ml_\ell$ is the $\ell$th row of $\M$ and $\A_{\setminus r,t}$ is the matrix $\A_t$ without its $r$th row. The rows of each variability matrix $\dM_t$ can be sampled in parallel to reduce the computational time of the sampler.
	
	\subsection{Sampling the label maps $\Z$ and the outliers $\Xb$} \label{sec:sample_X_Z}

According to \eqref{eq:likelihood} and Section \ref{sec:outlier_prior}, the outliers admit the following group-sparsity promoting conditional distributions
\begin{equation}
\begin{split}
p(\x_{n,t} &\mid \y_{n,t}, \Theta_{\setminus \{z_{n,t},\x_{n,t}\}}) = \\
&(1 - w_{n,t}) \delta(\x_{n,t}) + w_{n,t} \, \mathcal{N}_{\mathbb{R}_+^\nband} (\bmu_{n,t}^{(\X)}, \vartheta_t^2 \I_\nband)
\end{split}
\end{equation}
which are  mixtures of a Dirac mass at $\boldsymbol{0}$ and of truncated multivariate Gaussian distributions, where
\begin{align}
w_{n,t} &= \frac{\tilde{w}_{n,t}}{\tilde{w}_{n,t} + (1 - \omega_{n,t})}, \quad \vartheta_t^2 = \frac{\sigma_t^2 s_t^2}{\sigma_t^2 + s_t^2} \\
\tilde{w}_{n,t} &= \frac{\omega_{n,t}}{(s_t^2)^{\nband/2}} (\vartheta_t^2)^{\nband/2} \exp \left( \frac{1}{2 \vartheta_t^2} \Enorm{\bmu_{n,t}^{(\X)}} \right) \\
\bmu_{n,t}^{(\X)} &= \frac{s_t^2}{\sigma_t^2 + s_t^2} \bigl[ \y_{n,t} - (\M + \dM_t)\a_{n,t} \bigr].
\end{align}

In practice, the labels $z_{n,t}$ are first sampled according to a Bernoulli distribution to select one of the two models for $\x_{n,t}$, with probability \mbox{$\mathbb{P}[z_{n,t} = 1 \mid \y_{n,t}, \Theta_{\setminus \{z_{n,t},\x_{n,t}\}}] = w_{n,t}$}. Note that the labels $\z_{n,t}$ can be sampled in parallel by using a checkerboard scheme. In addition, the outliers $\x_{n,nt}$ can be sampled in parallel to decrease the computational time.

	\subsection{Sampling the outlier variances $\s$}	 \label{sec:sample_s}

According to Sections~\ref{sec:outlier_prior} and \ref{sec:hyperparameter_prior}, we can easily identify the conditional law of $s_t^2$ for $t = 1, \dotsc, \ntime$ as the following inverse gamma distribution
\begin{equation}
\begin{split}
s_t^2 \mid &\Theta_{\setminus \{s_t^2\}} \sim \\
& \mathcal{IG}\Bigl( \as + \frac{\sharp\{n:z_{n,t} = 1\} \nband}{2}, \bs + \frac{1}{2} \Fnorm{\X_t} \Bigr).
\end{split}
\end{equation}
	
	\subsection{Sampling the noise variances $\sig$} \label{sec:sample_sig}

Using Sections \ref{sec:noise_prior} and \ref{sec:hyperparameter_prior}, we obtain for $t = 1,\dotsc, \ntime$

\begin{equation}
\begin{split}
\sigma_t^2 \mid \Y_t, \Theta_{\setminus \{\sigma_t^2\}} \sim 
& \mathcal{IG}\Bigl( \asig + \frac{\nband\nbpix}{2}, \bsig + \\
& \frac{1}{2} \Fnorm{\Y_t - (\M + \dM_t)\A_t - \X_t} \Bigr)
\end{split}.
\end{equation}

	\subsection{Sampling the variability variances $\bPsi$} \label{sec:sample_psi}
Similarly, Section~\ref{sec:variability_prior} and \ref{sec:hyperparameter_prior} lead to
\begin{equation}
\begin{split}
\psi_{\ell,r}^2 \mid \Theta_{\setminus \{\psi_{\ell,r}^2\}}  \sim
&\mathcal{IG} \Bigl( \apsi + \frac{T-1}{2}, \bpsi + \\
& \frac{1}{2} \sum_{t = 2}^{\ntime} (dm_{\ell,r,t} - dm_{\ell,r,t-1})^2 \Bigr).
\end{split}
\end{equation}
	
	\subsection{Sampling the granularity parameters $\beta_t$} \label{sec:sample_beta}

Provided square images are considered, the partition functions $C(\beta_t)$ have the closed-form expressions \cite{Onsager1944,Giovannelli2010}
%
%
\begin{align*}
& \widetilde{C}(\beta_t) = \frac{1}{2} \log(2\sinh \beta_t) + \frac{1}{2\nbpix} \sum_{n = 1}^\nbpix \acosh \Delta_n (\beta_t) + \beta_t \\
&\Delta_n (\beta_t) = v(\beta_t) - C_n, \; v(\beta_t) = \frac{\cosh^2 \beta_t}{\sinh \beta_t} \\
&\widetilde{C}(\beta_t) = \frac{1}{\nbpix}\log C(\beta_t), \quad C_n = \cos\left( \frac{2n - 1}{2\nbpix} \pi \right).
\end{align*}

The exact partition function can then be used to sample the parameters $\beta_t$ using Metropolis-Hastings steps. In this work, new values of the granularity parameters have been proposed by the following Gaussian random walk
\begin{equation}
\beta_t^* = \beta_t^{(q)} + \varepsilon_t, \quad \varepsilon_t \sim \mathcal{N}\bigl( 0,\sigb(t) \bigr)
\end{equation}
where the parameters $\sigb (t)$ are adjusted during the burn-in iterations to yield acceptance rates in the interval $[0.4,0.6]$.

	\subsection{Computational complexity}
	
Assuming elementary arithmetic operations and scalar pseudo-random number generations are $O(1)$ operations, the overall computational complexity is dominated by matrix products needed to compute the parameters related to the conditional distribution of the variability vectors. Since $\nendm \ll \nband \ll  \nbpix$ and $\ntime \ll \nband$, the per-iteration computational cost of the proposed algorithm is $O(\nband \nendm^2 \nbpix \ntime)$ per iteration. As detailed in the preceding paragraphs, many parameters can be sampled in parallel to reduce the computational time of the proposed algorithm. In comparison, the computational complexity of VCA is $O(\nendm^2\nbpix)$ \cite{Nascimento2005} per image, and the per iteration complexity of the others algorithms for a single image are respectively: $O(\nbpix^2)$ for FCLS \cite{Bioucas2010}, $O(\nendm \nbpix)$ for SISAL \cite{Bioucas2009}, $O(\nband\nendm\nbpix)$ for rLMM \cite{Fevotte2015} and $O(\nendm^2(\nband + \nbpix))$ for OU~\cite{Thouvenin2015b}.


\input{true_endm}

\section{Experiments with synthetic data} \label{sec:exp}

The proposed method has been applied to an MTHS image composed of 10 acquisitions of size $50~\times~50$ with $\nband = 413$ bands. The first scenario deals with the appearance of a new material in specific regions of a few images. To this end, $4$ images out of the $10$ have been corrupted by spatially sparse outliers, corresponding to a new endmember extracted from a spectral library. Each image of the sequence corresponds to a linear mixture of 3 endmembers affected by smooth time-varying variability, and the synthetic abundances vary smoothly from one image to another.

First, so-called reference abundance maps corresponding to the first time instant have been generated (e.g., for Fig.~\ref{fig:A1}, we have taken the abundance maps obtained by VCA/FCLS on the widely studied Moffett dataset \cite{Dobigeon2009}). Then, the abundance maps corresponding to the remaining time instants have been generated by multiplying the reference maps with trigonometric functions to ensure a sufficiently smooth temporal evolution. For the first dataset composed of $\nendm = 3$ endmembers, the reference maps associated with the first two endmembers have been respectively multiplied by $\cos\Bigl(\frac{\pi}{100} + t \frac{48\pi}{100} \Bigr)$ and $\sin\Bigl(\frac{\pi}{100} + t \frac{48\pi}{100} \Bigr)$, with $t \in \{1, \dotsc ,\ntime \} $. The temporal evolution of the last abundance map has finally been obtained by leveraging the sum-to-one condition. With these abundance maps, the contribution of a given endmember (assumed to punctually disappear) has been replaced at specific time instants by a new endmember signature in pixels originally corresponding to its highest abundance coefficients (e.g., above 0.8).

 The mixtures have finally been corrupted by an additive white Gaussian noise to ensure a resulting signal-to-noise ratio (SNR) between $25$ and $\SI{30}{dB}$. Similarly, two complementary scenarii involving 5 HS images, of size $100\times 100$, composed of 6 and 9 endmembers, have been considered to analyze the performance of the method in the presence of a larger number of endmembers. Note that a larger image size has been considered for these two datasets to reflect the fact that a larger number of endmembers is expected to be observed in larger scenes. In addition, the images of this experiment do not satisfy the pure pixel assumption to assess the proposed method in challenging situations.

	Controlled spectral variability has been introduced by using the product of reference endmembers with randomly generated piecewise-affine functions as in \cite{Thouvenin2015}, where different affine functions have been generated for each endmember at each time instant. Typical instances of the signatures used in this experiment are depicted in Fig.~\ref{fig:endm_true}.
	The robustness of the proposed method to moderate spatial variability, i.e., endmember variability occurring within single images, has also been evaluated. The corresponding results can be found in the technical report \cite[Appendix E]{Thouvenin2017TR} due to space constraints.

\input{table_parameters}

	\subsection{Compared methods}

The results of the proposed algorithm have been compared to those of several unmixing methods from the literature, some of which are specifically designed to unmix a single HS image. In the following lines, the most relevant implementation details specific to each method are briefly recalled.

\begin{enumerate}
\item VCA/FCLS (no variability, single image): the endmembers are first extracted on each image using the vertex component analysis (VCA) \cite{Nascimento2005}, which requires pure pixels to be present. The abundances are then estimated for each pixel by solving a fully constrained least squares problem (FCLS) using the alternating direction method of multipliers (ADMM) \cite{Bioucas2010}. Note that the estimates provided by the VCA algorithm vary from one run to another, given its stochastic nature;
%
\item SISAL/FCLS (no variability, single image): the endmembers are extracted on each image by the simplex identification via split augmented Lagrangian (SISAL) \cite{Bioucas2009}, and the abundances are estimated for each pixel by FCLS. The tolerance for the stopping rule has been set to $10^{-3}$;
\item RLMM (no variability, single image): the unmixing method associated with the robust linear mixing model (RLMM) proposed in \cite{Fevotte2015} has been applied to each image of the series independently. The algorithm has been initialized with SISAL/FCLS, and the regularization parameter specific to this method is set as in~\cite{Fevotte2015};
%
\item OU: the endmembers are estimated using the online unmixing (OU) algorithm introduced in \cite{Thouvenin2015b} with endmembers initialized by the output of VCA applied to the first image of the sequence. The abundances are initialized by FCLS, and the variability matrices are initialized with all their entries equal to $0$. The other parameters are set to the same values as those given in~\cite[Table I]{Thouvenin2015b};
\item Proposed approach: the endmembers are initialized with VCA applied to the first image of the sequence, within which the observed materials are well represented (i.e., with sufficiently high abundance coefficients for each material). In this context the VCA algorithm, which requires pure pixels to be present in the data, has been observed to yield relevant results for the initialization. However, other endmember extraction techniques might be used to initialize the proposed algorithm if needed. The abundances are initialized by FCLS, and the variability matrices and label maps are initialized with all their entries equal to $0$ (\ie, the images are \ap assumed to contain no outlier). The values chosen for the other parameters are summarized in Table~\ref{tab:param1}. Further details on these values can be found in the supplementary material provided by the authors.

\end{enumerate}

Performance assessment has been conducted in terms of
\begin{enumerate}[label=(\roman*)] 
\item endmember estimation through the average spectral angle mapper (aSAM) \vspace{-0.2cm}
\begin{equation}
\aSAM(\M) = \frac{1}{\nendm } \sum_{r=1}^\nendm   \arccos \left( \frac{ \t{\mathbf{m}_r} \widehat{\mathbf{m}}_r }{ \lVert \mathbf{m}_r \rVert_2 \lVert \widehat{\mathbf{m}}_r \rVert_2 } \right);
\end{equation}
\item abundance and variability estimation through the global mean square errors (GMSEs) \vspace{-0.2cm}
\begin{align}
\GMSE(\A)  & = \frac{1}{\ntime \nendm \nbpix} \sum_{t=1}^\ntime \lVert \A_t - \widehat{\A}_t
\rVert_{\text{F}}^2 \\
\GMSE(\dM) & = \frac{1}{\ntime \nband \nendm } \sum_{t=1}^\ntime \lVert \dM_t - \widehat{\dM}_t \rVert_{\text{F}}^2; \vspace{-0.2cm}
\end{align}
\item quadratic reconstruction error (RE)
\begin{align}
\label{eq:RE}
\RE &= \frac{1}{\ntime \nband \nbpix} \sum_{t=1}^\ntime \lVert \Y_t - \widehat{\Y}_t \rVert_{\text{F}}^2
%
\end{align}
where $\widehat{\Y}_t$ is the matrix composed of the pixels reconstructed with the estimated parameters.
\end{enumerate}

\input{synth_endm}
\input{synth_abundances}

	\subsection{Results}
The endmembers estimated by the proposed algorithm are compared to those of VCA/FCLS, SISAL/FCLS, RLMM and OU in Fig.~\ref{fig:synth_endm}, whereas the corresponding abundance maps are displayed in Fig.~\ref{fig:A1}.
Note that the abundance maps and the endmembers obtained for the mixtures of 6 and 9 endmembers are included in a separate technical report \cite[Appendix~D]{Thouvenin2017TR} due to space constraints (see Figs.~17--24 for $\nendm = 6$, and Figs.~26--37 for $\nendm = 9$). The unmixing performance of each method, reported in Table \ref{tab:results_synth}, leads to the following conclusions.

\begin{figure}[h!]
	\centering
	\includegraphics[keepaspectratio,height=0.25\textheight , width=0.45\textwidth]{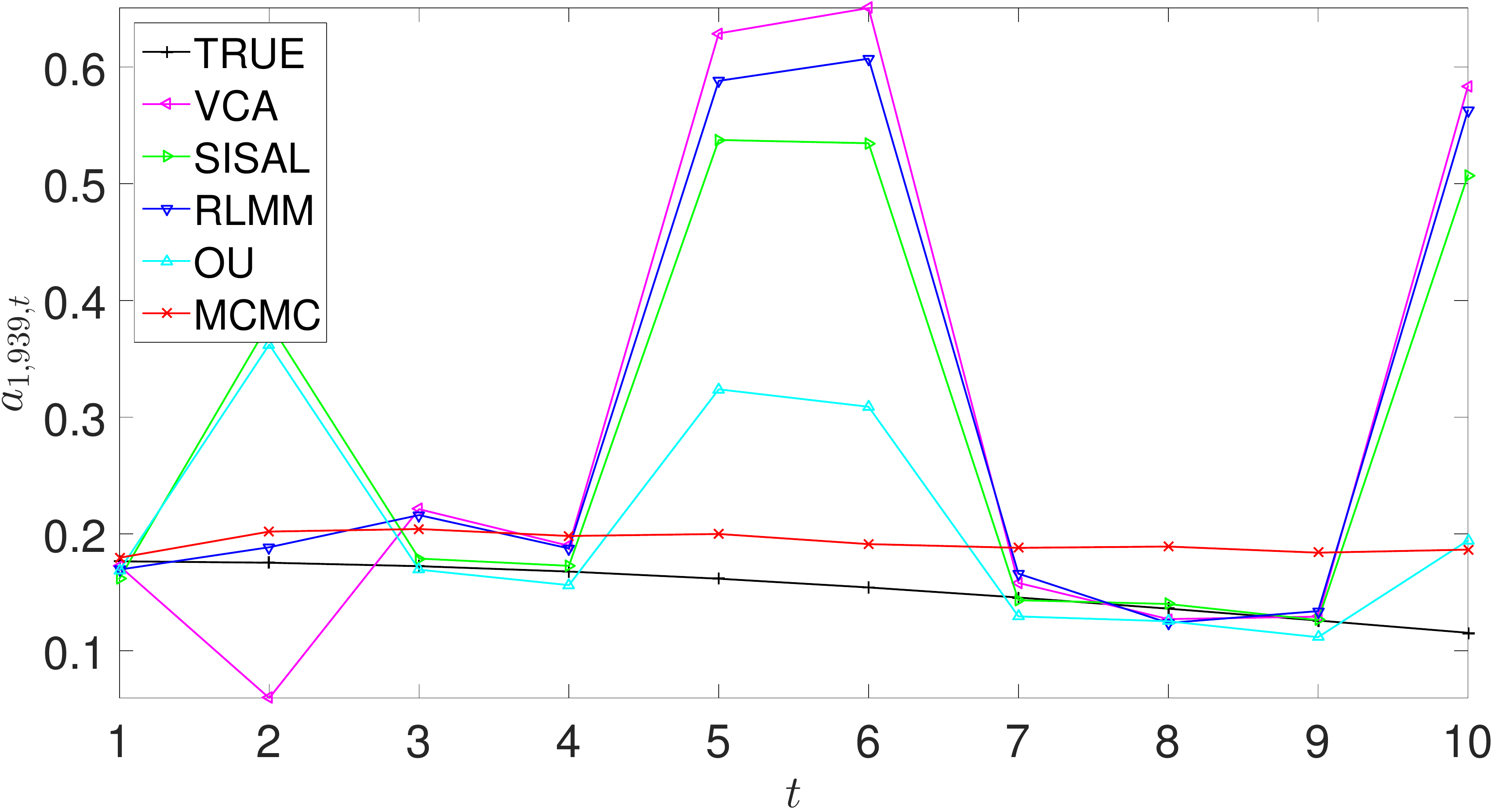}
	\caption{Evolution over time of the abundance associated with the first endmember in a given pixel. The similarity between the recovered result and the ground truth illustrates the relevance of the proposed abundance prior to mitigate the errors induced by the presence of outliers in the image (time instants 2, 5, 6 and 10).}
\label{fig:At1}
\end{figure}

\begin{itemize}
\item \textbf{Endmember estimation:} the proposed method shows an interesting robustness \wrt spatially sparse outliers in the sense that the estimated signatures (Figs. \ref{fig:endm2_mcmc}, see the supplementary material for the two other endmembers) are very close to the corresponding ground truth (Fig. \ref{fig:endm_true}). In comparison, the shape of the endmembers recovered by VCA, SISAL and RLMM and the variability extracted by OU are significantly affected by outliers, as exemplified in Figs.~\ref{fig:endm2_vca}, \ref{fig:endm2_sisal}, \ref{fig:endm2_rlmm} and \ref{fig:endm2_ou} respectively. These qualitative results are confirmed by the quantitative performance measures of each method provided in Table~\ref{tab:results_synth}. Note that the endmembers recovered by the SISAL and RLMM methods are very sensitive to the VCA initialization, as illustrated by the similarity between the signatures estimated by these methods (Figs. \ref{fig:endm2_vca} to \ref{fig:endm2_rlmm}).
\item \textbf{Abundance estimation:} the abundance maps estimated by FCLS, RLMM and SISAL reflect the high sensitivity of VCA (used to initialize SISAL and RLMM) to the presence of outliers (see the figures delineated in red in Fig.~\ref{fig:A1}). On the contrary, the abundances recovered by OU and the proposed approach are much closer to the ground truth. These observations are confirmed by the abundance estimation performance reported in Table \ref{tab:results_synth}. The proposed abundance smoothness prior appears to mitigate the errors induced by the presence of outliers as can be seen in Fig.~\ref{fig:At1}. More precisely, for images corrupted by outliers, the abundance coefficients estimated by the proposed approach are closer to the ground truth than the results of the other methods. A more detailed version of Table~\ref{tab:results_synth}, along with a complementary figure illustrating the interest of the proposed abundance prior can be found in the supplementary material.
\item \textbf{Overall performance:} the performance measures reported in Table \ref{tab:results_synth} are globally favorable to the proposed approach. It is important to mention that the price to pay with the good performance of the proposed method is its computational complexity, which is common with MCMC methods.
\end{itemize}
As a complementary output, the proposed algorithm is able to recover the location of the outliers within each image, as illustrated in Fig. \ref{fig:label_synth}. Up to a few false detections, the estimated labels are very close to the ground truth. The label errors observed for $t = 7,8$ and $9$ partly result from the different abundance constraints considered when an outlier is detected or not.

\input{tab_results_synth}
\input{synth_label}

\section{Experiment with real data} \label{sec:expr}

	\subsection{Description of the dataset}

We consider a real sequence of AVIRIS HS images acquired over the Lake Tahoe region (California, United States of America) between 2014 and 2015\footnote{The images from which the scene under study is extracted are freely available from the online AVIRIS flight locator tool at \url{http://aviris.jpl.nasa.gov/alt_locator/}.}. The scene of interest ($100 \times 100$), composed of a lake and a nearby field, has been unmixed with $\nendm = 3$ endmembers based on the results of the noise-whitened eigengap algorithm (NWEGA) \cite{Halimi2016} applied to each image of the series (see Table \ref{tab:tab_ega}). This choice is further supported by results obtained from a previous analysis conducted on the same dataset \cite[Appendix E]{Thouvenin2015TR}. For $\nendm = 4$ and $5$, the signatures of water, soil and vegetation were split into two or more components by the different algorithms, suggesting $\nendm = 3$ is more appropriate for this study. Note that prior studies led in \cite{Thouvenin2015b} revealed that this dataset contains outliers (area delineated in red in Fig. \ref{fig:cube5}). After removing the seemingly corrupted bands and the water absorption bands, 173 out of the 224 spectral bands were finally exploited.
The initial parameters used for the proposed algorithm are given in Table \ref{tab:param1}. The other methods have been run with the same parameters as in Section~\ref{sec:exp}. Note that the VCA results reported in this section are representative of those obtained over multiple runs (no significant differences have been observed from one run to another).

\input{time_series}

	\subsection{Results}
In the absence of any ground truth, the performance of the unmixing methods is assessed in terms of RE (Table \ref{tab:results_real}) while taking into account the consistency of the estimated abundance maps reported in Figs.~\ref{fig:A1_real}, \ref{fig:A2_real} and \ref{fig:A3_real}. More precisely, the abundances associated with the vegetation area are expected to be very high for $t = 1$, $3$, $5$ (corresponding to Figs. \ref{fig:cube1}, \ref{fig:cube3} and \ref{fig:cube5}) where the vegetation visually appears to be sufficiently irrigated (hence well represented). On the contrary, the abundance coefficients are supposed to be much lower for $t = 2$, $4$, $6$ (corresponding to Figs. \ref{fig:cube2}, \ref{fig:cube4} and \ref{fig:cube6}), where the vegetation is visually drier or almost absent. Concerning the presence of water in the bottom left-hand corner of the images, the latent variables introduced in Section \ref{sec:outlier_prior} are expected to reflect the abrupt variations in the presence of water observed at $t = 3$, $4$ and $5$. These observations, combined with the extracted signatures (Fig. \ref{fig:real_endm}) and the estimated abundances (Figs. \ref{fig:A1_real} to \ref{fig:A3_real}) lead to the following comments.
\begin{itemize}
\item \textbf{Endmember estimation:} the signature recovered for the soil by VCA, SISAL and RLMM  at time $t = 5$ shows an amplitude which is significantly greater than the amplitude of the signatures extracted at the other time instants, and a shape incompatible with what can be expected based on physical considerations (see the black signatures in Figs. \ref{fig:rendm1_vca}, \ref{fig:rendm1_sisal} and \ref{fig:rendm1_rlmm}). This is a clear indication that outliers are present in the corresponding image. A similar observation can be made for the vegetation signature obtained by VCA, SISAL and RLMM at time $t = 5$. On the contrary, the endmembers recovered by OU and the proposed approach are much more consistent from this point of view.
\item \textbf{Abundance estimation:} the estimated abundances globally reflect the previous comments made on the extracted endmembers. Notably, the abundance coefficients estimated at $t = 5$ by VCA, SISAL and RLMM (delineated in red in Figs. \ref{fig:A1_real} to \ref{fig:A3_real}) are visually inconsistent with the temporal evolution of the materials observed in the true color composition given in Fig. \ref{fig:cube}. More explicitly, the soil is not supposed to be concentrated on a few pixels as suggested by the corresponding abundance maps in Fig. \ref{fig:A1_real}. Similarly, the water is not supposed to be present in high proportions in all the pixels of the image as indicated in Fig. \ref{fig:A2_real}. These results, in contradiction with Fig. \ref{fig:cube}, suggest that outliers are present at $t = 5$. In addition, the abundance maps estimated at $t = 4$ and 6 by FCLS for the water and the vegetation (delineated in green in Figs. \ref{fig:A1_real} and \ref{fig:A2_real}) suggest that the water contribution has been split into two spectra. The corresponding signatures are represented in green in Figs. \ref{fig:rendm1_vca} and \ref{fig:rendm3_vca}. On the contrary, the results reported for OU and the proposed method are consistent with the expected evolution of water and vegetation over time (abundance values close to 1 at time $t = 1$, $3$, $5$, lower values at time $t = 2$, $4$, $6$). Finally, the vegetation abundance maps estimated by the proposed method globally presents a better contrast than those obtained with OU (Fig. \ref{fig:A3_real}). 
\end{itemize}
The previous comments, along with the lower reconstruction error reported in Table \ref{tab:results_real}, suggest that the proposed approach is robust to spatially sparse outliers while allowing smooth temporal variations to be exploited. Indeed, the pixels corresponding to abrupt variations of the water signature have been properly detected. Furthermore, the outliers previously detected in this dataset \cite{Thouvenin2015b} for $t = 5$ (highlighted in red in Fig.~\ref{fig:cube5}) are well captured by the latent variables $\Z$ (see Fig. \ref{fig:map_real}). In addition, the spatial distribution of the estimated outlier labels (Fig.~\ref{fig:map_real}) is in agreement with the results of the RLMM (in terms of the spatial distribution of the outlier energy) and with the non-linearity detector \cite{Altmann2013icassp} applied to each image of the sequence with the SISAL-estimated endmembers (see Fig. \ref{fig:Z_detector}). Concentrated on regions where non-linear effects can be reasonably expected, the active latent variables $\Z$ tend to capture the spatial distribution of the non-linearities possibly occurring in the observed scene.

\input{tab_ega}
\input{tab_results_real}
\input{real_endm}

\section{Conclusion and future work} \label{sec:conclusion}

This paper introduced a Bayesian model accounting for both smooth and abrupt variations possibly occurring in multitemporal hyperspectral images. The adopted model was specifically designed to handle datasets in which mostly the same materials were expected to be observed at different time instants, thus allowing information redundancy to be exploited.
An MCMC algorithm was derived to solve the resulting unmixing problem in order to precisely assess the performance of the proposed approach on multitemporal HS images of moderate size (\ie, moderate spatial and temporal dimensions). This algorithm was used to sample the posterior of the proposed hierarchical Bayesian model and to use the generated samples to build estimators of the unknown model parameters. Given its computational cost, the proposed approach is not intended to be applied to large datasets, for which different unmixing methods can provide a rougher analysis at a smaller computational cost. The proposed approach is rather meant to be used as a complementary tool to carry out an in-depth analysis of scenes of moderate size. Future research perspectives include the use of relaxation methods to the Ising field to tackle similar problems with online optimization techniques, and the development of distributed unmixing procedures to efficiently unmix larger datasets. Designing unmixing methods scaling with the problem dimension while simultaneously accounting for temporal and spatial endmember variability is another interesting prospect.

\begin{appendices}

\section{Sampling the projected endmembers $\E$} \label{app:subspace}
When using a PCA as a preprocessing step, the projected endmembers $\e_r$, for $r = 1,\dotsc,\nendm$, are distributed according to the following truncated Gaussian distributions
\begin{equation}
\e_r \mid \Yb, \Theta_{\setminus \{ \e_r \}} \sim \mathcal{N}_{\mathscr{E}_r} (\bmu_r^{(\E)}, \bLambda_r)
\end{equation}
with $\mathscr{E}_r = [c_{1,r},d_{1,r}] \times \dotsc \times [c_{K,r},d_{K,r}]$, and for $k = 1, \dotsc, K$
\begin{equation*}
c_{k,r} = \max_{\ell \in \mathcal{U}_k^+} \left( - \frac{\check{y}_\ell + \sum_{j \neq k} u_{\ell,j} e_{j,r} + b_{\ell,r}}{u_{\ell,k}} \right)
\end{equation*}
\begin{equation*}
d_{k,r} = \min_{\ell \in \mathcal{U}_k^-} \left( - \frac{\check{y}_\ell + \sum_{j \neq k} u_{\ell,j} e_{j,r} +  b_{\ell,r}}{u_{\ell,k}} \right)
\end{equation*}
\begin{equation*}
b_{\ell,r} = \min \Bigl\{ 0,\min_t (dm_{\ell,r,t}) \Bigr\}
\end{equation*}
\begin{equation}
\bLambda_r^{-1} = \Biggl[ \frac{1}{\xi} + \sum_{n,t} \frac{a_{r,n,t}^2}{\sigma_t^2} \Bigr] \I_{\nendm - 1} 
\end{equation}
\begin{equation*}
\begin{split}
\bmu_r^{(\E)} &= \bLambda_r \t{\U} \biggl[ \sum_{t,n} \frac{1}{\sigma_t^2} \Bigl(  \y_{n,t} - \x_{n,t} - \dM_t \a_{n,t} \Bigr. \biggr. \\
& \biggl. \Bigl. - \check{\y} a_{r,n,t} - \sum_{j \neq r} a_{j,n,t} \m_j \Bigr) a_{r,n,t} \biggr].
\end{split}
\end{equation*}

\input{real_abundances}
\begin{figure}[t!]
\centering
\includegraphics[keepaspectratio,width=0.5\textwidth]{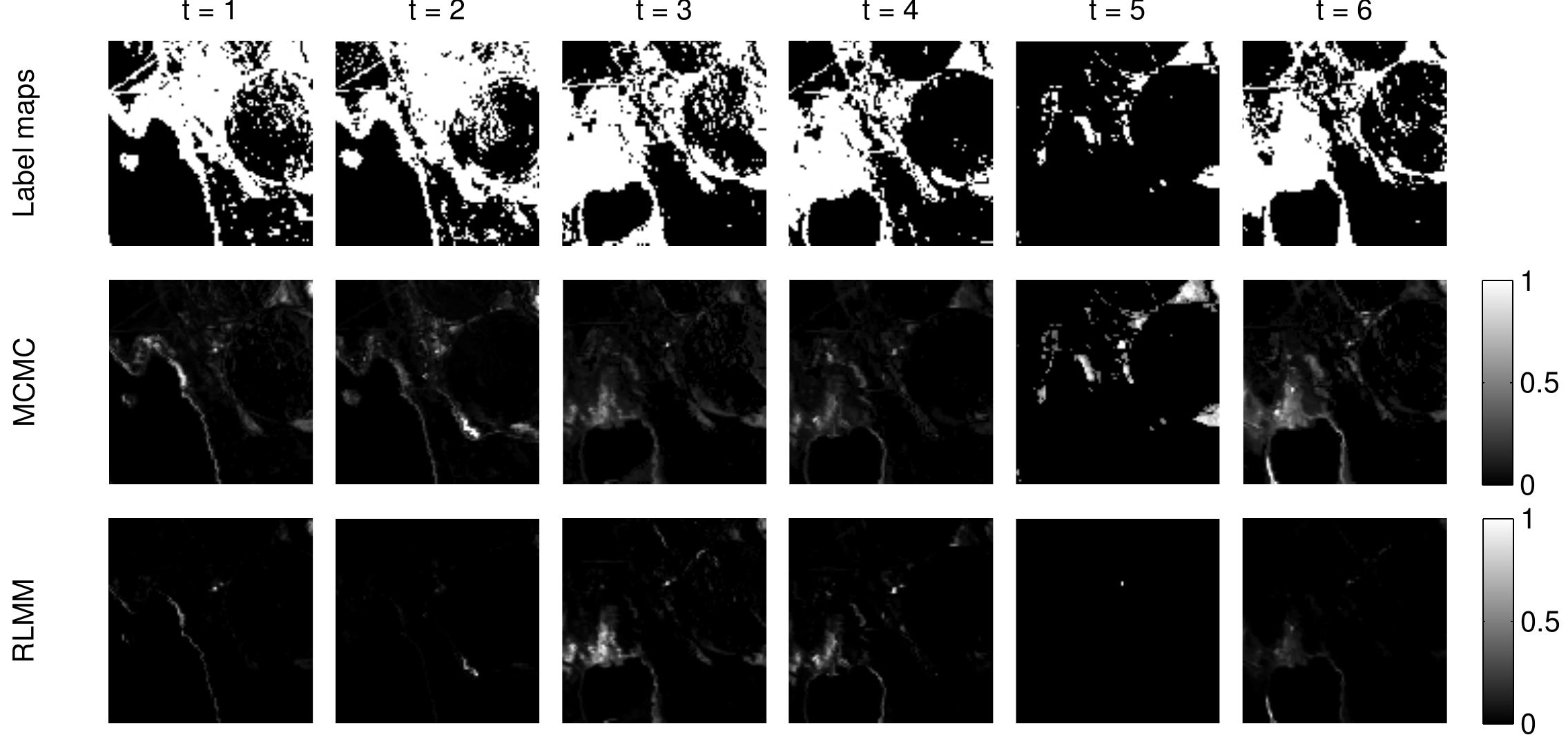}
\caption{mMAP estimates of the label maps recovered by the proposed approach, displayed at each time instant (the different rows correspond to: the estimated label map (pixels detected as outliers appear in white), the outlier energy map re-scaled in the interval $[0,1]$ obtained by the proposed method, and by RLMM).}
\label{fig:map_real}
\end{figure}

\begin{figure}[t!]
\centering
\includegraphics[keepaspectratio,width=0.5\textwidth]{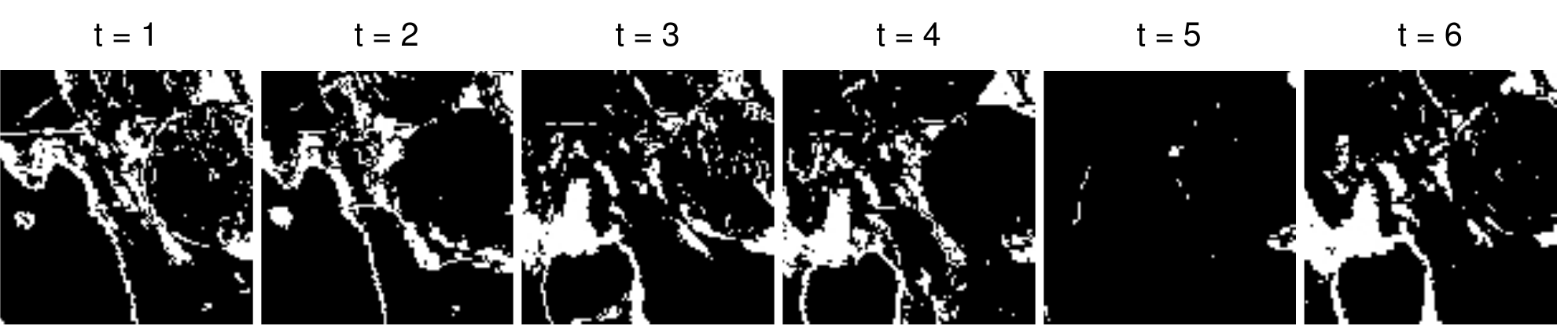}
\caption{Non-linearity maps estimated by the detector \cite{Altmann2013icassp} applied to each image with the SISAL-extracted endmembers, with a probability of false alarm of $10^{-3}$ (pixels detected as non-linearities appear in white).}
\label{fig:Z_detector}
\end{figure}
\end{appendices}


\bibliographystyle{IEEEtran}
\bibliography{strings_all_ref,all_ref}


\begin{IEEEbiographynophoto}{Pierre-Antoine Thouvenin} (S'15--M'17) received the state engineering degree in electrical engineering from ENSEEIHT, Toulouse, France, and the M.Sc. degree in signal processing from the National Polytechnic Institute of Toulouse (INP Toulouse), both in 2014. At the time of writing this paper, he was working towards the Ph.D. degree within the Signal and Communications Group of the IRIT Laboratory, Toulouse, France. His research interests include statistical modeling, optimization techniques and hyperspectral unmixing.
\end{IEEEbiographynophoto}

\begin{IEEEbiographynophoto}{Nicolas Dobigeon} (S'05--M'08--SM'13) received the state engineering degree in electrical engineering from ENSEEIHT, Toulouse, France, and the M.Sc. degree in signal
processing from the National Polytechnic Institute of Toulouse (INP Toulouse), both in June 2004, as well as the Ph.D. degree and Habilitation {\`a} Diriger des Recherches in Signal Processing from the INP Toulouse in 2007 and 2012, respectively.
He was a Post-Doctoral Research Associate with the Department of Electrical Engineering and Computer Science, University of Michigan, Ann Arbor, MI, USA, from 2007 to 2008.

Since 2008, he has been with the National Polytechnic Institute of Toulouse (INP-ENSEEIHT, University of Toulouse) where he is currently a Professor. He conducts his research within the Signal and Communications Group of the IRIT Laboratory and he is also an affiliated faculty member of the Telecommunications for Space and Aeronautics (T{\'e}SA) cooperative laboratory.
His current research interests include statistical signal and image processing, with a particular interest in Bayesian inverse problems with applications to remote sensing, biomedical imaging and genomics.
\end{IEEEbiographynophoto}

\begin{IEEEbiographynophoto}{Jean-Yves Tourneret} (SM'08) received the ing{\'e}nieur degree in electrical engineering from the Ecole Nationale Sup{\'e}rieure d'Electronique, d'Electrotechnique, d'Informatique, d'Hydraulique et des T{\'e}l{\'e}communications (ENSEEIHT) de Toulouse in 1989 and the Ph.D. degree from the National Polytechnic Institute from Toulouse in 1992. He is currently a professor in the university of Toulouse (ENSEEIHT) and a member of the IRIT laboratory (UMR 5505 of the CNRS). His research activities are centered around statistical signal and image processing with a particular interest to Bayesian and Markov chain Monte-Carlo (MCMC) methods. He has been involved in the organization of several conferences including the European conference on signal processing EUSIPCO'02 (program chair), the international conference ICASSP'06 (plenaries), the statistical signal processing workshop SSP'12 (international liaisons), the International Workshop on Computational Advances in Multi-Sensor Adaptive Processing CAMSAP 2013 (local arrangements), the statistical signal processing workshop SSP'2014 (special sessions), the workshop on machine learning for signal processing MLSP'2014 (special sessions). He has been the general chair of the CIMI workshop on optimization and statistics in image processing hold in Toulouse in 2013 (with F. Malgouyres and D. Kouam{\'e}) and of the International Workshop on Computational Advances in Multi-Sensor Adaptive Processing CAMSAP 2015 (with P. Djuric). He has been a member of different technical committees including the Signal Processing Theory and Methods (SPTM) committee of the IEEE Signal Processing Society (2001-2007, 2010-present). He has been serving as an associate editor for the IEEE Transactions on Signal Processing (2008-2011, 2015-present) and for the EURASIP journal on Signal Processing (2013-present).
\end{IEEEbiographynophoto}

\end{document}

%% file: preamble_article.tex
\usepackage{amsmath}
\usepackage{amsthm}
\usepackage{amssymb}
\usepackage{mathrsfs} 

\usepackage{siunitx} 
\usepackage{dsfont} 
\usepackage{mathtools}

\usepackage{stmaryrd} 
\SetSymbolFont{stmry}{bold}{U}{stmry}{m}{n}
\usepackage{enumitem} 

\usepackage[ruled,vlined]{algorithm2e} 

\usepackage{booktabs}
\usepackage[caption=false,font=footnotesize]{subfig} 
\usepackage{multirow} 
\usepackage[figuresleft]{rotating} 

\usepackage{booktabs}
\usepackage[caption=false,font=footnotesize]{subfig} 
\usepackage{multirow} 
\usepackage[figuresleft]{rotating} 

\usepackage[noadjust]{cite}
\usepackage{graphicx}
\DeclareGraphicsExtensions{.pdf,.jpg,.png}
\usepackage{textcomp} 
\usepackage{xspace} 

\usepackage{url}
\usepackage{hyperref}  
\hypersetup{colorlinks,%
            citecolor=red,%
            filecolor=black,%
            linkcolor=blue,%
            urlcolor=blue}
\newcommand{\ie}{i.e.\xspace}
\newcommand{\eg}{e.g.\xspace}
\newcommand{\ap}{a priori\xspace}
\newcommand{\wrt}{with respect to\xspace}

\theoremstyle{definition}

\newtheorem*{remark}{Remark}


%% file: notations.tex
\newcommand{\ubar}[1]{\text{\b{$#1$}}} 

\def\Y{\mathbf{Y}}
\def\Yb{\ubar{\mathbf{Y}}}
\def\y{\mathbf{y}}
\def\yl{\widetilde{\mathbf{y}}}

\def\M{\mathbf{M}}
\def\ml{\widetilde{\mathbf{m}}}
\def\m{\mathbf{m}}
\def\Mmse{\widehat{\mathbf{M}}^{\text{MMSE}}}

\def\A{\mathbf{A}}
\def\Ab{\ubar{\mathbf{A}}}
\def\a{\mathbf{a}}
\def\ar{\widetilde{\mathbf{a}}}
\def\Amse{\widehat{\mathbf{A}}^{\text{MMSE}}}

\def\X{\mathbf{X}}
\def\Xb{\ubar{\mathbf{X}}}
\def\x{\mathbf{x}}
\def\xl{\widetilde{\mathbf{x}}}
\def\Xmse{\widehat{\mathbf{X}}^{\text{MMSE}}}

\def\dM{\mathbf{dM}}
\def\dMb{\ubar{\mathbf{dM}}}
\def\dml{\widetilde{\mathbf{dm}}}
\def\dm{\mathbf{dm}}
\def\dMmse{\widehat{\mathbf{dM}}^{\text{MMSE}}}

\def\U{\mathbf{U}}

\def\E{\mathbf{E}}
\def\e{\mathbf{e}}

\def\I{\mathbf{I}}
\def\Z{\mathbf{Z}}
\def\z{\mathbf{z}}
\def\zmap{\widehat{z}_{n,t}^{\text{mMAP}}}
\def\s{\mathbf{s}^2}

\def\sig{\boldsymbol{\sigma}^2}
\def\sigb{\sigma_\beta^2}

\def\apsi{a_{\Psi}}
\def\bpsi{b_{\Psi}}

\def\asig{a_{\sigma}}
\def\bsig{b_{\sigma}}

\def\as{a_{\mathbf{s}}}
\def\bs{b_{\mathbf{s}}}

\def\eps{\boldsymbol{\varepsilon}^2}
\def\bPsi{\boldsymbol{\Psi}^2}
\def\bbpsi{\boldsymbol{\psi}^2}

\def\bbeta{\boldsymbol{\beta}}

\def\bmu{\boldsymbol{\mu}}
\def\bLambda{\boldsymbol{\Lambda}}

\def\Tf{\mathscr{T}_{n,t}^1}
\def\Ts{\mathscr{T}_{n,t}^2}

\def\tf{\tau_{n,t}^1}
\def\ts{\tau_{n,t}^2}

\def\nmc{N_{\text{MC}}}
\def\nbi{N_{\text{bi}}}

\newcommand{\zero}[1]{\mathbf{0}_{#1}}
\def\1#1{\mathbf{1}_{#1}}

\def\id{\mathds{1}}

\def\Sk{\mathcal{S}_{\nendm}}

\def\fnorm#1{\left\lVert #1 \right\rVert_{\text{F}}}%
\def\Fnorm#1{\left\lVert #1 \right\rVert_{\text{F}}^2}%
\def\Enorm#1{\lVert #1 \rVert_{2}^2}%

\def\argmax#1{\underset{#1}{\arg \max \,}}
\def\t#1{#1^{\text{T}}}

\DeclareMathOperator{\aSAM}{aSAM}
\DeclareMathOperator{\GMSE}{GMSE}
\DeclareMathOperator{\RE}{RE}

\DeclareMathOperator{\acosh}{acosh}

\newcommand\nbpix{N}
\newcommand\nendm{R}
\newcommand\nband{L}
\newcommand\ntime{T}

%% file: dag2.tex
\begin{figure}
\centering
\resizebox{0.45\textwidth}{!}{
\begin{tikzpicture}
\footnotesize{
\node[data] (Y) at (0,0) {$\Yb$};
\node (M) at (-3,1) {$\M$};
\node[hp] (xi) at (-3.5,2) {$\xi$};
\node (dM) at (-2,1) {$\dMb$};
\node[hp] (nu) at (-1.5,2) {$\nu$};
\node (Psi) at (-2.5,2) {$\bPsi$};
\node[hp] (apsi) at (-3,3) {$\apsi$};
\node[hp] (bpsi) at (-2,3) {$\bpsi$};
\node (Sig) at (0,1) {$\sig$};
\node[hp] (asig) at (-0.5,2) {$\asig$};
\node[hp] (bsig) at (0.5,2) {$\bsig$};
\node (A) at (2,1) {$\Ab$};
\node[hp] (eps) at (1.5,2) {$\eps$};
\node (X) at (3,1) {$\Xb$};
\node (Z) at (2.5,2) {$\Z$};
\node (s) at (3.5,2) {$\s$};
\node (beta) at (2,3) {$\mathbf{\bbeta}$};
\node[hp] (as) at (3,3) {$\as$};
\node[hp] (bs) at (4,3) {$\bs$};

\draw[<-,>=latex] (Y) edge[out=180, in=-75] (dM)
                      edge (Sig); 
\draw[<-,>=latex] (Y.east) to[out=0, in=-90] (A.south);
\draw[<-,>=latex] (Y.east) to[out=0, in=225] (X);                                            
\draw[<-,>=latex] (A) edge (eps);
\draw[<-,>=latex] (A.east) to[in=180,out=0] (X.west);

\draw[<-,>=latex] (X) edge (Z)
                      edge (s);
\draw[->,>=latex] (beta) to[bend left] (Z);                                           
\draw[<-,>=latex] (Sig) edge (asig)
                        edge (bsig);
\draw[<-,>=latex] (s) edge (as)
                      edge (bs);
\draw[<-,>=latex] (dM) edge (Psi)
                       edge (nu);                     
\draw[<-,>=latex] (Psi) edge (apsi)
                        edge (bpsi);
\draw[->,>=latex] (M.east) to[in=180,out=0] (dM.west);
\draw[->,>=latex] (M) to[out=-45, in=180] (Y.west);                                            
\draw[->,>=latex] (xi) -- (M);
}
\end{tikzpicture}
}
\caption{Directed acyclic graph associated with the proposed Bayesian model (fixed parameters appear in boxes).}
\label{fig:dag}
\end{figure}

%% file: gibbs.tex
\begin{algorithm}[!t]
\footnotesize{
 \KwData{$\nbi$, $\nmc$, $\M^{(0)}$, $\Ab^{(0)}$, $\dMb^{(0)}$, $\Xb^{(0)}$, $\boldsymbol{\sigma}^{2(0)}$, $\Z^{(0)}$, $\bbeta^{(0)}$, $\mathbf{s}^{2(0)}$, $\boldsymbol{\Psi}^{2(0)}$.}
 \Begin{
  \For{$q = 1$ \KwTo $\nmc$}{
    Sample the endmembers $\M^{(q)}$, cf. \S \ref{sec:sample_M} \smallskip\;
    Sample the variability terms $\dMb^{(q)}$, cf. \S \ref{sec:sample_dM}\smallskip\;
    Sample the abundances $\Ab^{(q)}$, cf. \S \ref{sec:sample_A} \smallskip\;
    Sample the labels and the outliers $(\Z^{(q)},\Xb^{(q)})$, cf. \S \ref{sec:sample_X_Z}\smallskip\;
    Sample the outlier variances $\mathbf{s}^{2(q)}$, cf. \S \ref{sec:sample_s}\smallskip\;
    Sample the noise variances $\boldsymbol{\sigma}^{2(q)}$, cf. \S \ref{sec:sample_sig}\smallskip \;
    Sample the variability variances $\boldsymbol{\Psi}^{2(q)}$, cf. \S \ref{sec:sample_psi}\smallskip\;
    Sample the granularity parameters $\bbeta^{(q)}$, cf. \S \ref{sec:sample_beta}\smallskip\;  
   }
 }
 \KwResult{$\Bigl\{\M^{(q)},\dMb^{(q)},\Ab^{(q)},\Z^{(q)},\Xb^{(q)},\boldsymbol{\sigma}^{2(q)}, \Z^{(q)}, \bbeta^{(q)}, \mathbf{s}^{2(q)},\Bigr. $\smallskip\\
 $\Bigl. \boldsymbol{\Psi}^{2(q)} \Bigr\}_{q = 1}^{\nmc}$.}
}
 \caption{Proposed hybrid Gibbs sampler.}
 \label{alg:gibbs} 
\end{algorithm}

%% file: true_endm.tex
\begin{figure}[t]
\centering
\resizebox{0.48\textwidth}{!}{%
\subfloat[4][Endmember 1]{
\includegraphics[keepaspectratio,height=0.15\textheight , width=0.15\textwidth]{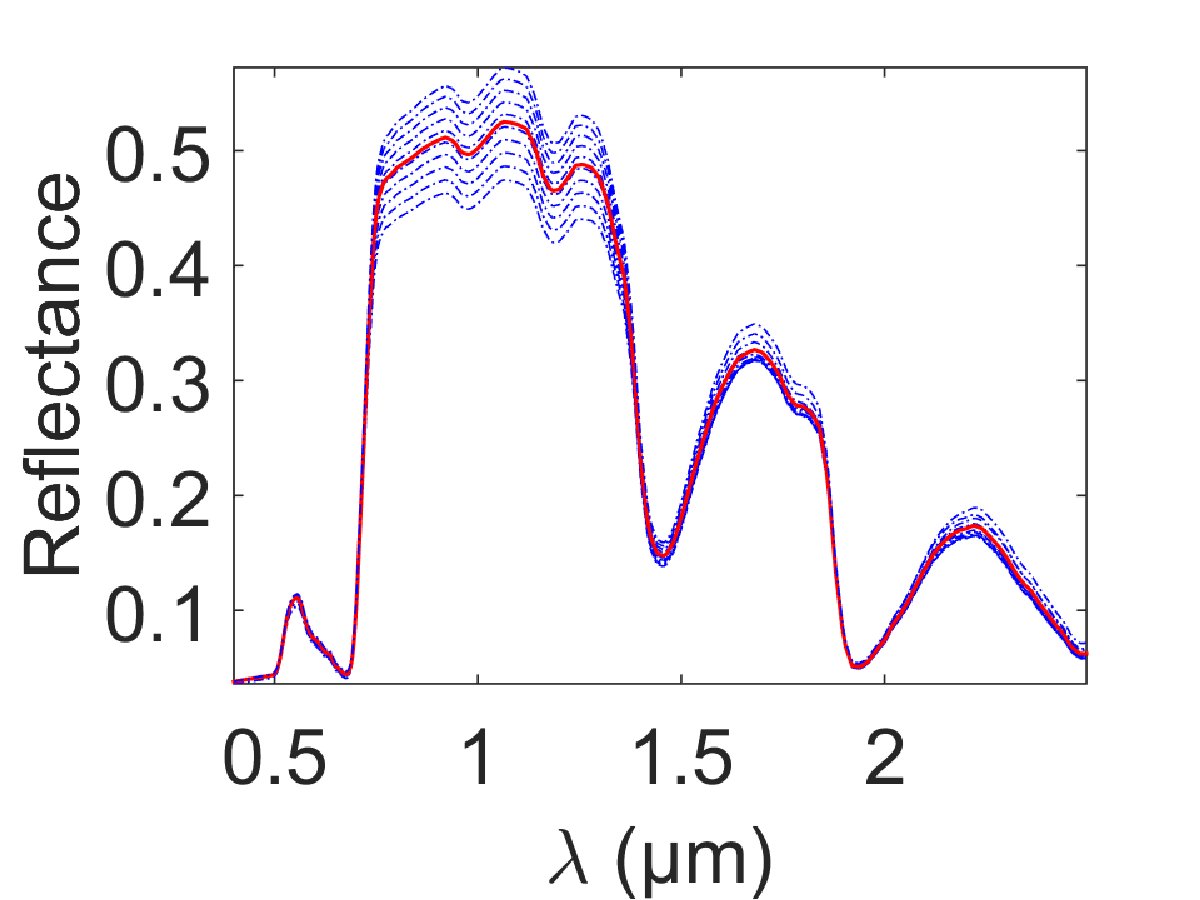}
\label{fig:endm1_true}}
\subfloat[5][Endmember 2]{
\includegraphics[keepaspectratio,height=0.15\textheight , width=0.15\textwidth]{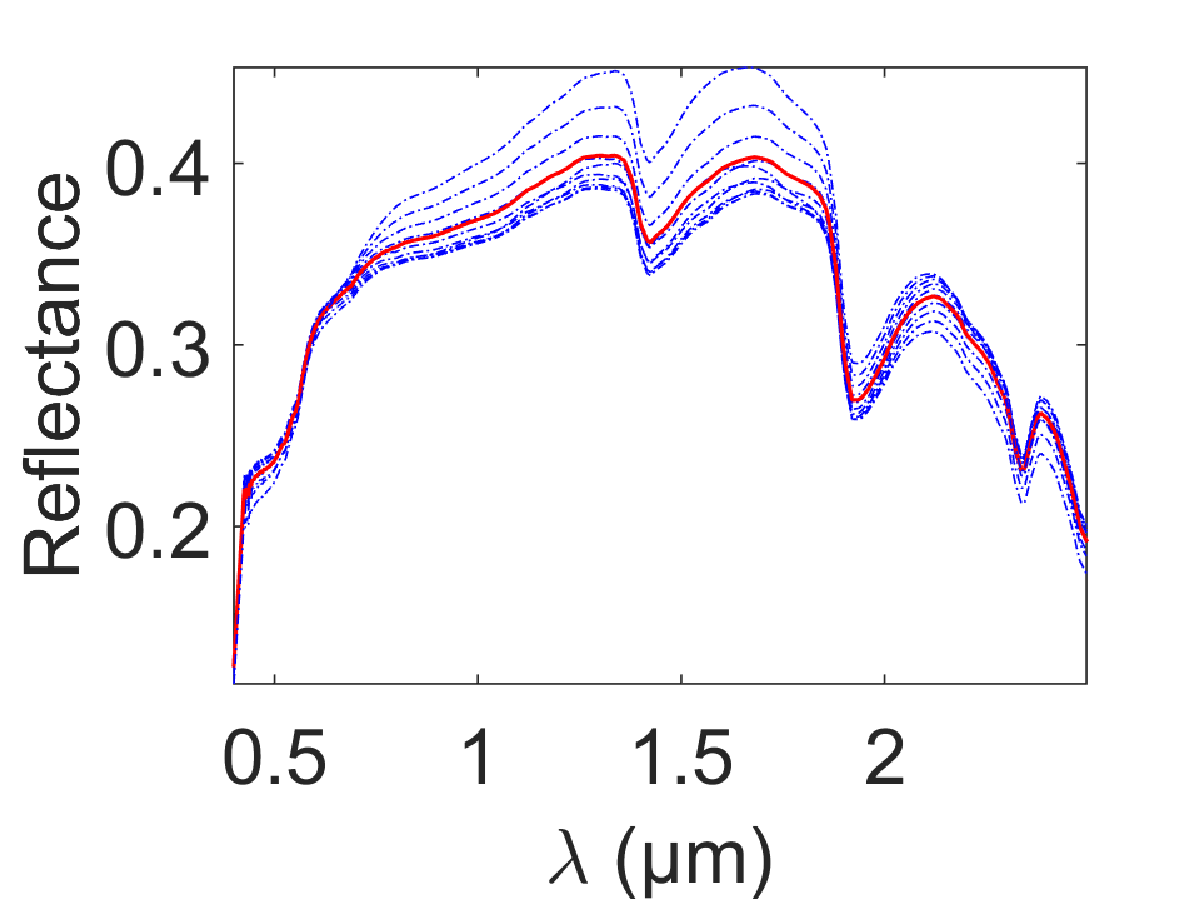}
\label{fig:endm2_true}}
\subfloat[6][Endmember 3]{
\includegraphics[keepaspectratio,height=0.15\textheight , width=0.15\textwidth]{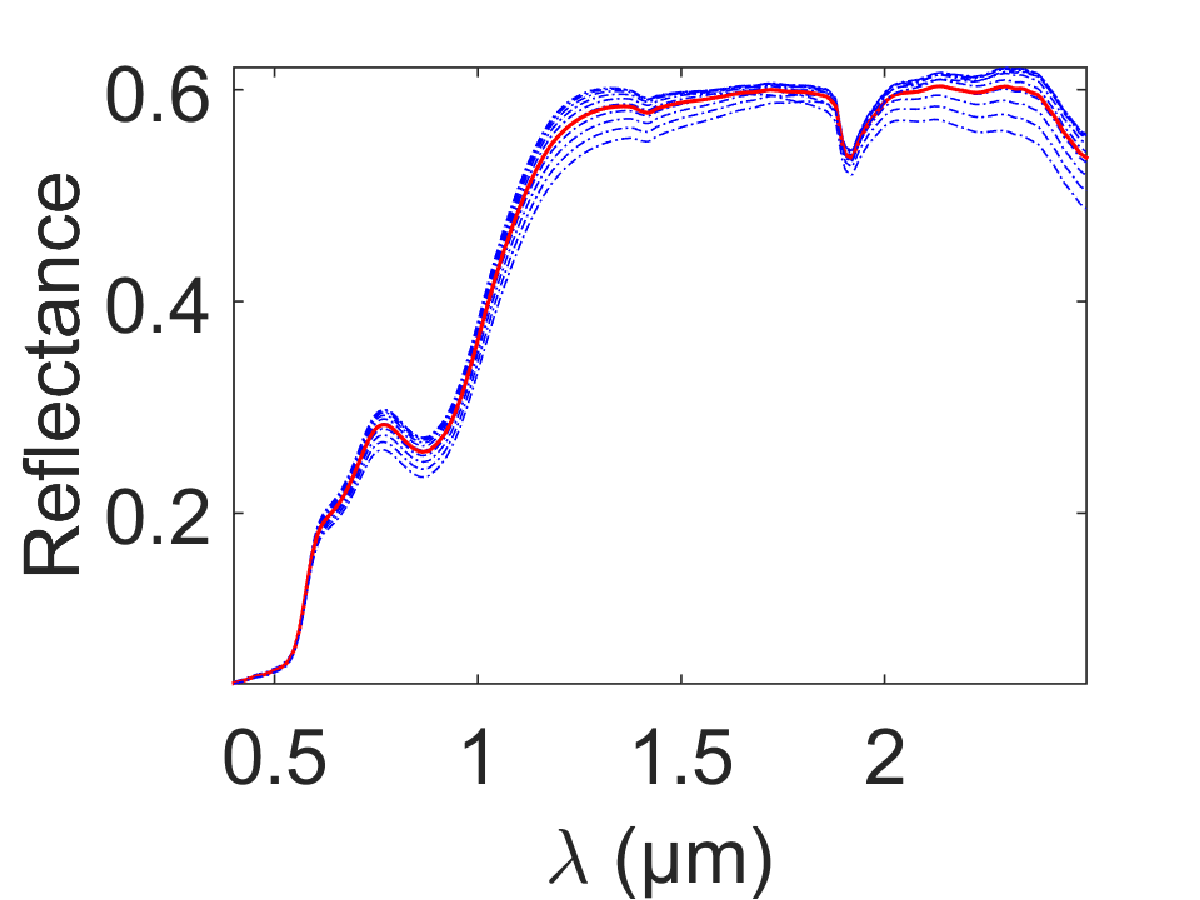}
\label{fig:endm3_true}}
}
\caption{Endmembers ($\m_r$, red lines) and their variants affected by variability ($\m_r + \dm_{r,t}$, blue dotted lines) used to generate the synthetic mixtures with $\nendm = 3$. Signatures corresponding to different time instants are represented in a single figure to better appreciate the variability introduced in the data.}
\label{fig:endm_true}
\end{figure}

%% file: table_parameters.tex
\begin{table}[!t] 
\caption{Fixed parameters, and initial values associated in the experiments to parameters later inferred from the model.}
\label{tab:param1}
	\begin{center}
		\begin{tabular}{@{}llll@{}} \toprule
& 		 					& Synthetic data	& Real data     \\ \cmidrule{3-4} 	
\multirow{7}{*}{\rotatebox{90}{Fixed parameters}}
&$\varepsilon_n^2$			& $10^{-3}$  & $10^{-2}$		 	\smallskip \\
&$\xi$			 			& 1		& 1		  	\\	 
&$\nu$						& $10^{-3}$	& $10^{-5}$	 	\\
&$\as,\apsi,\asig$			& $10^{-3}$  	& $10^{-3}$		 	\\ 
&$\bs,\bpsi,\bsig$			& $10^{-3}$  	& $10^{-3}$		 	\\  
&$\nbi$			 			& 350	   & 450	  	 	\\
&$\nmc$ 						& 400	   & 500	  	 	\\
\cmidrule{1-4}
\multirow{4}{*}{\rotatebox{90}{Initial values}} 
&$\sigma_t^2$				& $10^{-4}$	   & $10^{-4}$		 		\smallskip \\ 
&$s_t^2$						& $5 \times 10^{-3}$	   & $5 \times 10^{-3}$		 		\smallskip \\ 
&$\psi_{\ell,r}^2$			& $10^{-3}$  & $10^{-2}$		 	\smallskip \\	
&$\beta_t$					& 1.7 	& 1.7		 	\\ 
 \bottomrule
		\end{tabular}
	\end{center}
\end{table}

%% file: synth_endm.tex
\begin{figure}[t]
\centering
\subfloat[2][VCA]{
\includegraphics[keepaspectratio,height=0.15\textheight , width=0.15\textwidth]{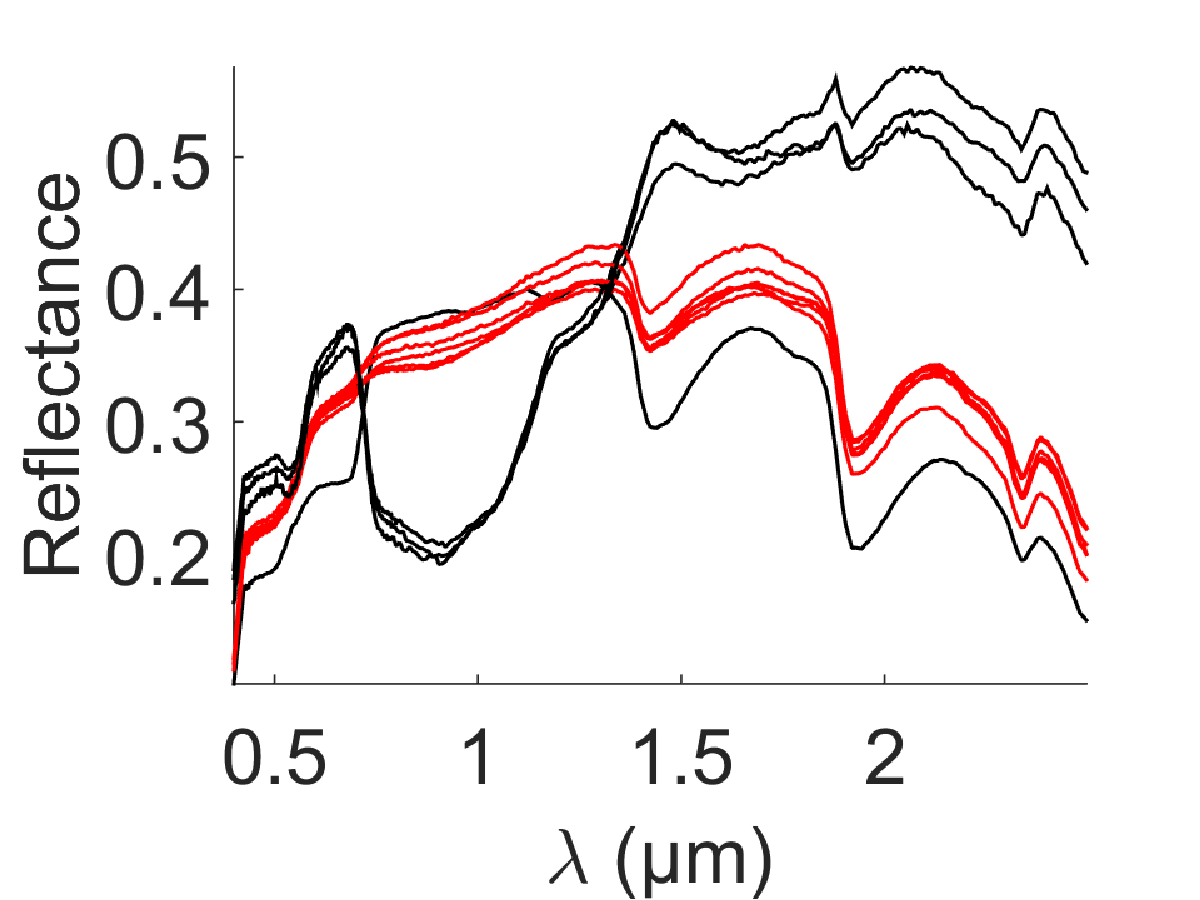}
\label{fig:endm2_vca}}
\subfloat[2][SISAL]{
\includegraphics[keepaspectratio,height=0.15\textheight , width=0.15\textwidth]{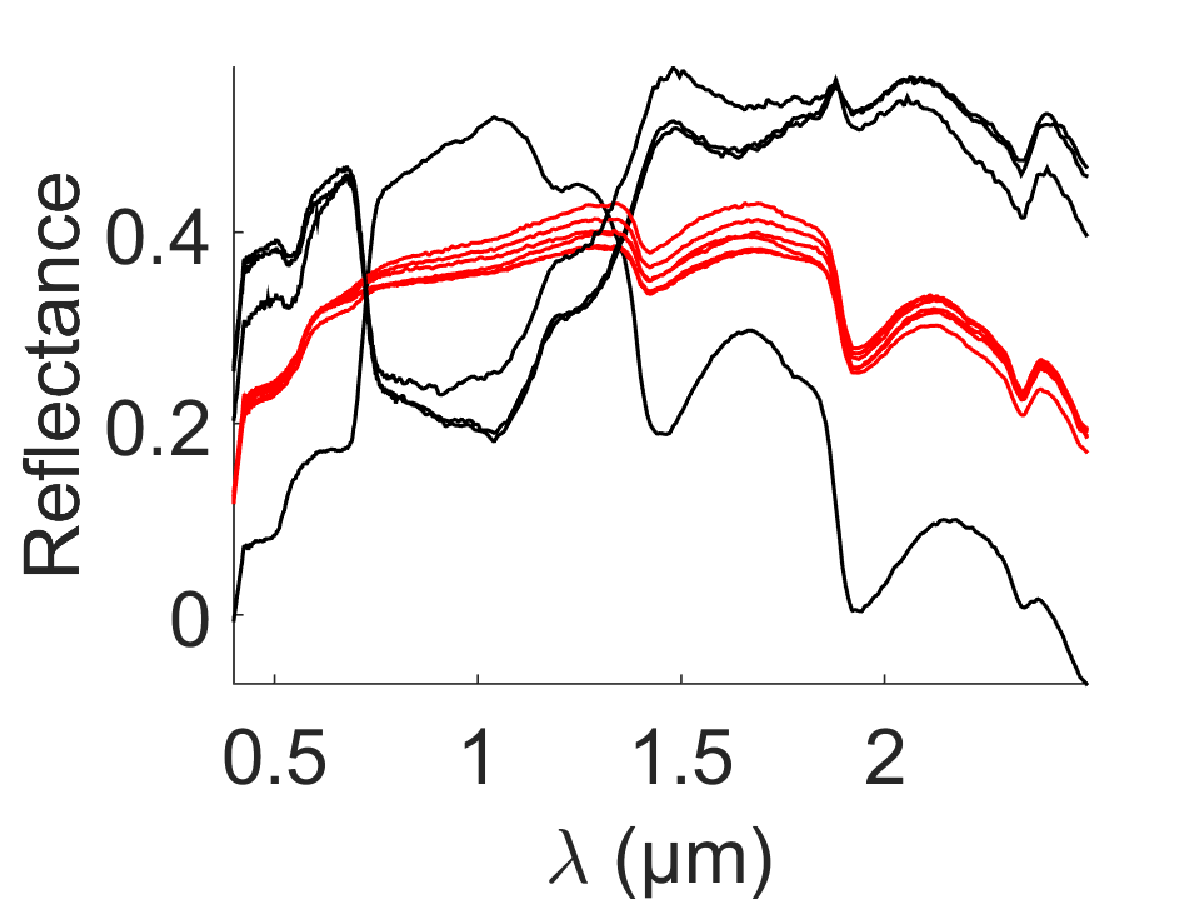}
\label{fig:endm2_sisal}}
\subfloat[2][RLMM]{
\includegraphics[keepaspectratio,height=0.15\textheight , width=0.15\textwidth]{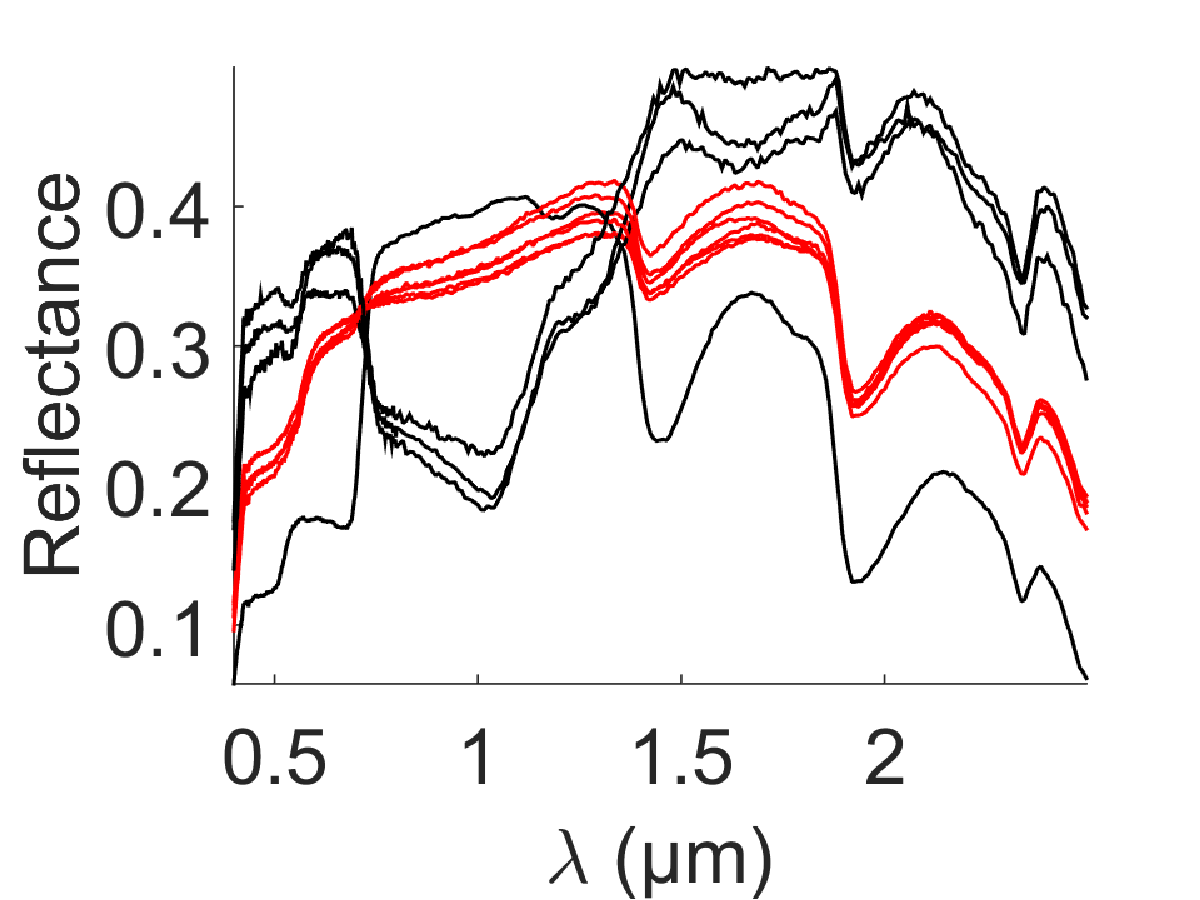}
\label{fig:endm2_rlmm}}
\\
\subfloat[2][OU]{
\includegraphics[keepaspectratio,height=0.15\textheight , width=0.15\textwidth]{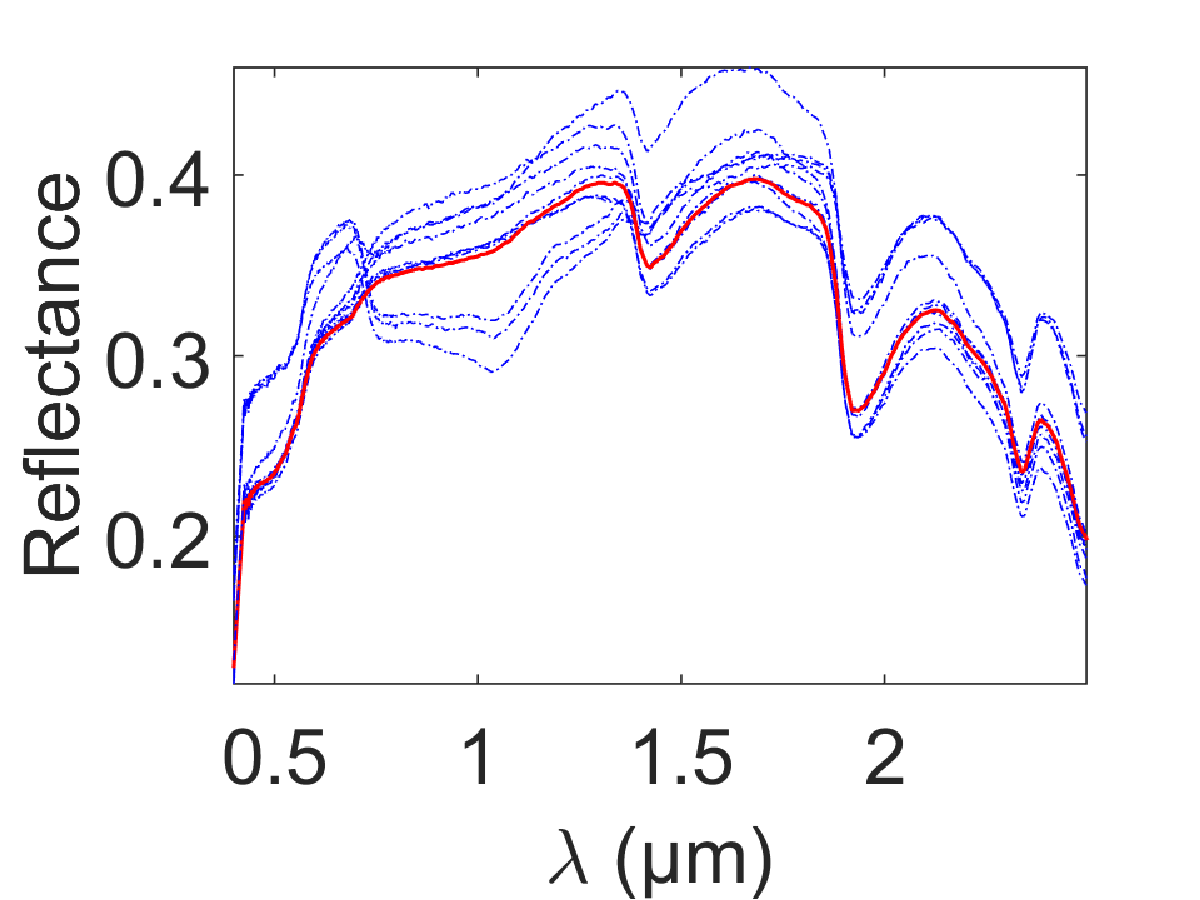}
\label{fig:endm2_ou}}
\subfloat[2][Proposed]{
\includegraphics[keepaspectratio,height=0.15\textheight , width=0.15\textwidth]{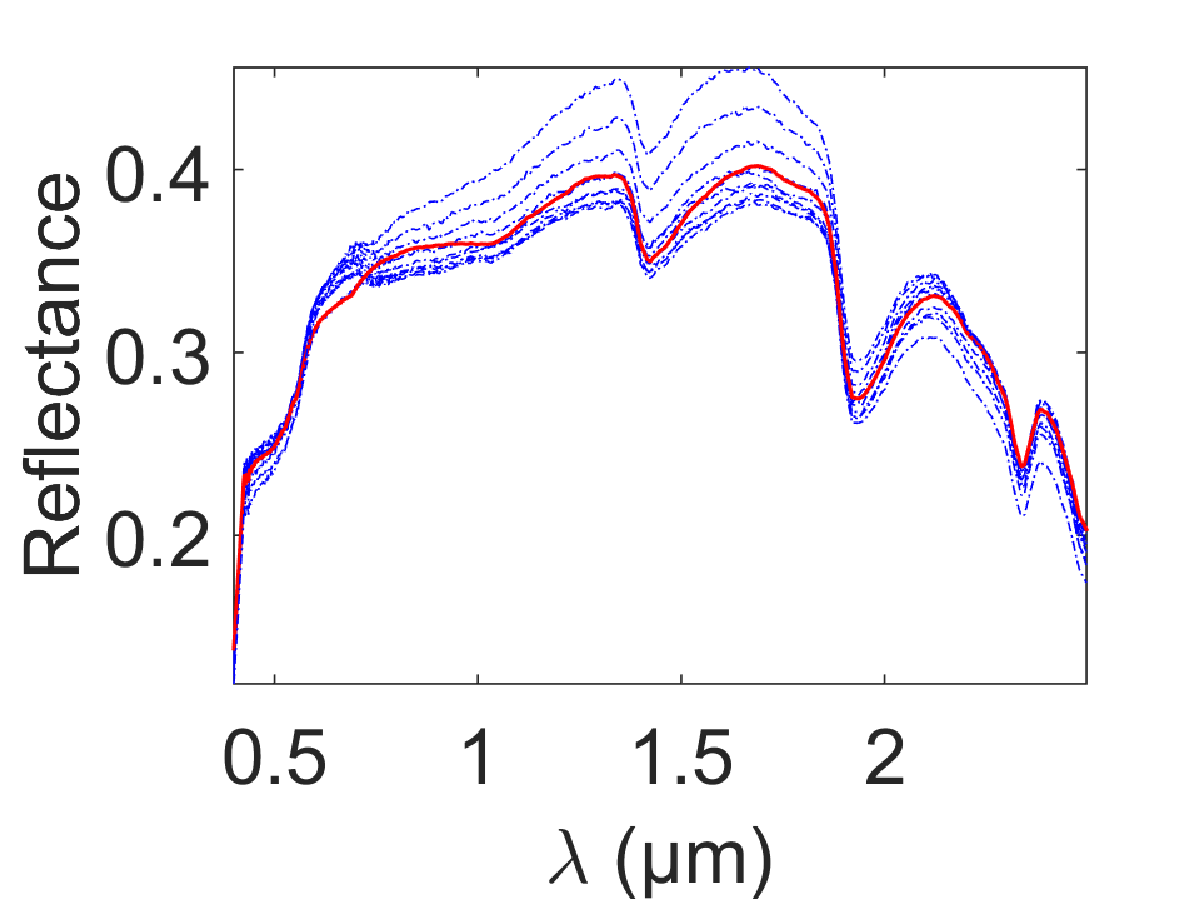}
\label{fig:endm2_mcmc}}
\caption{Second endmember ($\m_2$, red lines) and its variants affected by variability ($\m_2 + \dm_{2,t}$, blue dotted lines) recovered by the different methods from the synthetic mixtures with $\nendm = 3$. Due to space restrictions, the signatures extracted for the other two endmembers have been included in the associated supplementary material. Signatures corresponding to different time instants are represented on a single figure to better appreciate the variability recovered from the data. The spectra represented in black correspond to signatures corrupted by outliers.}
\label{fig:synth_endm}
\end{figure}

%% file: synth_abundances.tex
\begin{figure}[tbhp]
\centering
\resizebox{0.48\textwidth}{!}{
\includegraphics[keepaspectratio,width=0.5\textwidth]{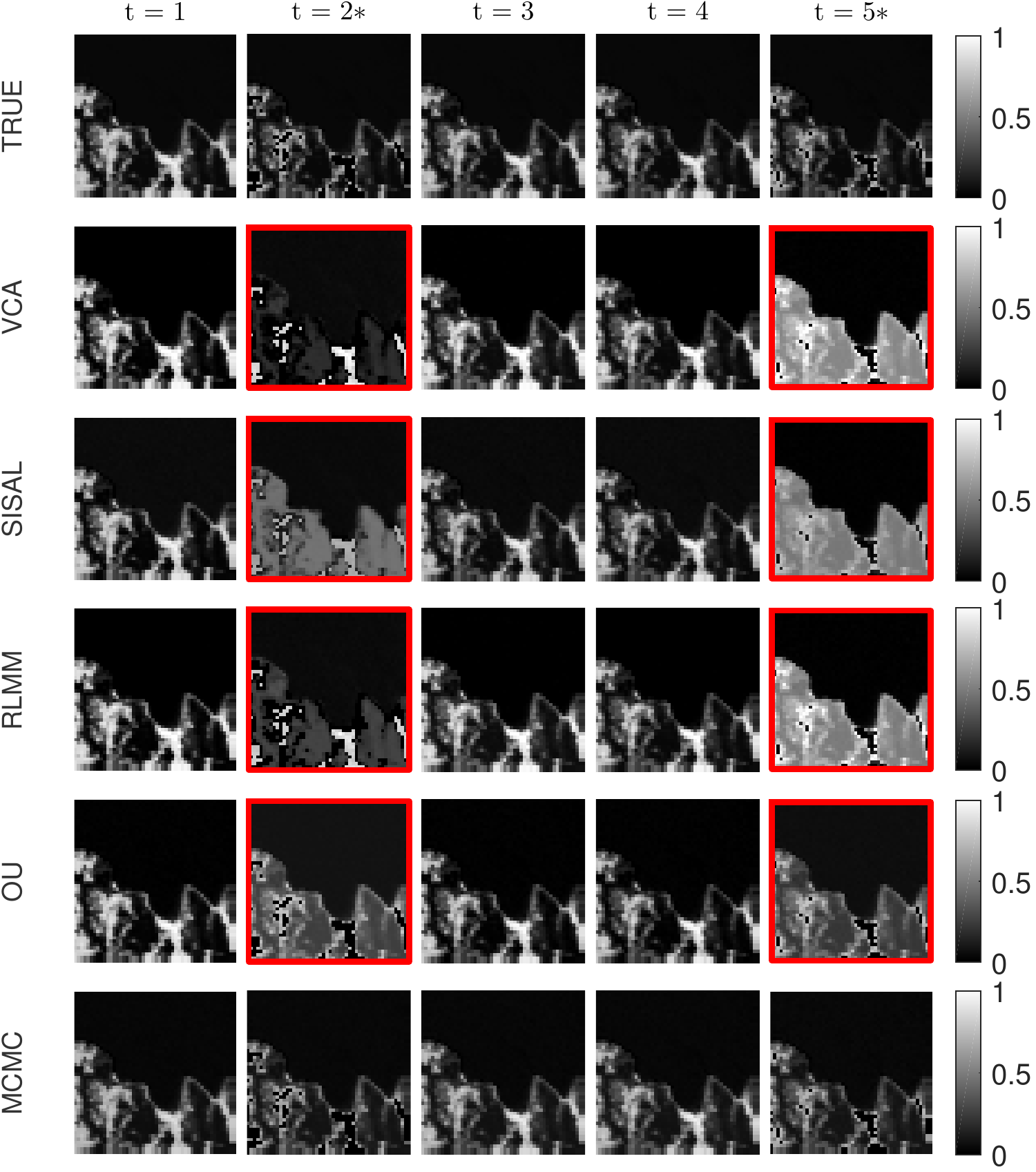}}
\caption{Abundance map of the first endmember recovered by the different methods (in each row) at the first five time instants (given in column) for the experiment with $\nendm = 3$ [the different rows correspond to the true abundances, VCA/FCLS, SISAL/FCLS, RLMM, OU and the proposed method]. The images delineated in red show that several methods are highly sensitive to the presence of outliers, and the time instants represented with $\ast$ denote images containing outliers. Due to space restrictions, the abundance maps obtained at each time instant for each endmember have been included in the supplementary material.}
\label{fig:A1}
\end{figure}

%% file: tab_results_synth.tex
\setlength\columnsep{0.1pt}
\begin{table}[t!] 
\centering
\caption{Simulation results on synthetic data (aSAM($\M$) in (\textdegree), GMSE($\mathbf{A}$)$\times 10^{-2}$, GMSE($\mathbf{dM}$)$\times 10^{-4}$, RE $\times 10^{-4}$, time in (s)).}
	\begin{center}
	\resizebox{0.48\textwidth}{!}{%
		\begin{tabular}{@{}llccccc@{}} \toprule
&		   	  & aSAM($\M$) & GMSE($\A$) & GMSE($\dM$) & RE & time \\ \cmidrule{1-7}
\multirow{5}{*}{\rotatebox{90}{$\nendm = 3$}}
&VCA/FCLS      & 6.07  & 2.32 & / & 3.91 & \textbf{1} \\
&SISAL/FCLS	   & 5.07  & 1.71  & / & 2.28 & 2	\\
&RLMM	       & 5.13 & 2.04 & / & \textbf{0.31} & 463 \\
&OU  		   & \textbf{1.90} & 0.42 & 3.22 & 2.61 & 98 \\
&Proposed 	   & 2.03 & \textbf{0.15} & \textbf{1.85} & 2.00 & 2530 \\ 
\cmidrule{1-7}
\multirow{5}{*}{\rotatebox{90}{$\nendm = 6$}}
&VCA/FCLS      & 3.81  & 1.57 & / & 3.09 & \textbf{2} \\
&SISAL/FCLS	   & 5.76  & 0.91  & / & 4.49 & 3	\\
&RLMM	       & 2.73 & 1.26 & / & \textbf{0.29} & 1453	\\
&OU  		   & 2.74 & 0.38 & 3.70 & 1.13 & 420 \\
&Proposed 	   & \textbf{1.48} & \textbf{0.16} & \textbf{2.84} & 0.51 & 8691 \\ 
\cmidrule{1-7}
\multirow{5}{*}{\rotatebox{90}{$\nendm = 9$}} 
&VCA/FCLS      & 3.74  & 0.65 & / & 6.83 & \textbf{4} \\
&SISAL/FCLS	   & 5.91  & 0.36 & / & 5.56 & 5	\\
&RLMM	       & 2.48  & 0.54  & / & \textbf{0.31} & 1447  \\
&OU  		   & 6.08  & 0.47  & \textbf{2.19} & 0.89 & 1024 \\
&Proposed 	   & \textbf{2.23} & \textbf{0.15} & 8.38 & 0.82 & 17151 \\ 
    \bottomrule
		\end{tabular}
}
	\end{center}
\label{tab:results_synth}
\end{table}

%% file: synth_label.tex
\begin{figure*}[t!]
\centering
\includegraphics[keepaspectratio,height=0.15\textheight , width=0.95\textwidth]{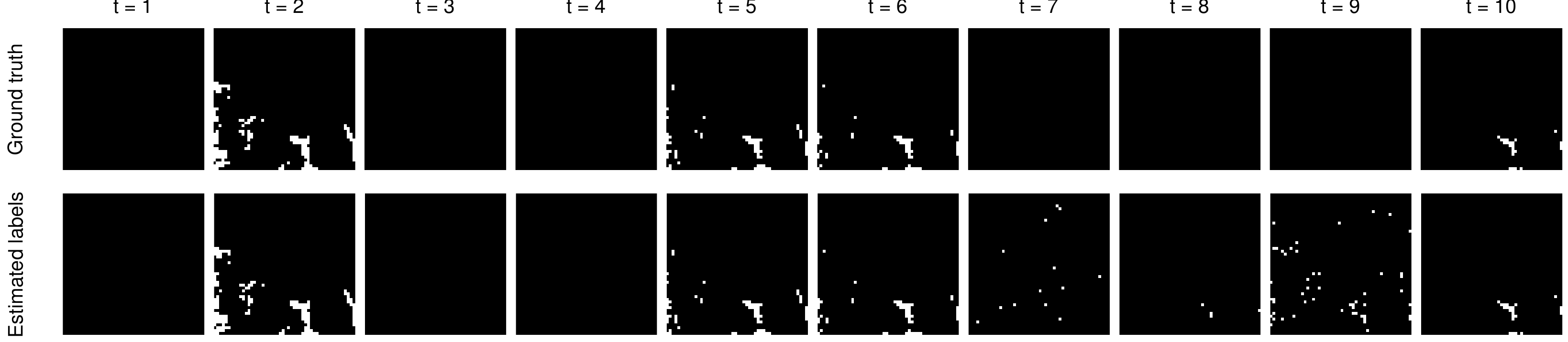}
\caption{Outlier labels $\z_t$ estimated for each image of the synthetic dataset with 3 endmembers (the different rows correspond to the true labels, and the estimated labels) [0 in black, 1 in white].}
\label{fig:label_synth}
\end{figure*}

%% file: time_series.tex
\begin{figure*}[tbhp!]
\centering
\resizebox{0.9\textwidth}{!}{
\subfloat[1][04/10/2014]{
\includegraphics[keepaspectratio,height=0.15\textheight , width=0.15\textwidth]{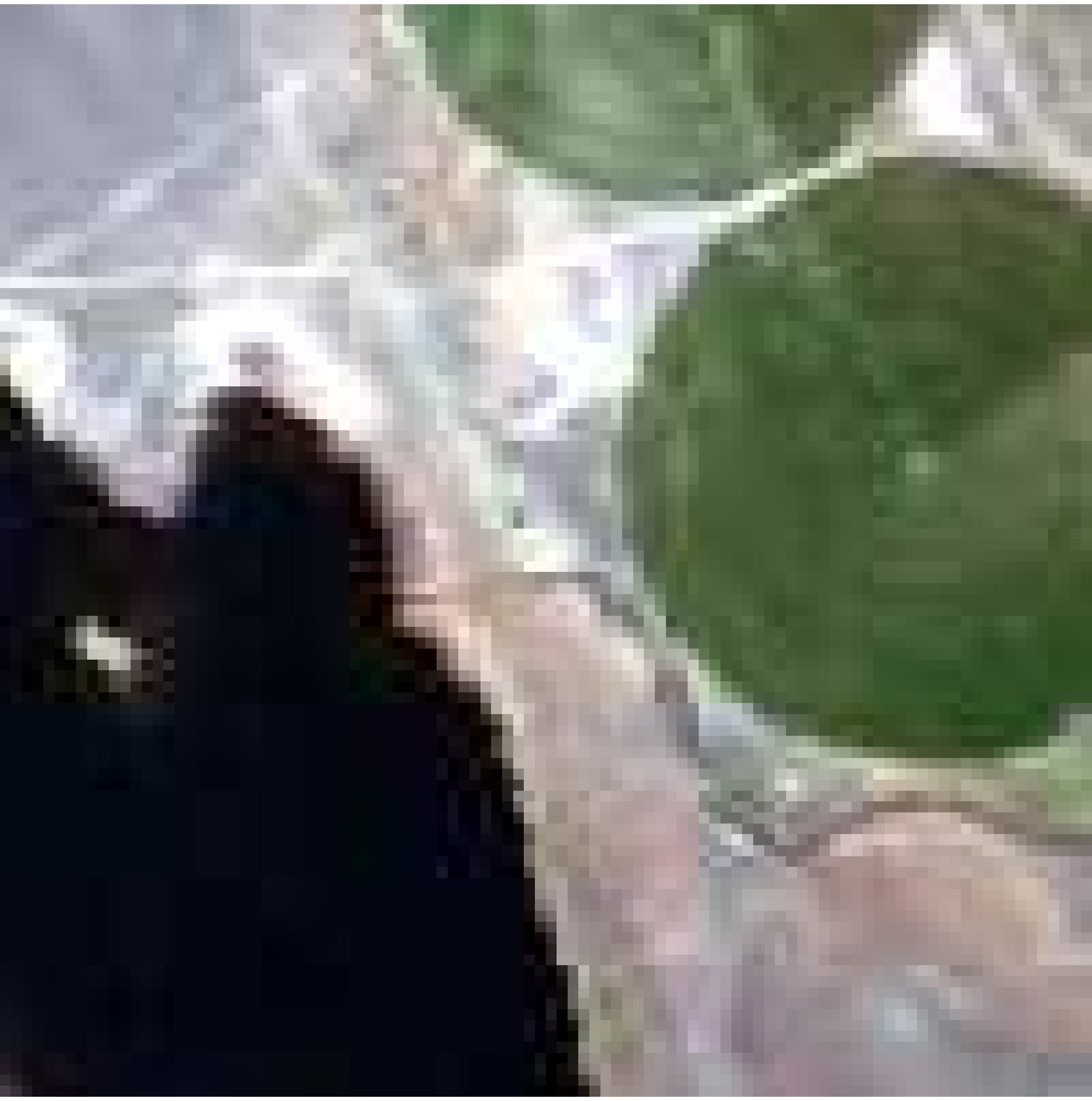}
\label{fig:cube1}}
\subfloat[2][06/02/2014]{
\includegraphics[keepaspectratio,height=0.15\textheight , width=0.15\textwidth]{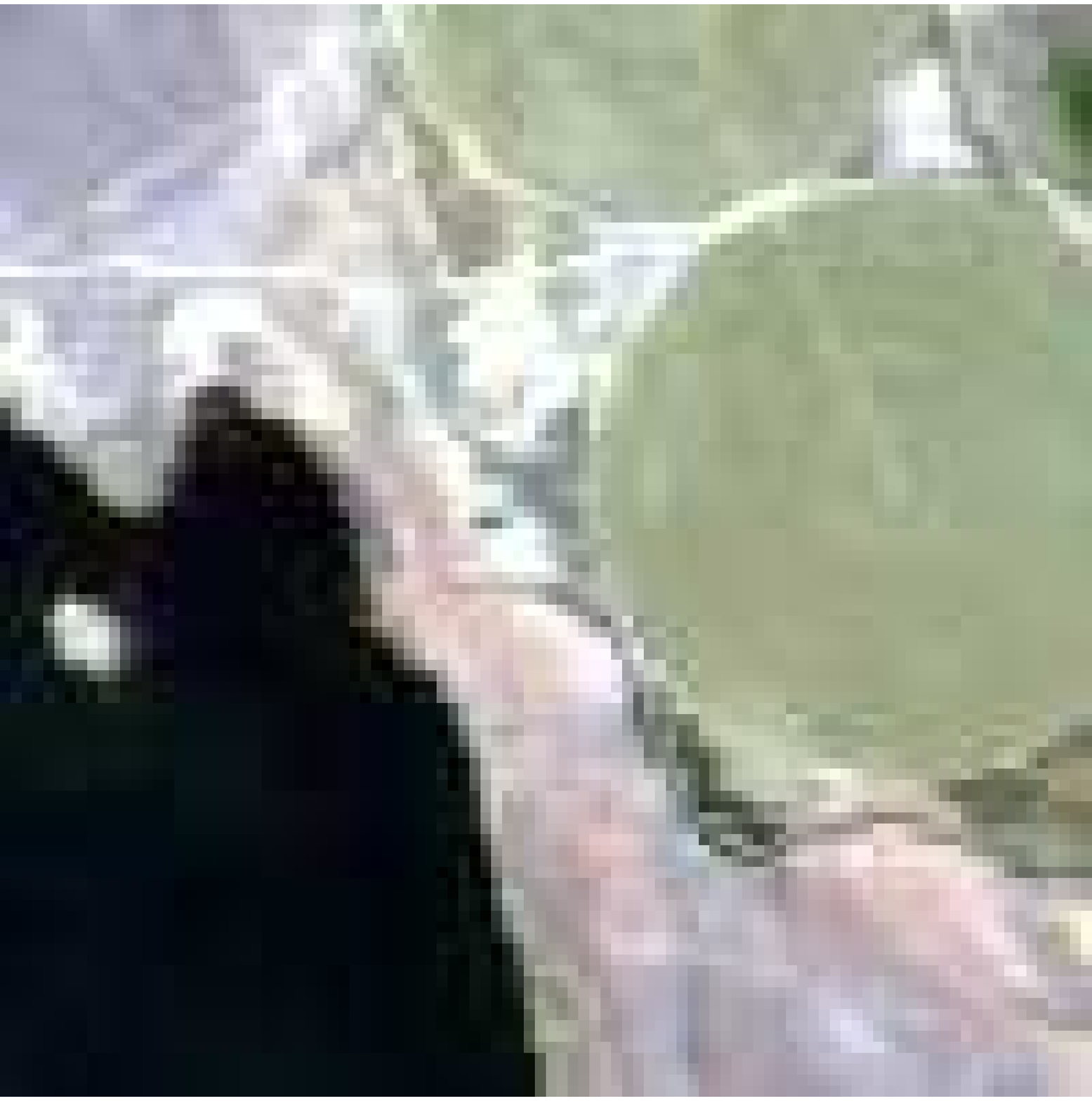}
\label{fig:cube2}}
\subfloat[3][09/19/2014]{
\includegraphics[keepaspectratio,height=0.15\textheight , width=0.15\textwidth]{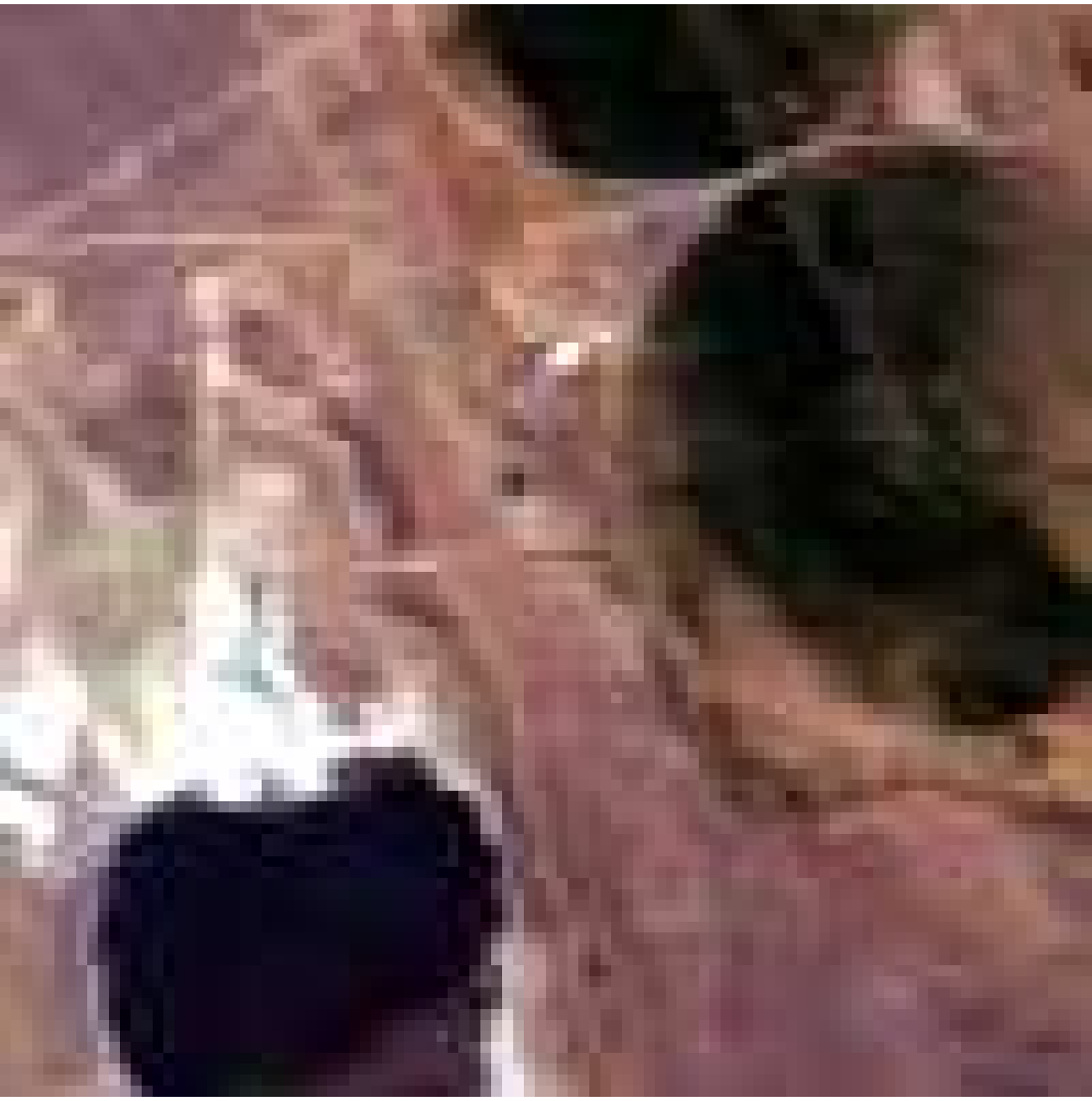}
\label{fig:cube3}}
\subfloat[4][11/17/2014]{
\includegraphics[keepaspectratio,height=0.15\textheight , width=0.15\textwidth]{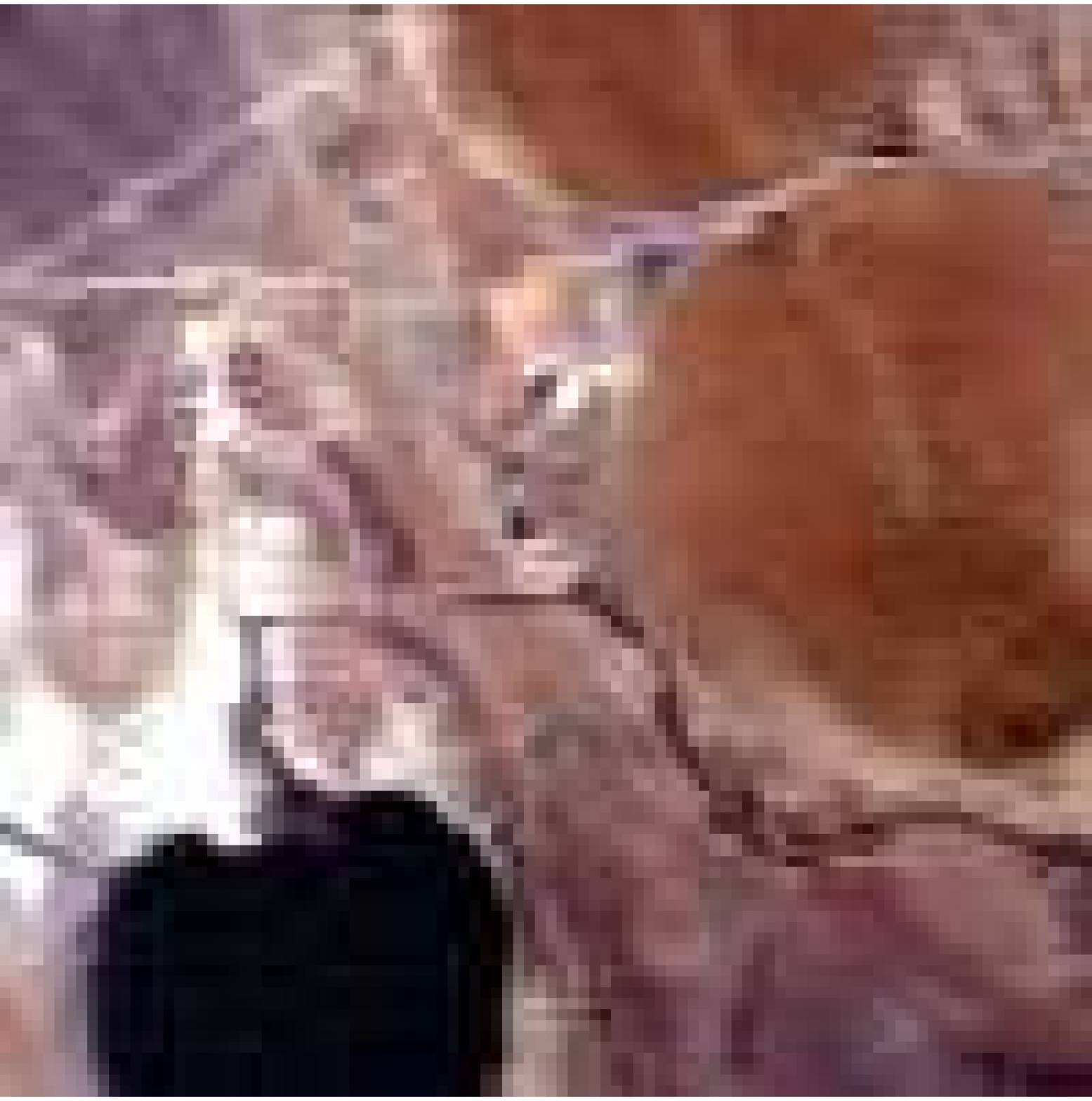}
\label{fig:cube4}}
\subfloat[5][04/29/2015]{
\includegraphics[keepaspectratio,height=0.15\textheight , width=0.15\textwidth]{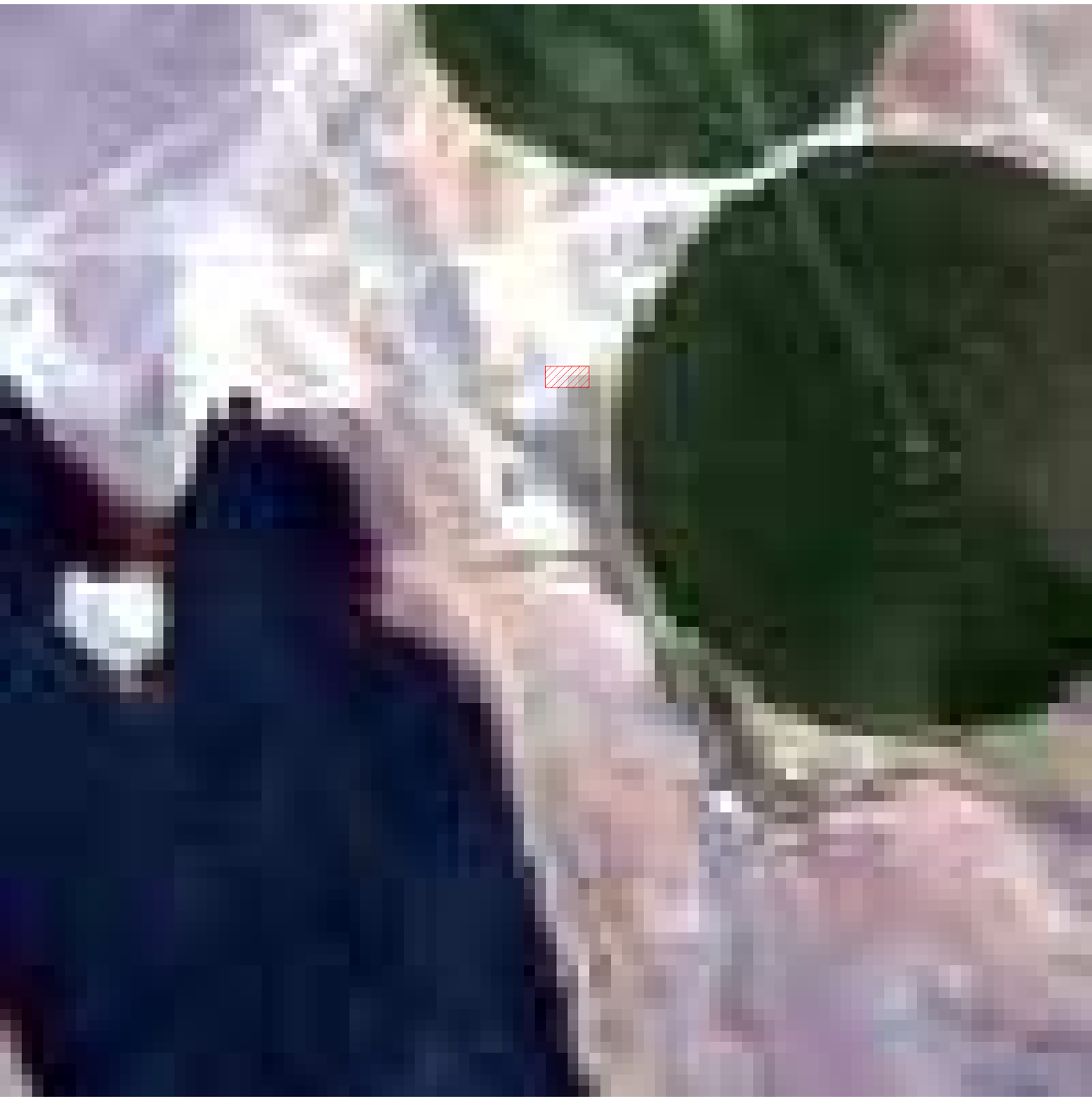}
\label{fig:cube5}}
\subfloat[6][10/13/2015]{
\includegraphics[keepaspectratio,height=0.15\textheight , width=0.15\textwidth]{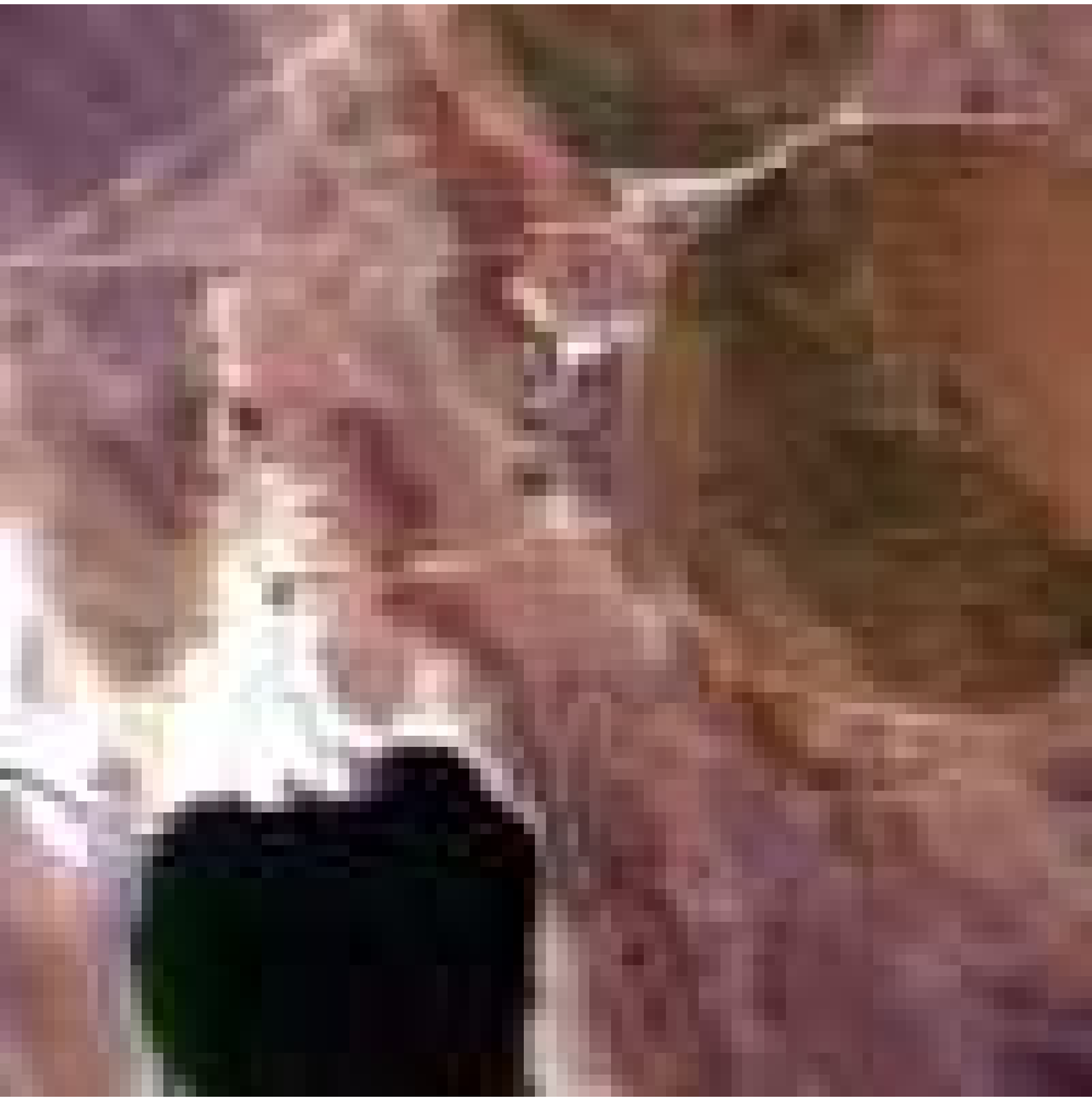}
\label{fig:cube6}}}
\caption{Scenes used in the experiment, given with their respective acquisition date. The area delineated in red in Fig. \ref{fig:cube5} highlights a region known to contain outliers (this observation results from a previous analysis led on this dataset in \cite{Thouvenin2015b}).}
\label{fig:cube}
\end{figure*}

%% file: tab_ega.tex
\begin{table}[t]
\vspace{-0.5cm}
\caption{Endmember number $\nendm$ estimated by NWEGA \cite{Halimi2016} on each image of the real dataset.} 
	\begin{center}
	\resizebox{0.48\textwidth}{!}{%
		\begin{tabular}{@{}lcccccc@{}} \toprule
			& 04/10/2014 & 06/02/2014 & 09/19/201 & 11/17/2014 & 04/29/2015 & 10/13/2015  \\ \cmidrule{2-7}
NWEGA       & 3 & 3 & 3 & 4 & 3 & 4 \\ \bottomrule	
		\end{tabular}
	}
	\end{center}
\label{tab:tab_ega} \vspace{-0.3cm}
\end{table}

%% file: tab_results_real.tex
\setlength\columnsep{0.1pt} 
\begin{table}[!t] 
\vspace{-0.3cm}
\caption{Simulation results on real data (RE $\times 10^{-4}$).}
	\begin{center}
		\begin{tabular}{@{}llcc@{}} \toprule
&		   	   & RE & time (s) \\ \cmidrule{1-4}
\multirow{5}{*}{\rotatebox{90}{$\nendm = 3$}}
&VCA/FCLS      &  45.05 & \textbf{1} \\
&SISAL/FCLS	   &  1.65  & 2 	\\
&rLMM		  & 2.51 & 390 \\
&OU			  & 2.50 & 508  \\
&Proposed 	  & \textbf{0.34} & 23608  \\ 
    \bottomrule
		\end{tabular}
	\end{center}
\label{tab:results_real} \vspace{-0.3cm}
\end{table}

%% file: real_endm.tex
\begin{figure}[t]
\centering
\resizebox{0.48\textwidth}{!}{%
\subfloat[1][Soil (VCA)]{
\includegraphics[keepaspectratio,height=0.15\textheight , width=0.15\textwidth]{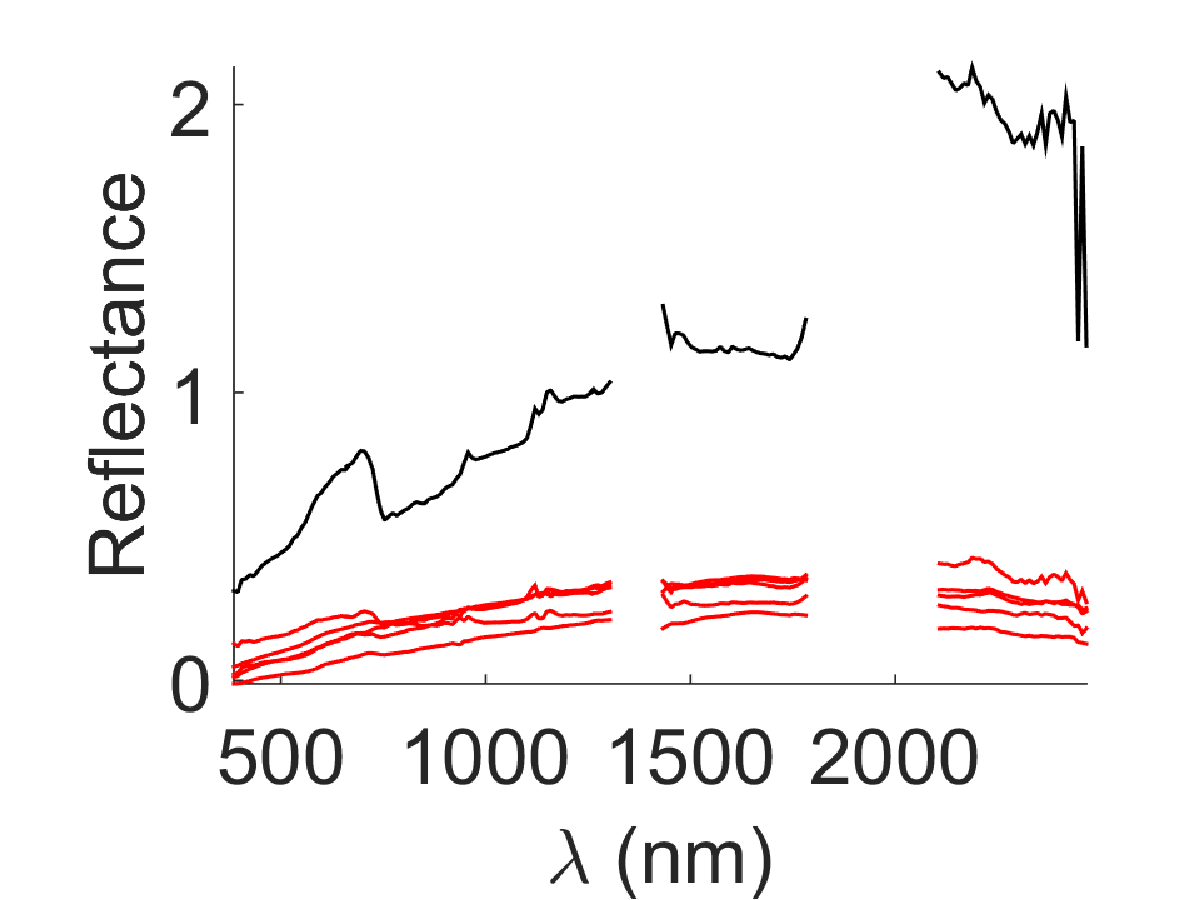}
\label{fig:rendm1_vca}}
\subfloat[2][Water (VCA)]{
\includegraphics[keepaspectratio,height=0.15\textheight , width=0.15\textwidth]{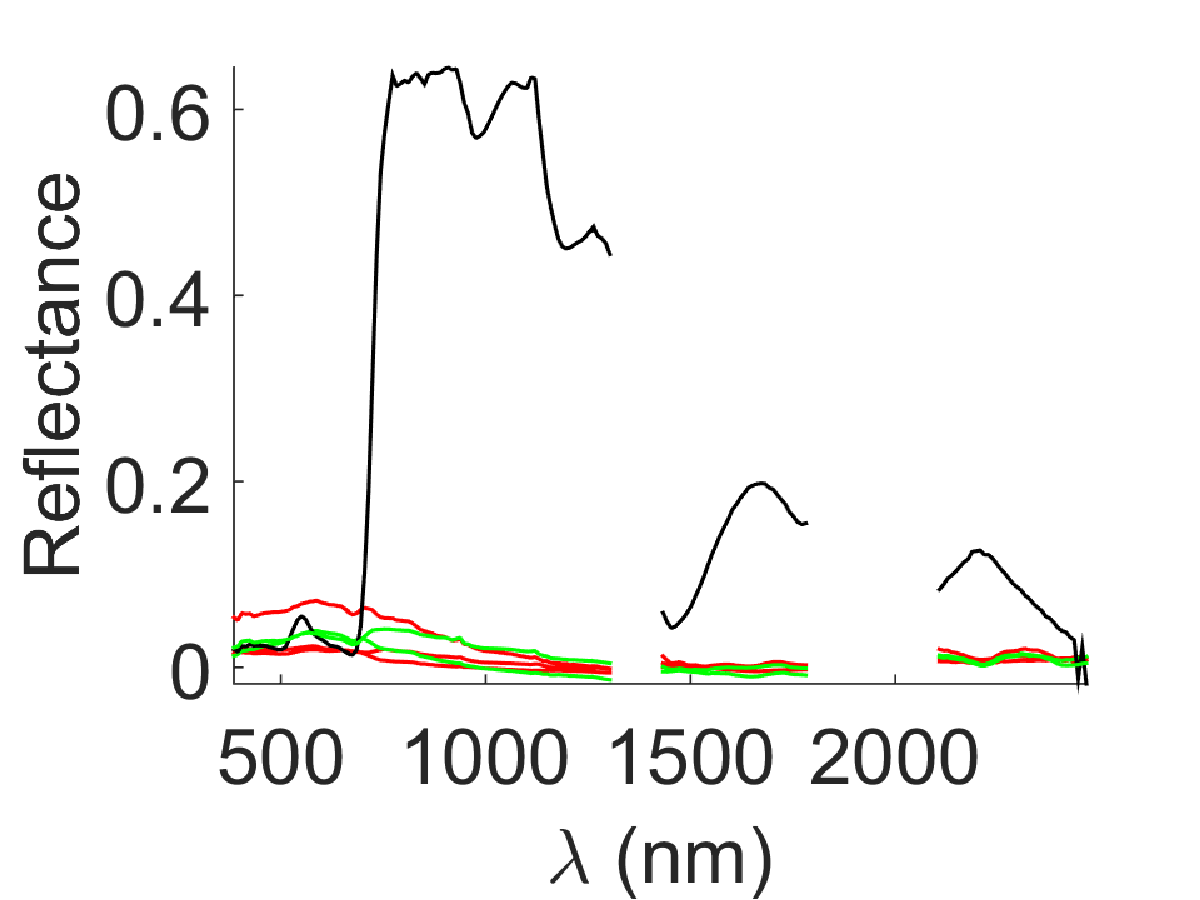}
\label{fig:rendm2_vca}}
\subfloat[3][Veg. (VCA)]{
\includegraphics[keepaspectratio,height=0.15\textheight , width=0.15\textwidth]{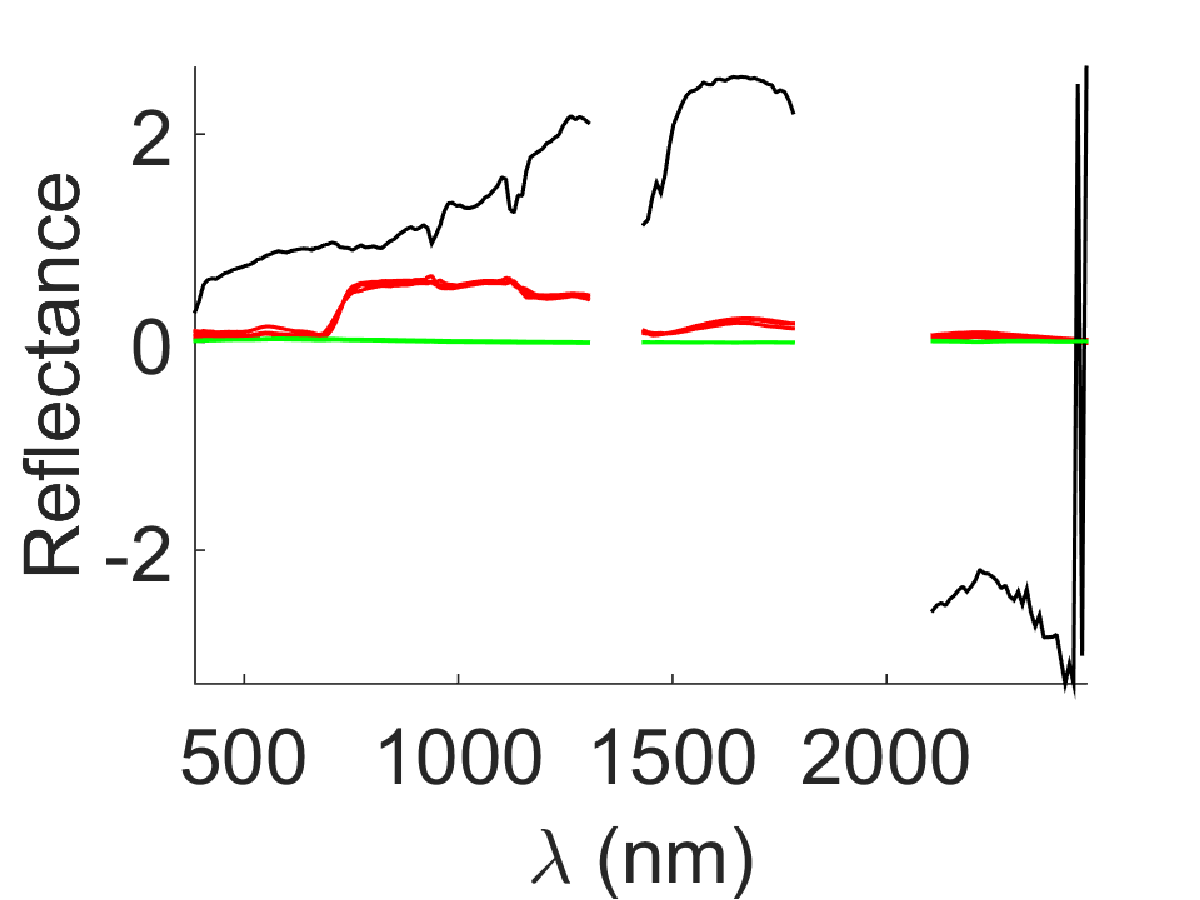}
\label{fig:rendm3_vca}}
}
\\
\resizebox{0.48\textwidth}{!}{%
\subfloat[1][Soil (SISAL)]{
\includegraphics[keepaspectratio,height=0.15\textheight , width=0.15\textwidth]{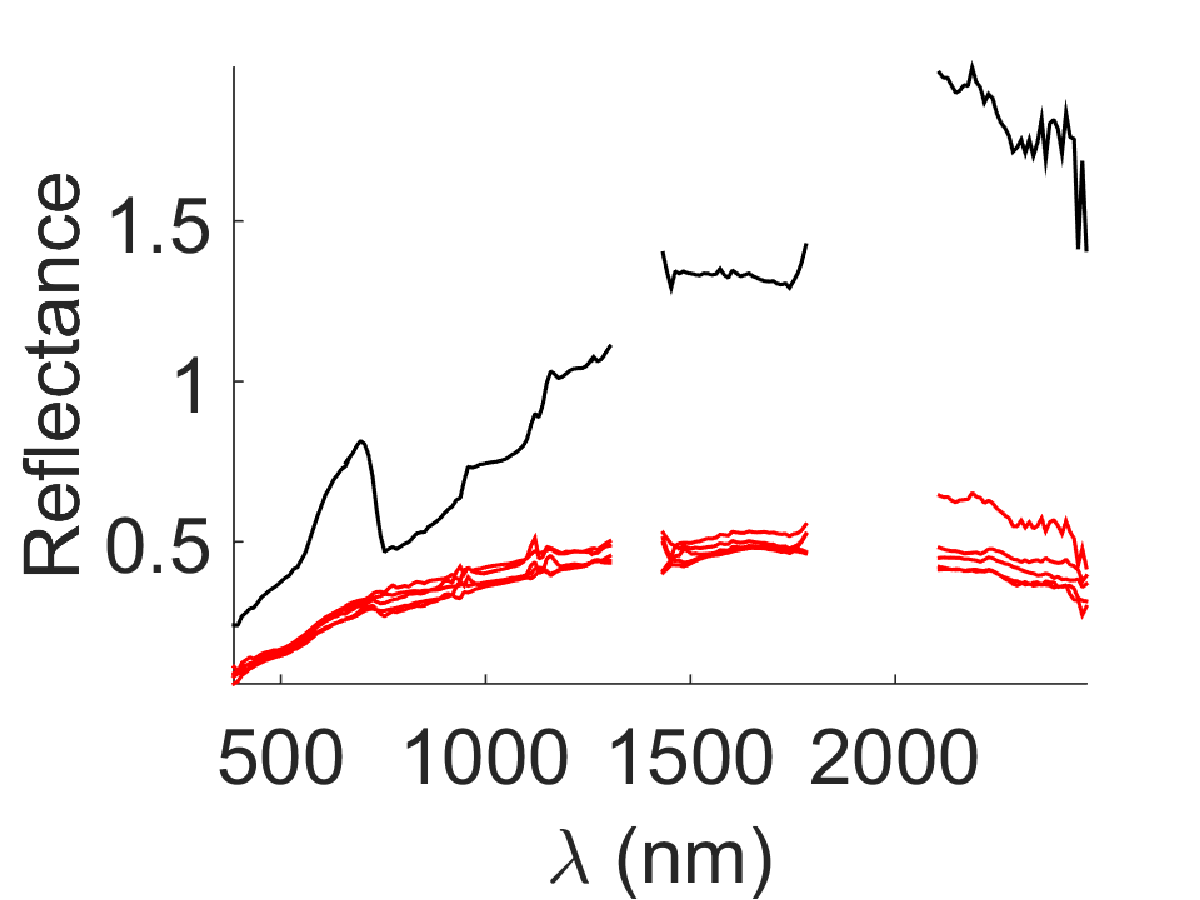}
\label{fig:rendm1_sisal}}
\subfloat[2][Water (SISAL)]{
\includegraphics[keepaspectratio,height=0.15\textheight , width=0.15\textwidth]{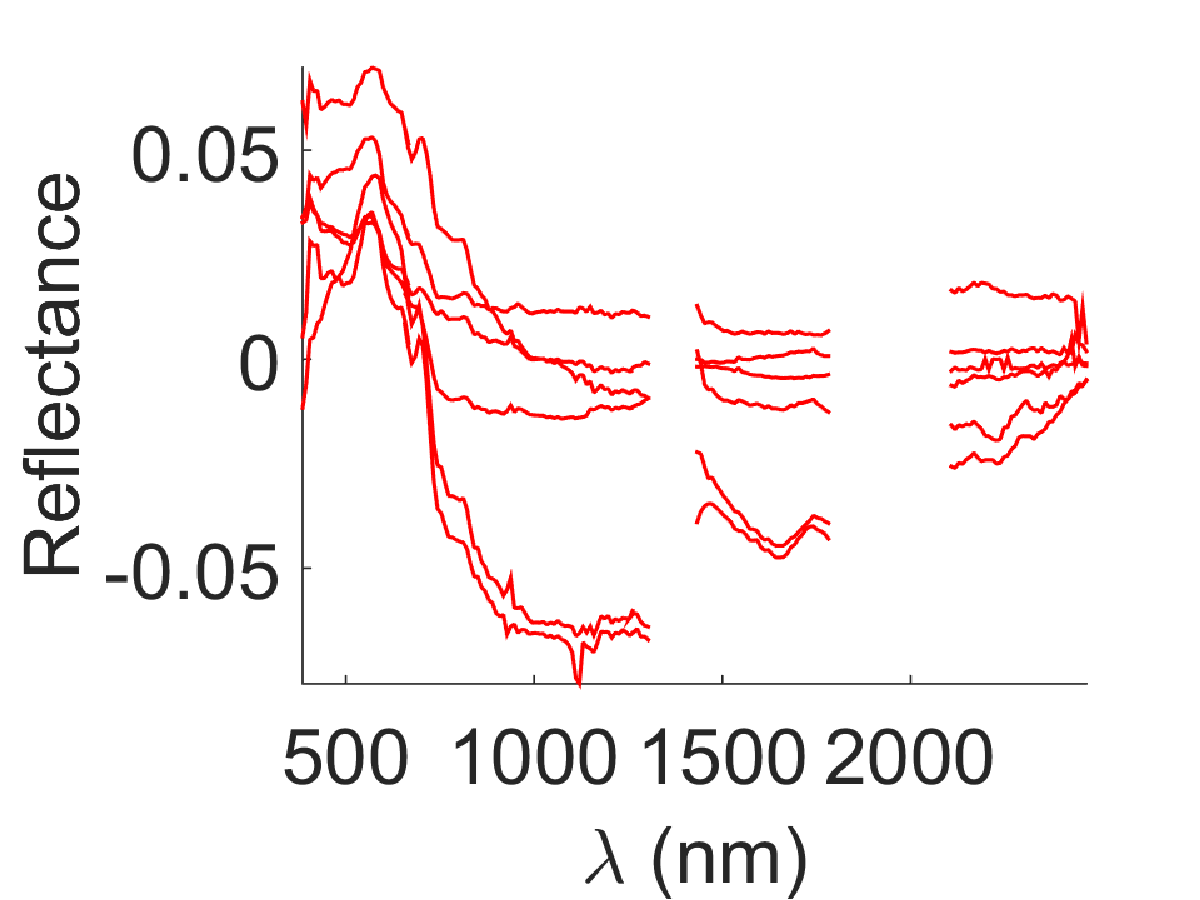}
\label{fig:rendm2_sisal}}
\subfloat[3][Veg. (SISAL)]{
\includegraphics[keepaspectratio,height=0.15\textheight , width=0.15\textwidth]{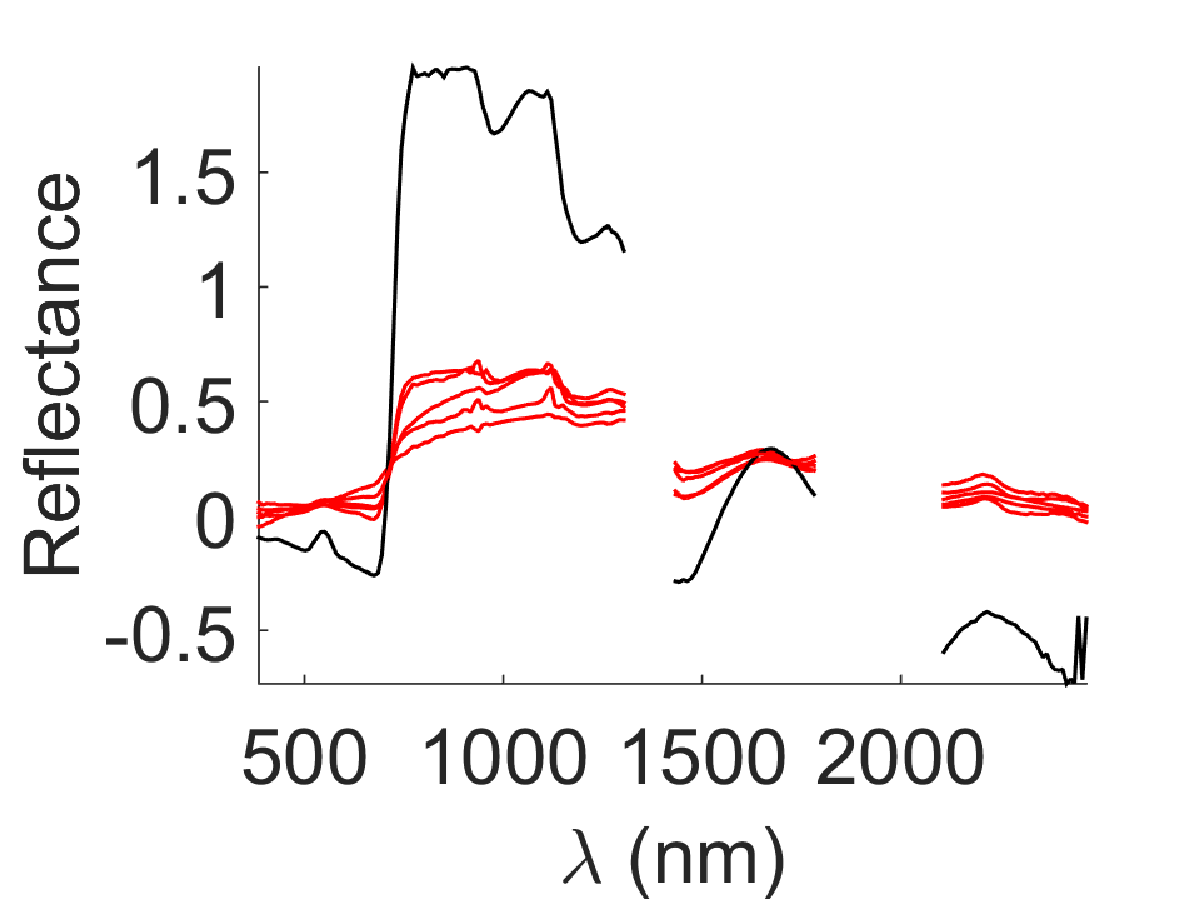}
\label{fig:rendm3_sisal}}
}
\\
\resizebox{0.48\textwidth}{!}{%
\subfloat[1][Soil (RLMM)]{
\includegraphics[keepaspectratio,height=0.15\textheight , width=0.15\textwidth]{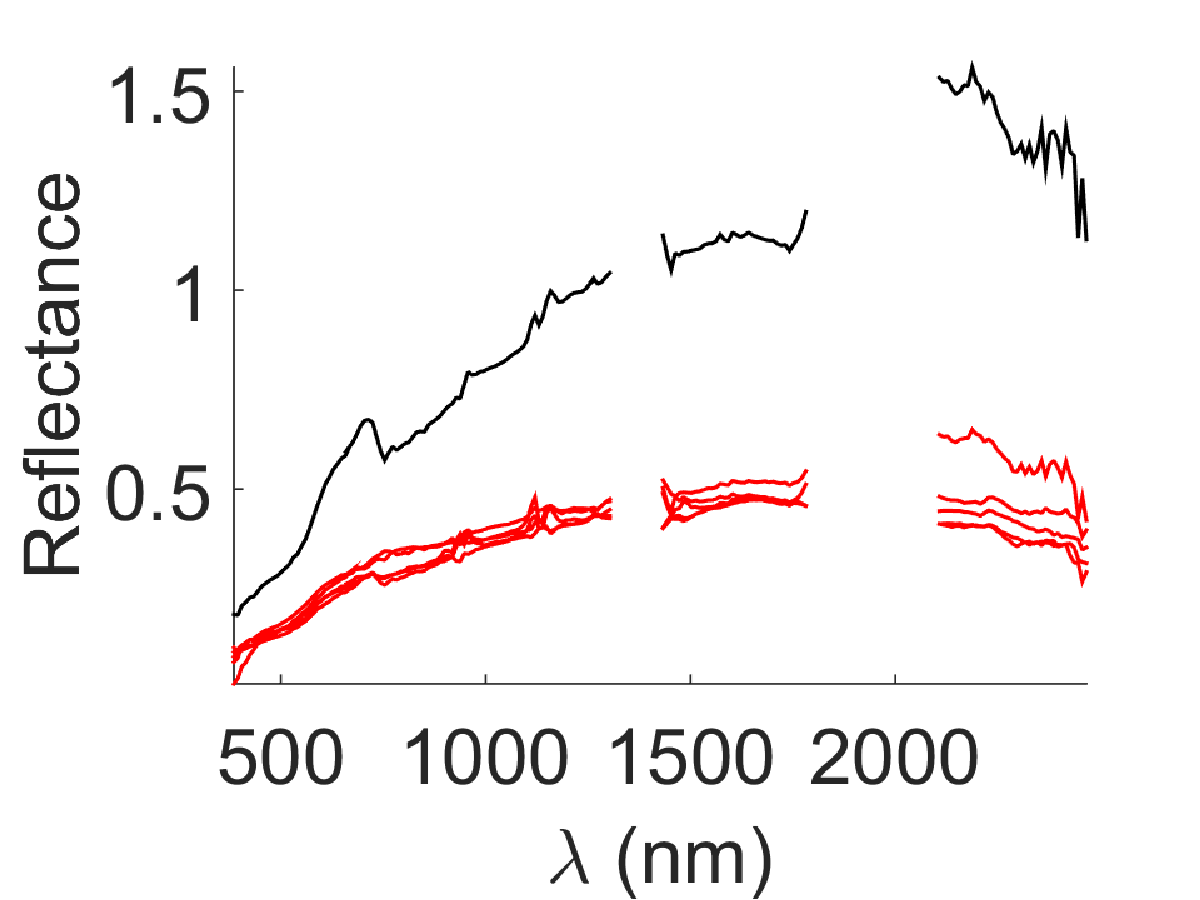}
\label{fig:rendm1_rlmm}}
\subfloat[2][Water (RLMM)]{
\includegraphics[keepaspectratio,height=0.15\textheight , width=0.15\textwidth]{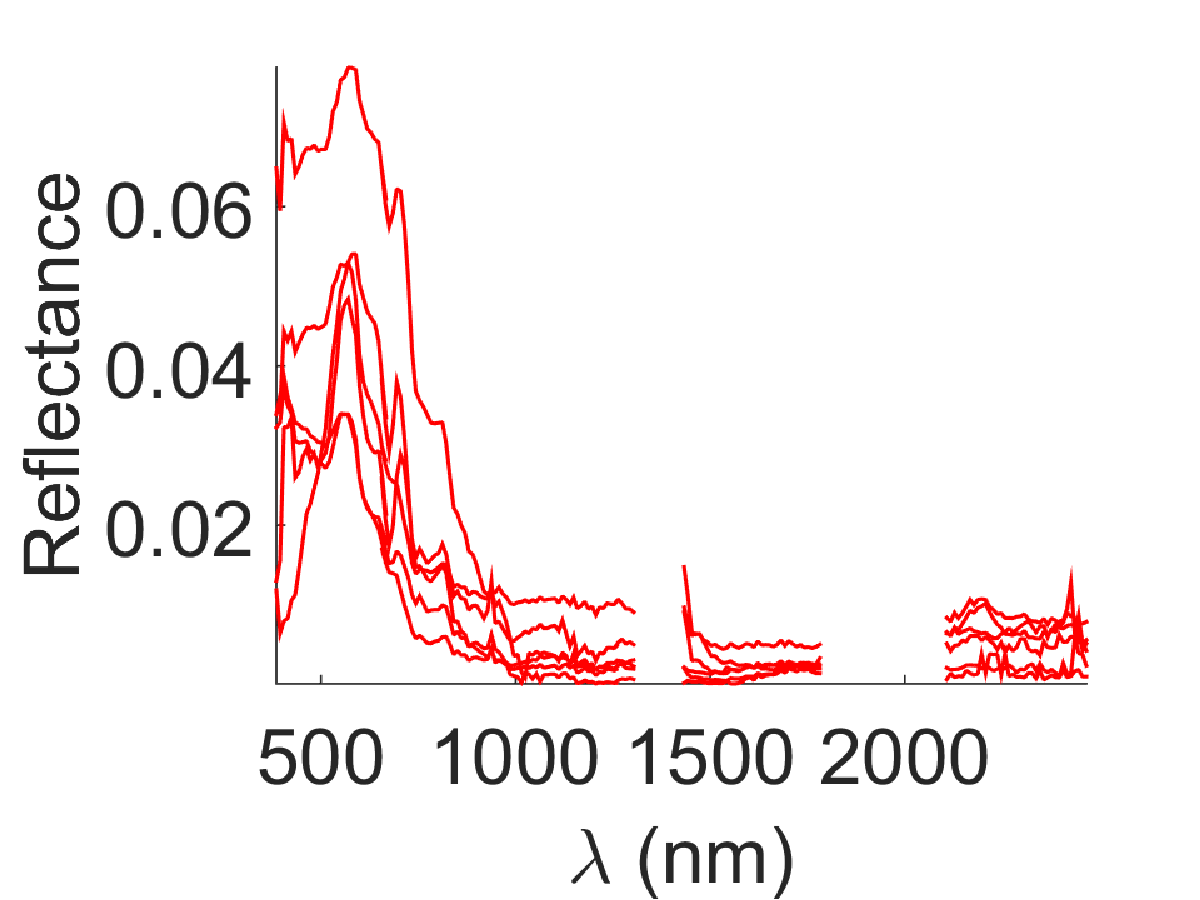}
\label{fig:rendm2_rlmm}}
\subfloat[3][Veg. (RLMM)]{
\includegraphics[keepaspectratio,height=0.15\textheight , width=0.15\textwidth]{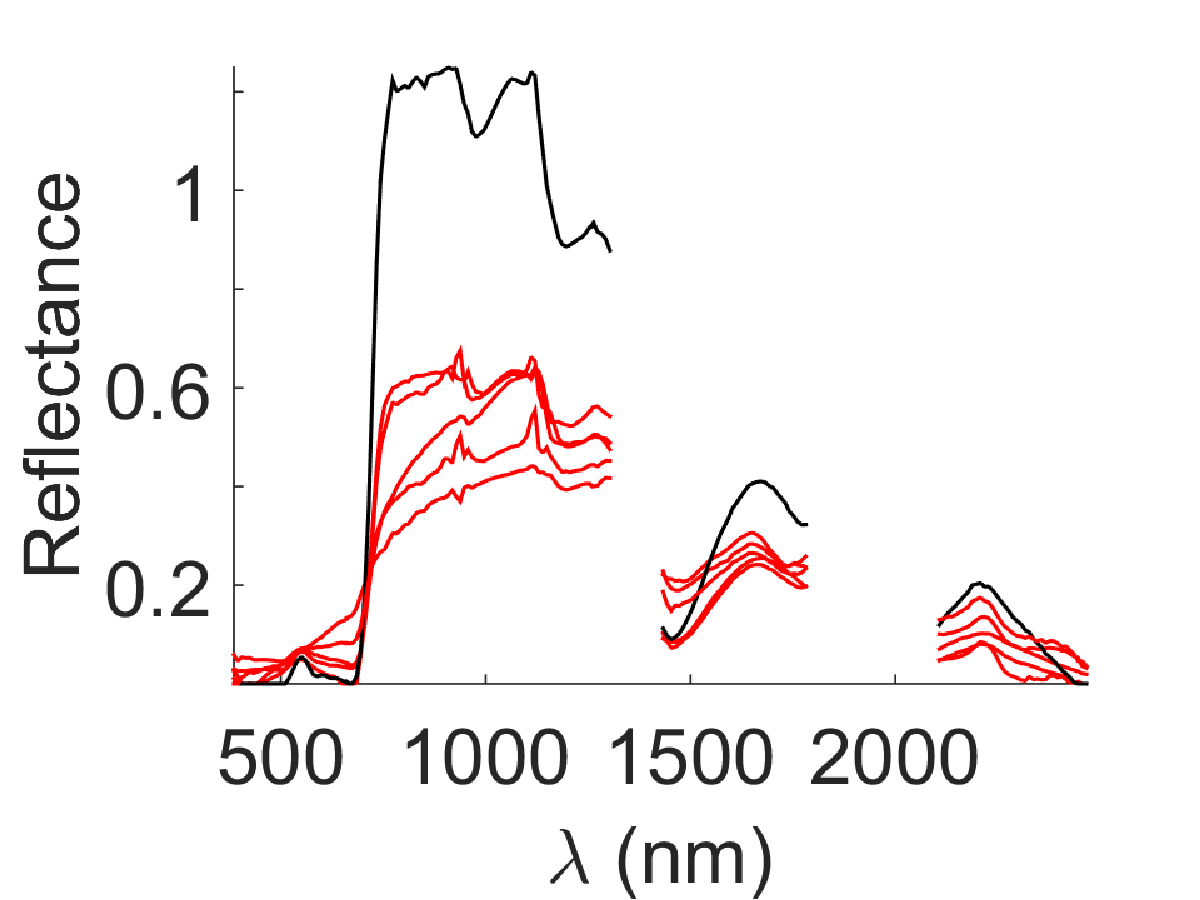}
\label{fig:rendm3_rlmm}}
}
\\
\resizebox{0.48\textwidth}{!}{%
\subfloat[1][Soil (OU)]{
\includegraphics[keepaspectratio,height=0.15\textheight , width=0.15\textwidth]{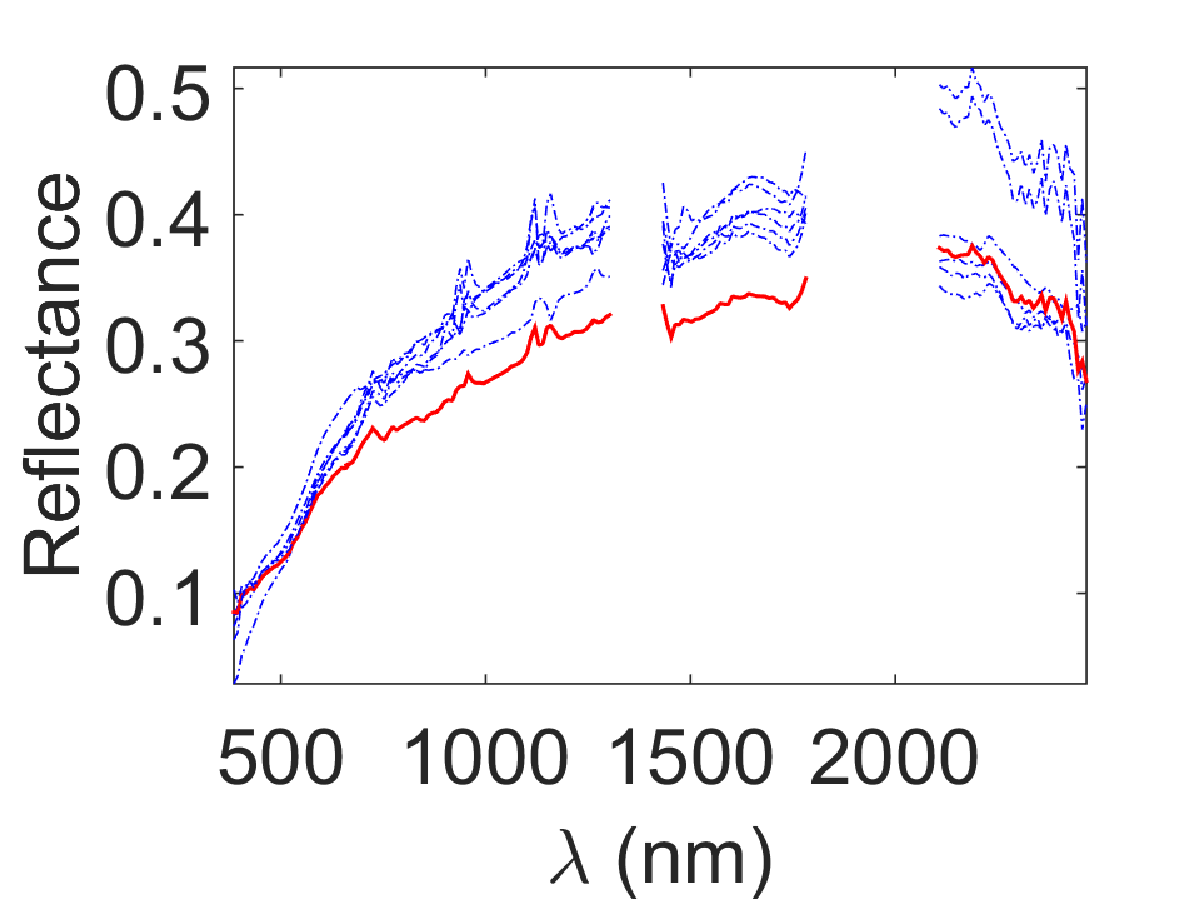}
\label{fig:rendm1_ou}}
\subfloat[2][Water (OU)]{
\includegraphics[keepaspectratio,height=0.15\textheight , width=0.15\textwidth]{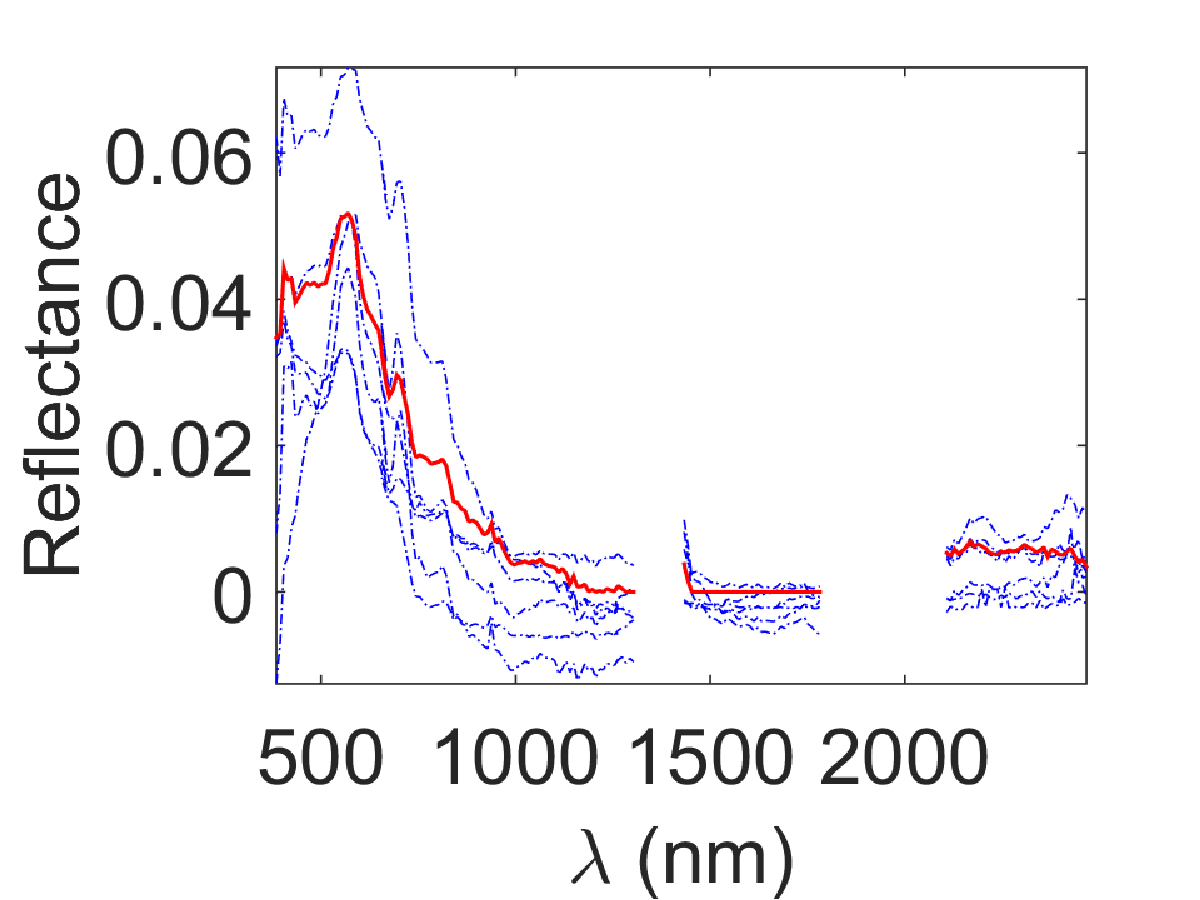}
\label{fig:rendm2_ou}}
\subfloat[3][Veg. (OU)]{
\includegraphics[keepaspectratio,height=0.15\textheight , width=0.15\textwidth]{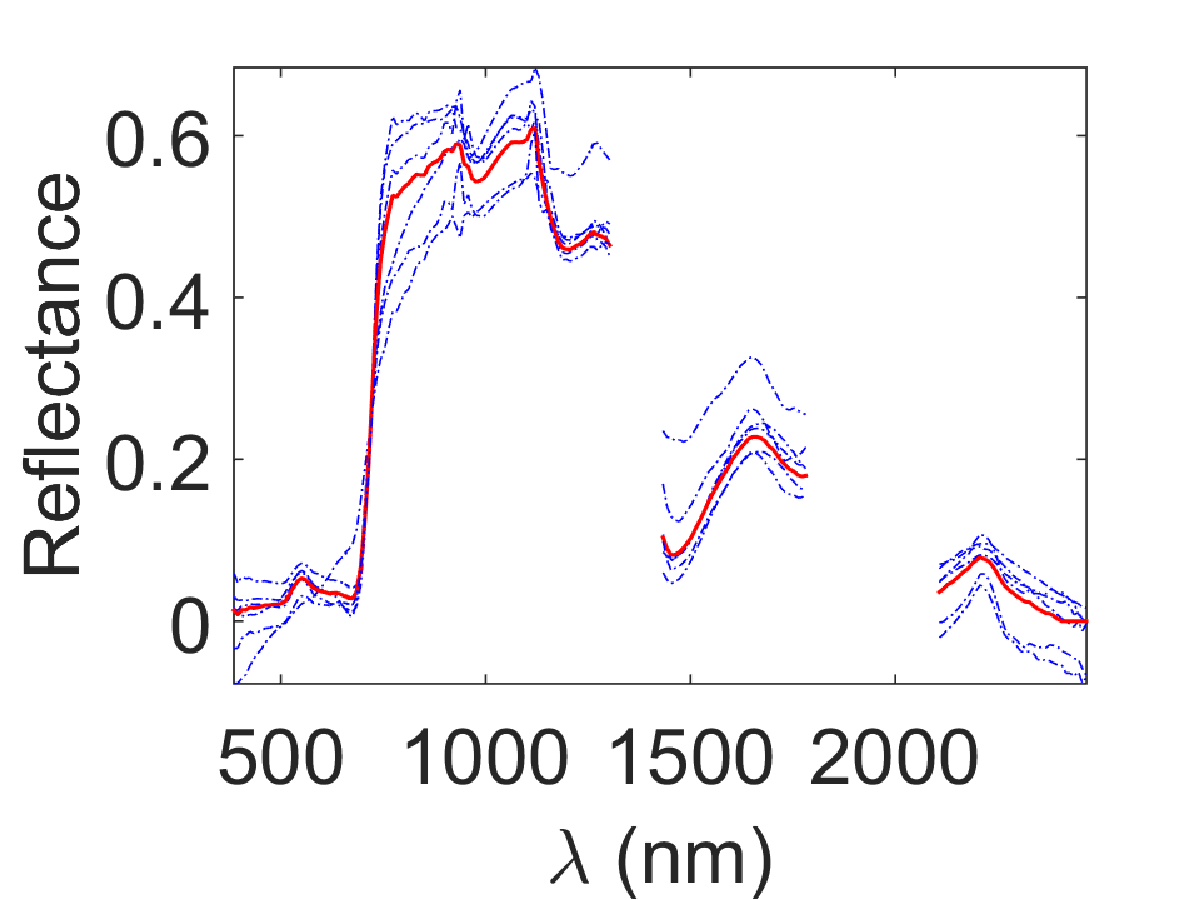}
\label{fig:rendm3_ou}}
}
\\
\resizebox{0.48\textwidth}{!}{%
\subfloat[1][Soil (Prop.)]{
\includegraphics[keepaspectratio,height=0.15\textheight , width=0.15\textwidth]{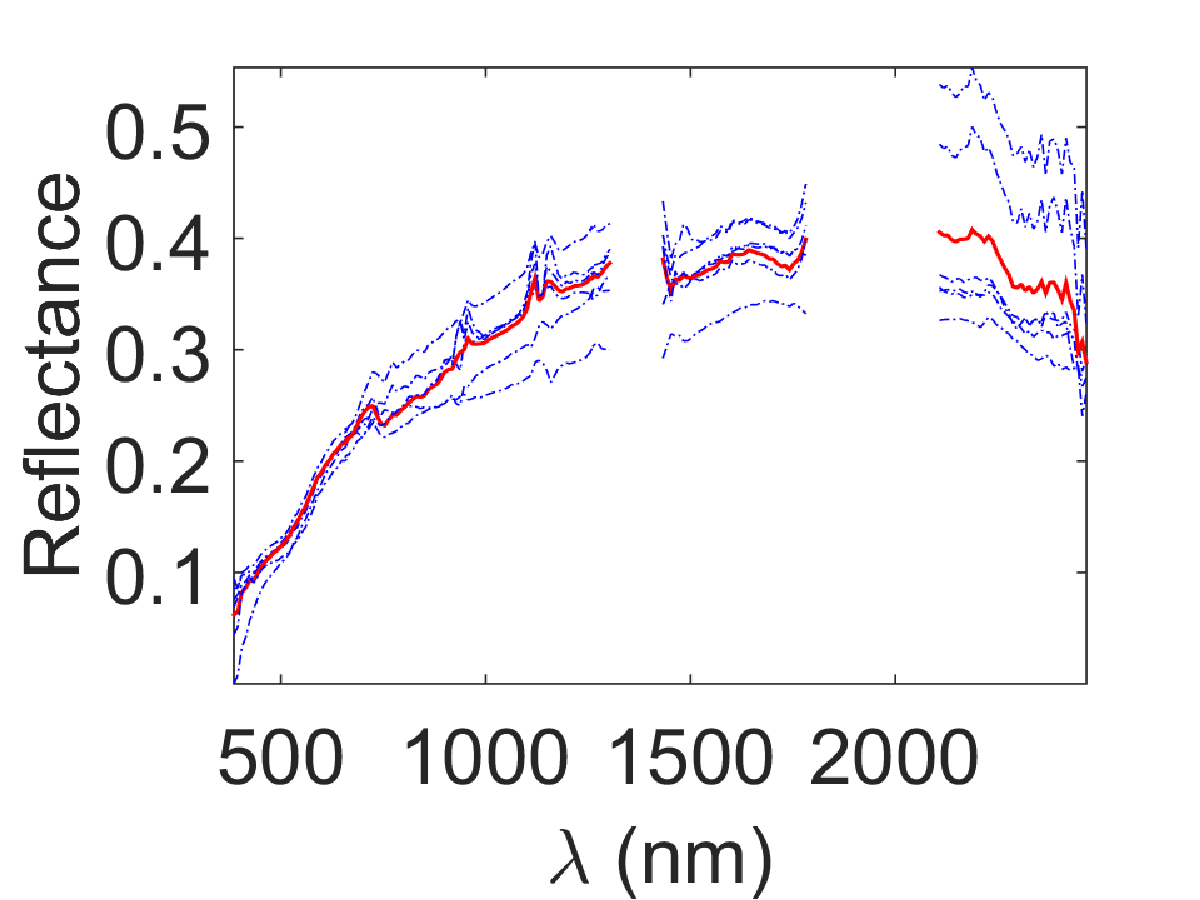}
\label{fig:rendm1_mcmc}}
\subfloat[2][Water (Prop.)]{
\includegraphics[keepaspectratio,height=0.15\textheight , width=0.15\textwidth]{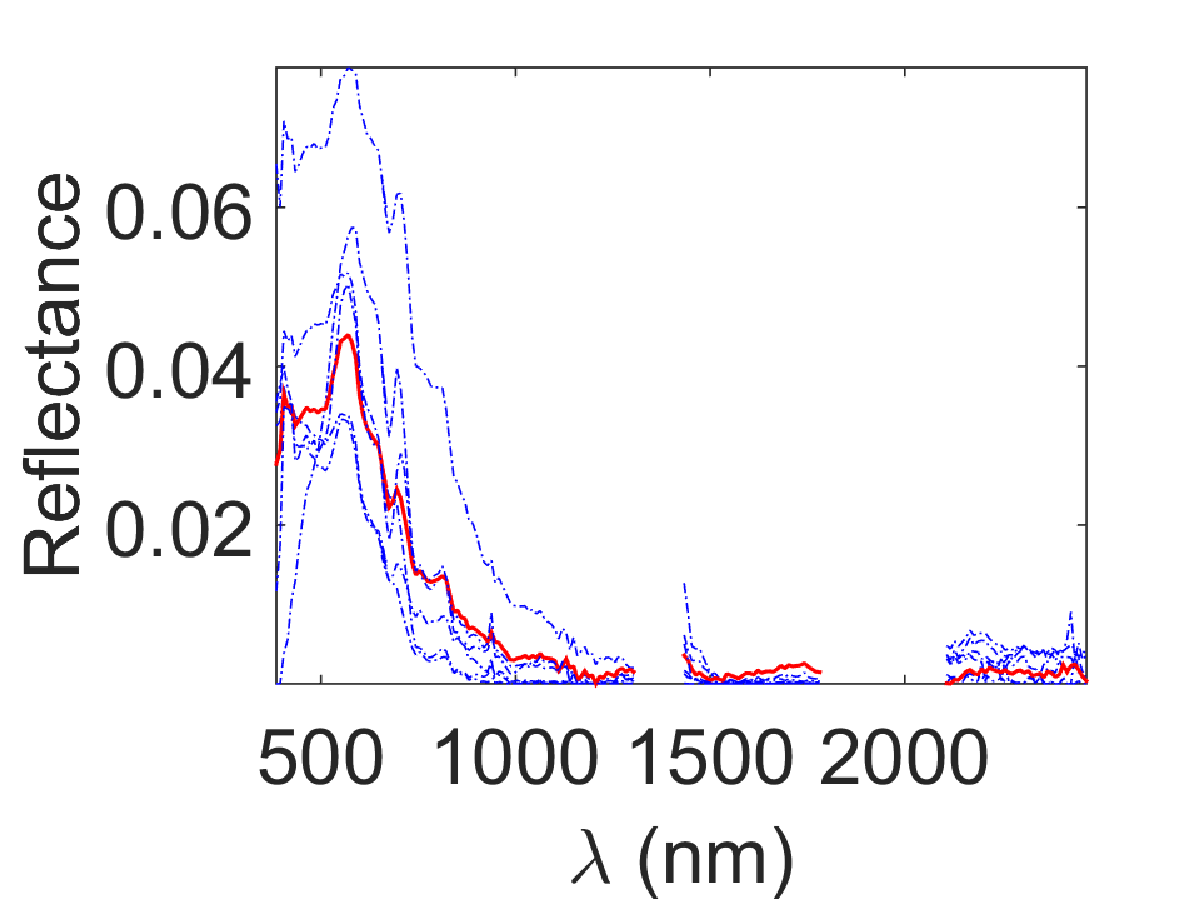}
\label{fig:rendm2_mcmc}}
\subfloat[3][Veg. (Prop.)]{
\includegraphics[keepaspectratio,height=0.15\textheight , width=0.15\textwidth]{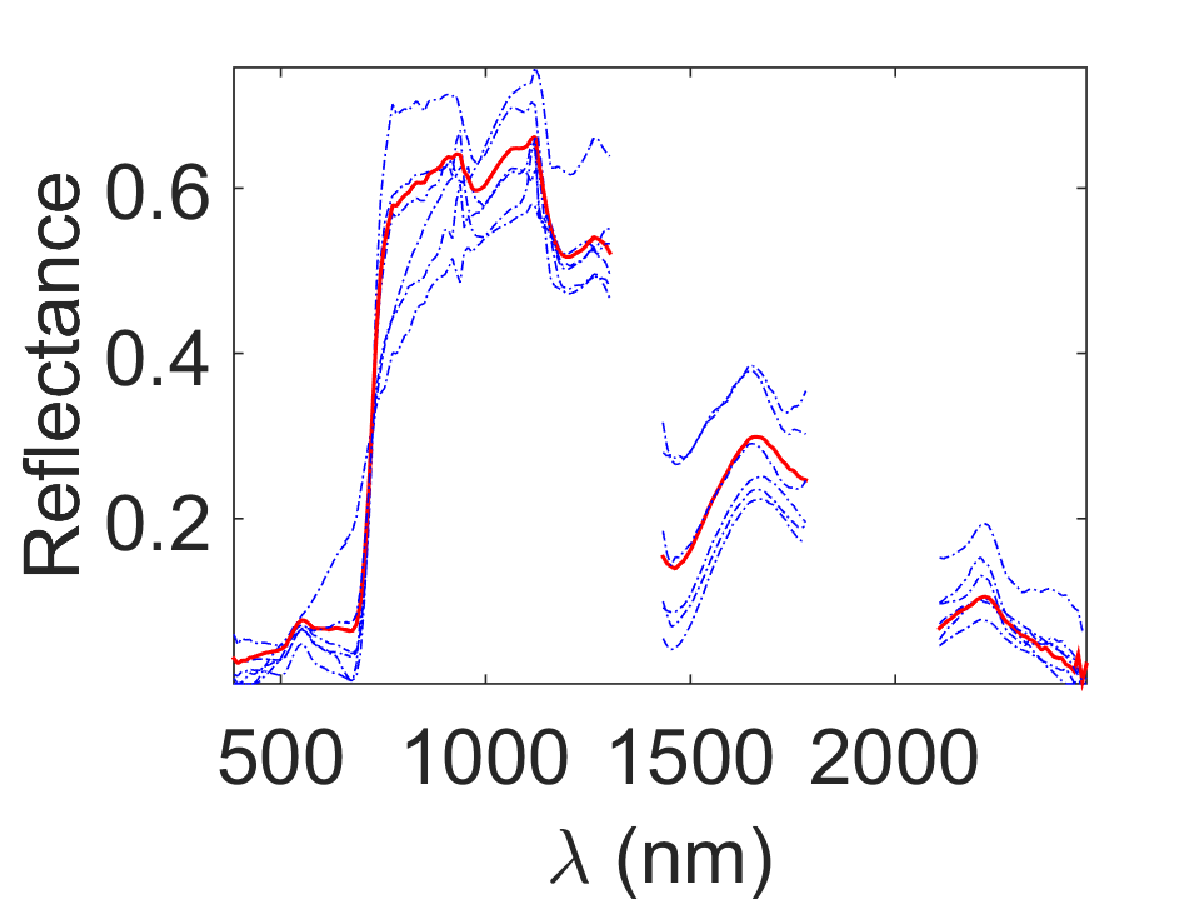}
\label{fig:rendm3_mcmc}}
}
\caption{Endmembers ($\m_r$, red lines) and their variants affected by variability ($\m_r + \dm_{r,t}$, blue dotted lines) recovered by the different methods from the real dataset depicted in Fig. \ref{fig:cube}. The spectral gaps in the recovered signatures correspond to the low SNR bands which have been removed prior to the unmixing procedure. Signatures corresponding to different time instants are represented in a single figure to better appreciate the variability recovered from the data. The spectra represented in black correspond to signatures corrupted by outliers, while those given in green represent endmembers which have been split into several components by the associated estimation procedure.}
\label{fig:real_endm}
\end{figure}

%% file: real_abundances.tex
\begin{figure}[t!]
\centering
\resizebox{0.48\textwidth}{!}{
\includegraphics[keepaspectratio,width=0.48\textwidth]{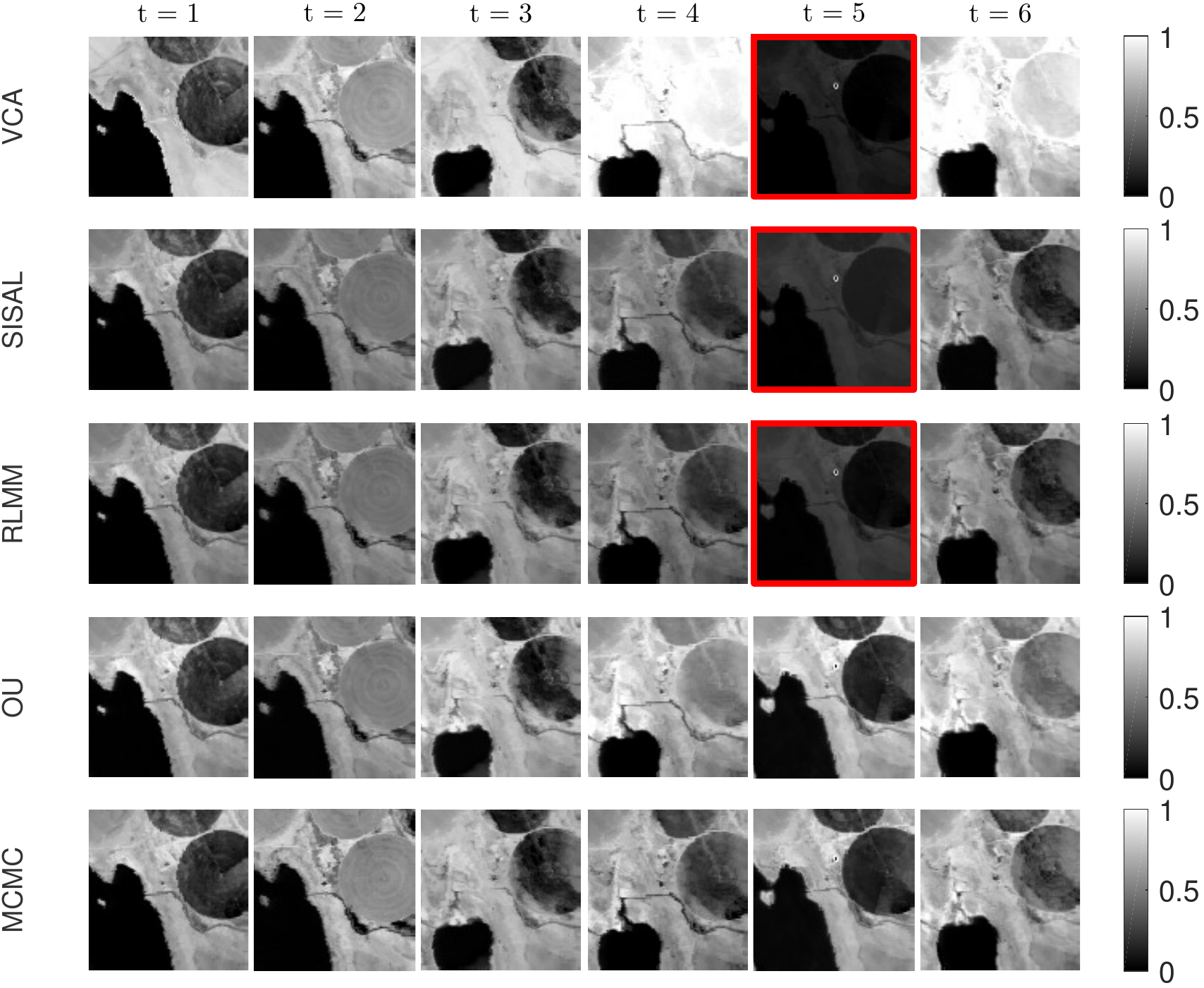}
}
\caption{Soil abundance map recovered by the different methods (in each row) at each time instant (given in column) for the experiment on the real dataset [the different rows correspond to VCA/FCLS, SISAL/FCLS, RLMM, OU, and the proposed method]. \pa{The images delineated in red suggest that some of the methods are particularly sensitive to the presence of outliers.}}
\label{fig:A1_real}
\end{figure}

\begin{figure}[t!]
\centering
\resizebox{0.48\textwidth}{!}{
\includegraphics[keepaspectratio,width=0.48\textwidth]{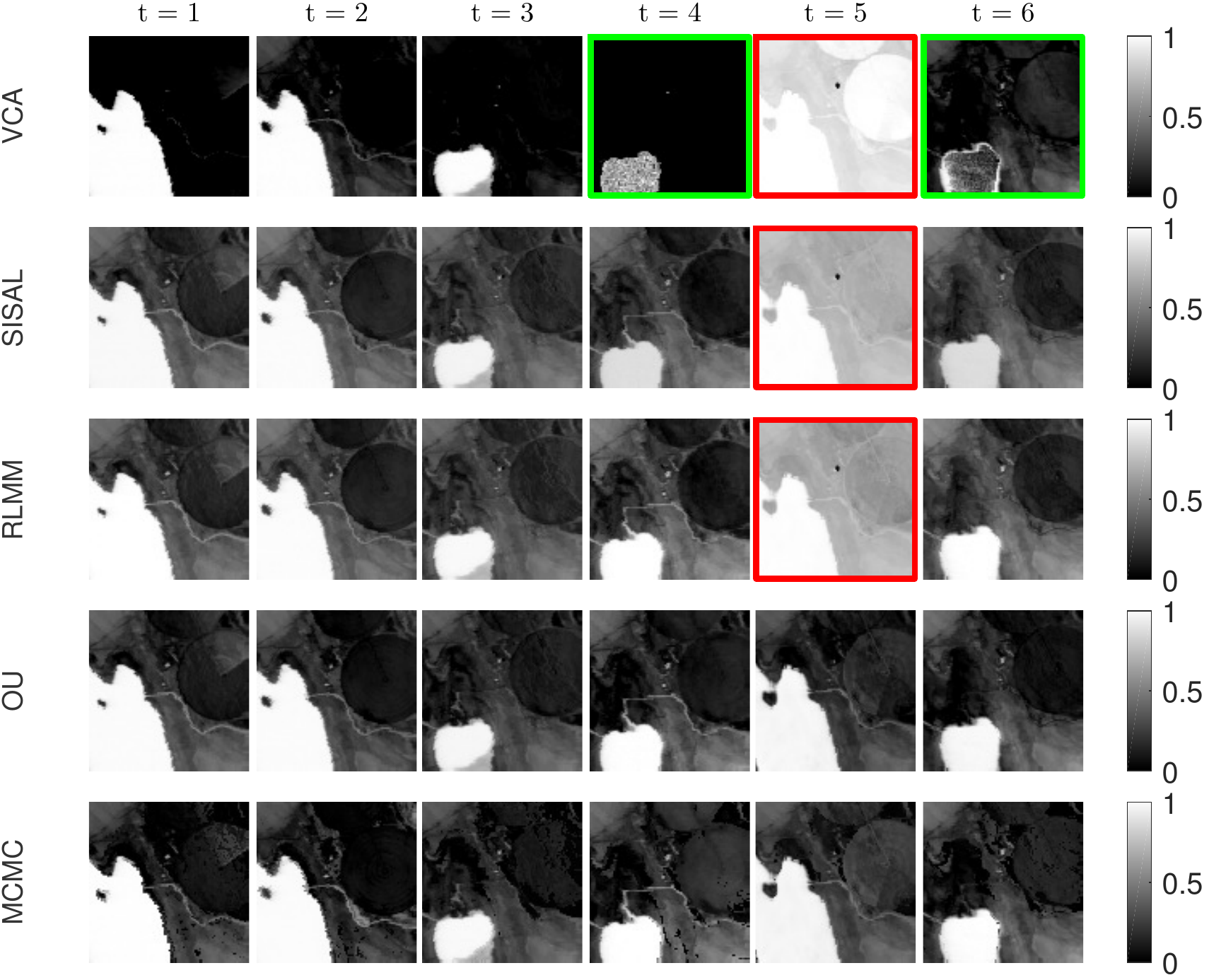}
}
\caption{Water abundance map recovered by the different methods (in each row) at each time instant (given in column) for the experiment on the real dataset [the different rows correspond to VCA/FCLS, SISAL/FCLS, RLMM, OU, and the proposed method]. \pa{On the one hand, the images delineated in red suggest that some of the methods are particularly sensitive to the presence of outliers. On the other hand, the images delineated in green represent the abundance maps associated with signatures which have been split into two components by the corresponding unmixing procedures.}}
\label{fig:A2_real}
\end{figure}

\begin{figure}[tbhp]
\centering
\resizebox{0.48\textwidth}{!}{
\includegraphics[keepaspectratio,width=0.48\textwidth]{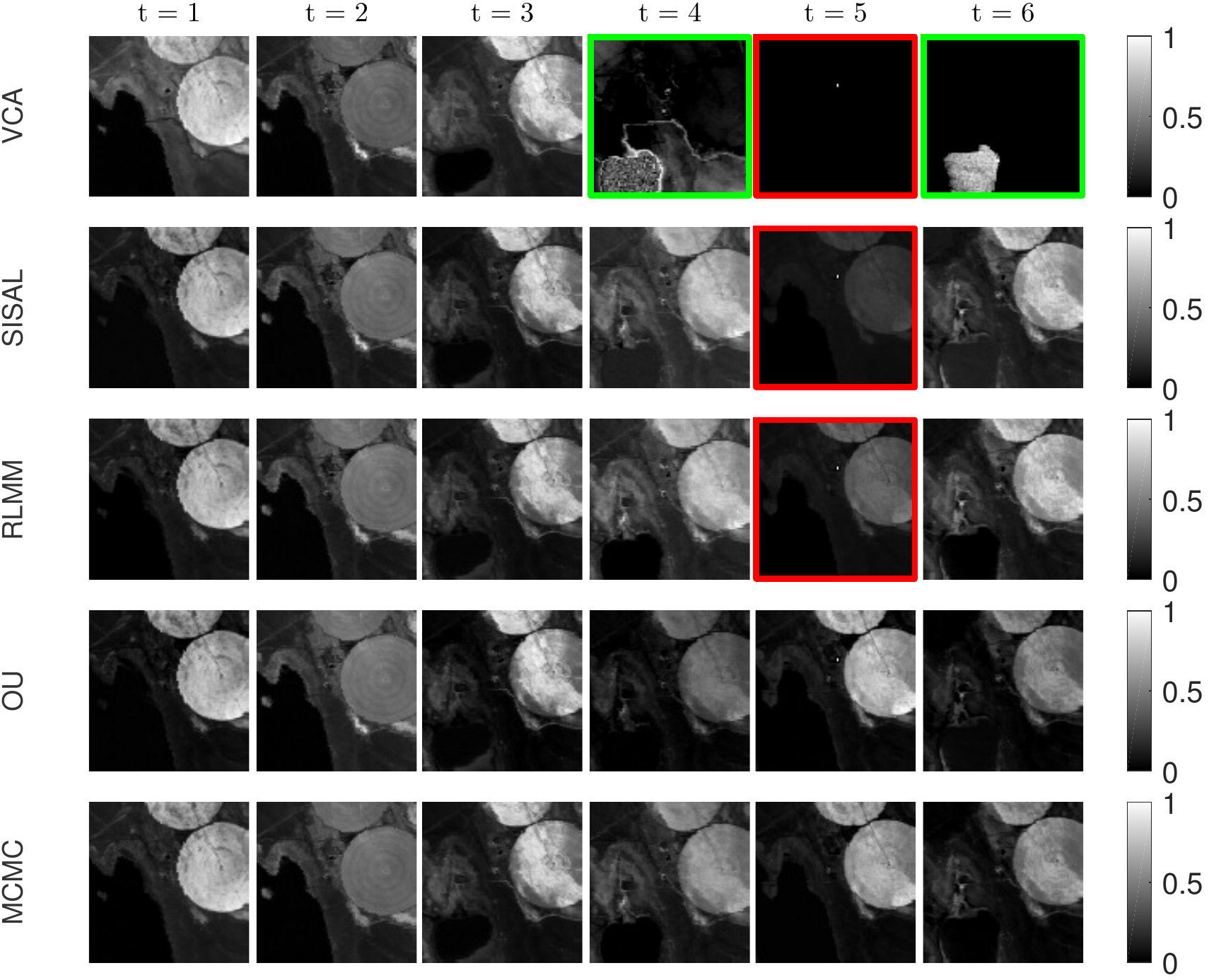}
}
\caption{Vegetation abundance map recovered by the different methods (in each row) at each time instant (given in column) for the experiment on the real dataset [the different rows correspond to VCA/FCLS, SISAL/FCLS, RLMM, OU, and the proposed method]. The images delineated in red suggest that some of the methods are particularly sensitive to the presence of outliers.}
\label{fig:A3_real}
\end{figure}